# Quantum Adiabatic Evolution for Global Optimization in Big Data

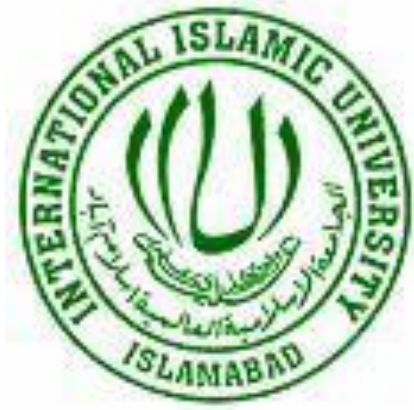

MS Thesis

**Sahil Imtiyaz**

Registration No. 882-FBAS/MSCS/F15

Supervisor:

Dr Jamal Abdul Nasir

Assistant Professor, DCS&SE, FBAS, IIUI

**Department of Computer Science and Software Engineering**

**Faculty of Basic and Applied Sciences**

**International Islamic University Islamabad**

**Sector H-10, Islamabad 44000, Pakistan**

**(2018)**

*A dissertation submitted to the
Department of Computer Science & Software Engineering,
International Islamic University, Islamabad as a partial fulfilment of the
requirements for the award of the degree of
Master of Science in Computer Science*

# Department of Computer Science and Software Engineering, International Islamic University Islamabad Pakistan

Date: 02-05-2018

# Final Approval

It is certified that we have examined the thesis report submitted by Mr. Sahil Imtiyaz, Registration No. 882-FBAS/MS (CS)/F15, and it is our judgment that this thesis is of sufficient standard to warrant its acceptance by the International Islamic University, Islamabad for the Master of Science in Computer Science I hereby endorse that the work done in this thesis is fully adequate to qualify, in scope and quality, for a dissertation of MS degree in Computer Science.

## Committee:

*External Examiner*

Dr. Waseem Shahzad
Assistant Professor
Department of Computer Science,
National University of Computer and Emerging Sciences
Islamabad

*Internal Examiner*

Dr. Muhammad Ahsan Qureshi
Assistant Professor
Department of Computer Science and Software Engineering
International Islamic University Islamabad

*Supervisor*

Dr. Jamal Abdul Nasir
Assistant Professor
Department of Computer Science and Software Engineering
International Islamic University, Islamabad

# Declaration

I hereby certify that this MS thesis entitled "**Quantum Adiabatic Evolution for Global Optimization in Big Data**" is my own work carried out at Department of Computer Science, International Islamic University Islamabad, Pakistan. Due credit has been given to the sources thereof. Further, no part of it, in any version, has been submitted at any Institute/University elsewhere.

<div align="right">**Sahil Imtiyaz**</div>

To my Parents who taught me how the clock of patience ticks and Professor Mario Rasetti for his guidance

*Everything and everyone is interconnected through an invisible web of stories. Whether we are aware of it or not, we are all in a silent conversation. Do no harm. Practice compassion. And do not gossip behind anyone's back—not even a seemingly innocent remark! The words that come out of our mouths do not vanish but are perpetually stored in infinite space, and they will come back to us in due time. One man's pain will hurt us all. One man's joy will make everyone smile.——Shams of Tabriz (Forty Rules of Love)*

# Acknowledgments

........And when (Solomon) saw it truly before him, he exclaimed:"This is [an outcome] of my Sustainers bounty, to test me as to whether I am grateful or ungrateful! However, he who is grateful [to God] is but grateful for his own good; and he who is ungrateful [should know that], verily, my Sustainers is self-sufficient, most generous in giving!"
——from the story of the Queen of Saba, Quran, 27:40

There is an association of a long journey that evolved gradually, passing through the different and undistinguished phases. The tradition of morality and philosophy of ethics revives me to pay sincere gratitude to all those who have been very instrumental so far in this untamed voyage of my thesis. I would like to thank my mother for taming my wild intellect to order during the darkness and failures and my father for having belief in my hard work. I am highly thankful to my supervisor Dr Jamal Abdul Nasir, IIUI Pakistan for supervising and guiding me in difficult times. I don't have words to thank and manifest in literal bounds of language the admiration for Prof Mario Rasetti, ISI Foundation in Italy for guiding me and continuously interacting on the evolution of my thoughts. I feel pleased to thank Dr Mir Faizal at University of Lethbridge, Alberta Canada for his academic suggestions. I am also thankful to my collaborator Dr Wail Mardini from Jordan University of Science and Technology for his valuable scientific thoughts. I would like to thank Charles Bennett IBM Fellow from Harvard University for suggesting me institutions for implementation of my idea. I am thankful to Jamia Millia Islamia, New Delhi, IIT Roorkee and Aligarh Muslim University for assisting me for carrying out simulations in the lab and its reductions. I am also thankful to Dr Qin Zhao at National University of Singapore for her encouraging and emotional support. Dr Andrei Kirilyuk at Institute of Metal Physics, Ukraine, Dr Kazuharu Bamba at Fukushima University, Japan, Junaid ul Haq at Jamia Millia Islamia, Muhammad Anas at IIT Kharagpur, and Syed Masood at IIUI for their suggestions and perspectives.

Lastly I am thankful once again to my mother and grandmothers for shaping instinctive and imaginative nature of my intellect via long continuous moulding and my father and my uncle for his support has helped much in continuation of my thesis work.

**Sahil Imtiyaz**

**Chapter 4**

**Sahil Imtiyaz**, Wail Mardini, Junaid ul Haq and Jamal Abdul Nasir, *Beyond Just Qubits: An Understanding of Information via Topological Field Theory of Data*, Emerging Trends in Advanced Computing and Technology February 2018 (Conference Paper)

**Chapter 6**

**Sahil Imtiyaz**, Jamal Abdul Nasir, Junaid ul Haq and Wail Mardini, *Implementation of Random Nature of Qubits for Random Number Generation via Simulations*, International Journal for Engineering Sciences, Vol 27 March 2018

**Chapter 4**

**Sahil Imtiyaz**, *"On Simulation of Big Data"* submitted to IEEE Access, submission no: Access-2018-03686 (**Under Review**)

**Chapter 3**

**Sahil Imtiyaz**, Syed Masood, Junaid ul Haq, Jamal Nasir, Reducing NP Hardness of Big Data via Quantum Matter submitted to European Physical Journal of Data Sciences submission no: EPDS-D-18-00095 (**Under Review**)

**Chapter 7**

**Sahil Imtiyaz,** *"Towards the Theory of Completeness of Brain: Insights to its mathematical Structure"* to International Journal of Theoretical Physics Submission no: IJTP-D-18-00349 (**Under Review**)

# List of Figures

# List of Graphs and Tables

# Abstract

Big Data is characterized by Volume, Velocity, Veracity and Complexity. The interaction between this huge data is complex with an associated 'free will' having dynamic and non linear nature. This hardness of big data has not yet been reduced to any physical in Hamiltonian formalism, neither any quantum formulation of big data has initiated. Additionally there is no specific simulator for optimization of this 'quantum space'. We reduced big data based on its characteristics, conceptually driven by quantum field theory and utilizing the physics of condensed matter theory. It is reduced to a complex non linear dynamic system in Hamiltonian formalism. The model is formulated from the dynamics and evolution of single datum, eventually defining the global properties and evolution of collective data space via action, partition function, green's propagators in almost polynomially solvable $O(n \log n)$ complexity. It is an initiation towards new understanding of information with epistemological shift. The simulation was carried out by mapping the dye laser parameters with the conditions of Quantum Adiabatic Evolution for global optimization. Energy and pulse width are mapped to the time-gap condition in terms of transition time, stability factor and adiabatic conditions. The algorithm was designed based on five main stages Reduction, Mapping, Evolution, Optimization and Simulation phase. The simulated results show that the time complexity of our algorithm for global optimization via quantum adiabatic evolution is almost $O(\log n)$. During simulation the excitation of rhodium atom is taken as a realizable qubit. Pertinently there is a quantum nature associated with the evolution and at some point it started decreasing and eventually got randomized in nature. As an inference, it is manifest that these information carriers are prone to noise effects and decoherence, pertinently decreasing the efficiency and quantumness of system. This behaviour was not seen in large time scale but in an infinitesimal time range between 0-1 fs, consequently affecting the tunnelling probability. Our algorithm first mines the space via greedy approach and makes a list of all ground state Hamiltonians, then utilizing the tunnelling property of quantum mechanics optimizes the algorithm unlike up hill and iterative techniques and doesn't let algorithm to get localized in local minima or sharp valley due to adiabatic evolution of the system. The loss in quantumness makes it inefficient by decreasing the tunnelling probability. Additionally, non realizable, no clone, noise, decoherence, splitting of energy states due to electric and magnetic fields, variant to perturbations and less life time of 23 microsecond makes it inefficient for practical implementation. As an application of 'reduction' it can be implemented in neural networks via mapping to quantum matter. Neural state mapped to fermionic states, neural superposition to fermionic braiding, synergetic order parameter to partition function, Spacio-temporal integration of neuronal signals to Feynman Schrödinger equation. The trend has changed from classical to quantum but due to unrealizable nature of qubits there must be a coherent form of physical information that can be well defined in fermionic system that is equivalent to qubits via Jordan Wigner Transformation. The inefficiencies of qubit can be well overcome via property that remains invariant to perturbation, Cartesian independence having well defined mathematical structure. It can be well addressed via topological field theory of data.

# Table of Contents

# Chapter 1

## 1.1 Introduction

*"But instinct is something which transcends knowledge. We have, undoubtedly, certain finer fibers that enable us to perceive truths when logical deduction, or any other wilful effort of the brain, is futile."——— Nikola Tesla*

In quest of unrevealing the seemingly gigantic, complex and inaccessible nature of Reality and to frame a model that can input our perceptions and compute efficiently the comprehensive picture of reality has inculcated and intrinsically moulded human mind in a vast and unending stretch of questions which has limited and exhausted the tireless efforts of philosophers. How does reality look like? Does it have any nature or form? Why perception-reality dualism? These seem little philosophical. Rephrasing the same basic notions of human curiosity in scientific formalism can be asked as; can we construct a universal Turing machine? Can we solve NP problems? What is the limit to computational capacities? Why experiments and measurements restrict and disorder the observation? Science based on experiments and observations is effectively probing these puzzles to define a Universal theory of everything and model a Universal Turing Machine for everything! The ultimate prototype of the unified picture that can defines every infinite and infinitesimal interaction in nature and also devising a computational model that can evolve the system comprehensively from any input to any output based on God's Algorithm is what will be a universal Computer.

Reducing the above seemingly more philosophical lines into a physical problem we state intuitively that there is some 'Grand algorithm' (Ultimate Reality) inaccessible to us in a 'black box' with six faces. Suppose this Grand algorithm is defined inside by two variables 'a' and 'b'. If we want to experiment what lies inside we need to look into it through its six faces (here). This formalism can give us six different definitions of this Grand Algorithms depending upon the variations in the two variables. These we call specific algorithms (Perceptions).In order to know this universal algorithm we integrate the all six different varying algorithms and resulting in a universal algorithm that will asymptotically be equal to the Grand Algorithm. The machine that we theoretically model is called Universal Turing machine that can computationally manipulate the data with the grand algorithm to desired results. This algorithm is expectely the Grand design of God playing a dice!

Computational Science is directly dependent on progression in electronics platform as technology is being implemented on it. Just like as prime as that of Charles Babbage mechanical prototype of computer known as difference engine that could be programmed by shifting levers and other mechanical parts. With the onset of electronic revolution from diode to transistors and from Bipolar Junction Transistors towards newer trends has revolutionized the computational speedup and efficiency as well. So as electronic industry progressed same happened to computational efficiency. This can be understood from the very similar framework of both platforms and that is storage of information in the form of bits, which can store in two possible states as 0 signifying OFF and 1 signifying ON. This is how a gate works and on higher hierarchy it's how a microprocessor works and in turns models the real functionality of how computers work in accordance with the electronic technological progression. With the advent of new challenges and increasing amount of data with huge volumes, complexities and interrelations there is nowadays challenges on the computational capacities of the classical computers based on electronic gates. Today's age is recognized as the age of Exabyte due the huge voluminous, complex, heterogeneous from multiple decentralized autonomous sources and is simply known as **'Big Data'**

## 1.2 End of Moore's Law

Gordon Moore postulated a law between the increase in the number of transistors and the time scale, keeping the chip area constant. He stated that every 12 months the number of transistors doubles imparting growth of computing power due to increase in transistor density. In meanwhile the time scale increased to first 24 months to 5 years! In his paper he has discussed the importance of integrated circuits in terms of reliability counts, two mil squares, cost and curves, heat dissipation, increase in yield and linear circuitry [1]. The trend has improvised to very large scale integration (VLSI) and embedding newer technologies like Field Programmable Gate Array (FPGA). In fact mid-2020s could bring monolithic 3D chips, where a single piece of silicon has multiple layers of components that are built up on a single die [2]. With the shrinking in size of the chip, current leaked and became hotter and since there is temperature dependence of semiconductor technology so it hampered the view of scaling up this view for future designs of chips [3]. Moreover, as the technology may proceed in scaling down there is a classical limit above which there is a paradigm shift towards Quantum effects [4]. As we may scale more down, imagine the size of an atom the quantum physics governs the system. At this level of operation the common sense of our conventional thinking ceases and things happen as per the 'common sense of an electron'! To quote a few Quantum tunnelling effects could be seen along with redefinition of the nature of operation in terms of waves and what not. If Imagine we reduced the size much more smaller than that of electron (although it can't happen so far) then the chip could feel the gravity strongly and it could sense one additional dimension! How amazing it would be like!

### 1.2.1 The Quantum Limit to Moore's Law and End of Theory

Thirty years ago George box and later stated at the Reilly Emerging Technology Conference this past March, Peter Norvig, Google's research director, offered an update to George Box's maxim: "All models are wrong, and increasingly you can succeed without them."[3] The era of finding model and inputting the data and getting result has now become old fashioned. As a ongoing progression started from the digital computers making information readable then Internet made it reachable. Then first search engine made it in a single database and now Google treats the massive data corpus as a laboratory for human conditions. In parallel to this progression, kilobytes were stored in floppy disks, megabytes in hard drives. Tera-byte in disk arrays and Peta-byte in cloud .At the Peta-byte scale, information is not a matter of simple three- and four-dimensional taxonomy and order but of dimensionally agnostic statistics. It calls for an entirely different approach, one that requires us to lose the tether of data as something that can be visualized in its totality. It forces us to view data mathematically first and establish a context for it later. For instance, Google conquered the advertising world with nothing more than applied mathematics and physics. In this scenario, correlations are more important than modelling the data. If we rely on very large scale integration for optimization there are following challenges [1, 2]

1. The localization of area of holding charge in silicon is due to the 'doping' impurities mixed with it. Smaller transistors still need to maintain the same charge, but increasing the concentration limits its electrical nature to non electrical clumps.

2. Gate controls the flow of electrons in chip but reductions of the size to smaller dimensions induce Quantum effects. Gates are simple on off mechanism but at quantum dominating level the electron due to its intrinsic wavy nature tunnels through the barrier of the gate in off conditions, known as tunnelling effects. So Quantum mechanics can render it useless. Packan says, "Quantum Mechanics is not like an ordinary manufacturing difficulty, we're running into the road block at the most fundamental level"

3. When size of the chip is bigger then there is a slight statistical fluctuations in the doping density but due to size it gets neglected but as size increase it induces many defects in the chip with unknown solutions.

4. The wider metal interconnections on one hand increases the 'transistor channel width' but the interconnect latency and energy dissipation dominates key metrics of transistor performance with energy dissipation being 5 times more than corresponding transistor.

5. Fundamental an unavoidable noise process, the thermal noise that breaks Moore's law. Noises due to switching of devices, excess thermal noise due to hot electrons and other noise phenomena (shot noise, 1/f noise).

The above constraint in designing a chip that can be used for the huge bulk of data seems unfeasible. In technological world impossible is simple rather it is unfeasible that is a problem! So we make use of mathematics and Quantum physics to find correlations to compute outputs. To conclude this we state that increasing density of transistors for high computations for Big data, the effects that constrain this dominate. So what is the solution to this impossible bound? Quantum Mechanics has a solution!

As conclusion to the above implications, we can quote in short that as the volume and complexity of data increases then the operations implemented on them in classical platform makes them infeasible. So there is a need of changing the definition, form and platform of data so that we can devise a robust machine that can make every operation feasible. By utilization of mathematical relations and inter-disciplinary ideas especially from quantum mechanics, we try to redefine the data formalism and then utilizing the quantum platform increases efficiency.

# 1.3 Big Data

With the expansion of data in an irreversibly accelerated manner has induced various characteristics to the data. As the volume, density and size of this data increased, its various aspects started getting more chaotic. This age is termed as age of Exabyte and the term denoting this is Big Data. The nature of big data can be characterized by HACE theorem, stating that Big Data starts with large-volume, heterogeneous, autonomous sources with distributed and decentralized control, and seeks to explore complex and evolving relationships among data [4] it denotes the versatility of the big data being huge and voluminous and ever expanding with heterogeneous and diverse dimensionality. The nature of data is heterogeneous in the manner that it can be defined in any format (video, audio and images) and each data point in the space has associated fields and attributes like if we have to define some person we add age, address and other fields associated with the concerned data element. This is included in the diverse dimensionality. Not only this, big data is produced from decentralized approach in which each node is autonomous in the sense that it can add any data irrespective of any constraint and its decentralized nature makes this more complex. Each data point in space has its own attributes and associated dimensionality but the data is huge and voluminous, this makes a interconnection between two data points at ground level and with all of the data points on higher levels of hierarchy. This forms the complex interrelationships of one data points with almost all of the data points. To derive a useful value of information from this complex interrelated heterogeneous and decentralized data is complex and hard problem in computer science.

As an example to the above fact Dr. Yan Mo won the 2012 Nobel Prize in Literature. This is probably the most controversial Nobel prize of this category, searching on Google with "Yan Mo Nobel Prize", we get 1,050,000 web pointers on the Internet (as of January 3, 2013). All the users all over the globe commented on this controversy, some being on his side and other not, adding this much of data in the space in different formats (image or videos and texts). This denotes the huge voluminous and heterogeneous data being added in the space. Now every user can comment from his local machine so it makes the process autonomous and decentralized in nature. Now since each comment has a relationship with all other comments in the way that it can be a reply to someone's viewpoints, hence deriving complex relationships between each data points. This depicts an intuitive way of perceiving the nature of big data.

# 1.3.1 Characteristics of Big Data

In order to understand big data in its essence and seek the challenges that are associated with it, big data has some characteristics that not only give a concrete picture of this but also the challenges. The main parameters include Data Velocity, Data Volume, Data Value, Data Veracity, Data Variety and Data Complexity. These are the main characteristics as well as challenging aspects of Big Data [4, 5]

Data Volume: It is the measure of the amount of data of some organization and need not be owned wholly by the organization rather can distribute it. As the volume of the data increases then the value decreases. The volume of big data is in Exabyte and it is correlated with heterogeneous nature and source and its diverse dimensionality. More volume data is outcome of the two paradigms defined as Heterogeneity and dimensionality. Peta-byte archives for remotely sensed imagery data, ever increasing volume of real time sensor observations and location-based social media data, vast amount of VGI data, etc., as well as continuous increase of these data, raise not only data storage issues but also a massive analysis issue (Dasgupta 2013).[5]

Data Velocity: It measures the speed of data creation, streaming, and aggregation of data. It may also include the rate of change of data values in the space. Imagery data with frequent revisits at high resolution, continuous streaming of sensor observations, Internet of Things (IoT), real-time GNSS trajectory and social media data all require matching the speed of data generation and the speed of data processing to meet demand (Dasgupta 2013).

Data Variety: Data variety is a measure of the richness of the data representation – text, images video, audio, etc. From an analytic perspective, it is probably the biggest obstacle to effectively using large volumes of data. Incompatible data formats, non-aligned data structures, and inconsistent data semantics represents significant challenges that can lead to analytic sprawl. Imagery data, geotagged text data, structured and unstructured data, raster and vector data, all these different types of data – many with complex structures – calls for more efficient models, structures, indexes and data management strategies and technologies, e.g., use of NoSQL.[4]

Data Value: It measures the usefulness of the data in making decisions. As we defined the complex and heterogeneous nature of big data so in order to evaluate the essence and meaning of the huge chaotic and unstructured data is the challenge the methodology of data mining and its newer techniques. Raw data without processing is raw not information and the value associated the information out of this huge big massive data.

Data Veracity: It measures the biases, noise and abnormality in the data. It addresses the question like, is the data that is being stored and mined meaningful to the problem being analyzed. Much of spatial big data are from unverified sources with low precision and accuracy, the level of accuracy varies depending on data sources, quality assessment of source data and how to "statistically" improve the quality of results.

Data Complexity: It measures the degree of huge interconnectedness and interdependence in big data. In this scenario, a very little change in few data points signifies a large aggregated change or even if it may be a small change that ripples across or through the system and substantially affects the behaviour, or not changing at all.

The other parameters that define characteristics and challenges of big data although subsidiary but are listed below:

Visualization: Imposing human thinking in big data analysis for identification and distinction of different patterns in data so that end users better grasp and communicate dominant patterns and relationships that emerge from big data analysis.

Visibility: The merge of cloud technology with that of big data is expected to resolve many constraints and threats such as data provenance. The big data will be geographically distributed in cloud with high computational capacities and storage.

# 1.4 Global Optimization

Informally we can define Optimization as an efficient output of a function with a specified objective and simultaneously subject to various constraints. The Objective defining function of the problem is the Objective function. Most of the daily decisions are based on the optimization problems in which we have to choose among the set of feasible solutions that gets formed from the specified Objective function and constraints of the problem. Formally An objective function f: X → Y with Y ⊆ R is a mathematical function which is subject to optimization. The set X defines the problem and Y is the set of feasible solutions to the problem [6]

Classification of Optimization Algorithms: Broadly there are two modes of classification of Optimization algorithms based on mode of operation it performs and on the basis of algorithmic properties (speed and accuracy) of algorithms. [7-10]

On the basis of former mode it is divided into two categories, one is deterministic algorithms and other being probabilistic algorithms. In first case there is a clear relation between the characteristics of the feasible solutions and the existence of utility of the given problem. These can be easily solved by simple search techniques like that of divide and conquer, dynamic programming and others. On the other side the probabilistic comes into work when the relation between a solution candidates formalism is too complicated and is not clear, or the data has high dimensionality, so it becomes harder to do a brute force or exhaustive search over this nature of data by the known techniques. Examples include Monte Carlo based approaches and heuristics can be utilized to find which of the solution to access next among the possible set of solutions like Simulated annealing, for example, decides which solution candidate to be evaluated next according to the Boltzmann probability factor of atom configurations of solidifying metal melts.

Other mode of analysis is on the basis of properties, in which the optimization is done as per their algorithmic structure and underlying principles and analyzes many factors including speed and accuracy. The two factors are complementary to each other and inversely related to each other so we need to trade off between the two depending upon the nature of our objectives. Pertinently, we can divide it in two categories one in which time matters and calculations and algorithms need to be faster and other is in which accuracy matters not the time, they being on line optimization and of-line optimization respectively.

Single Objective Function: When Optimization is based on getting single criteria either maxima or minima we term it as single objective function. The problem we are studying in our thesis is to find the minimum of a function in case of big data under various constraints utilizing the techniques of Quantum adiabatic evolution. So we have a single objective to analyze here too. [43]

# 1.5 Adiabatic Evolution

In order to construct a machine based on some ruling mechanism, it is important to have some fundamental functional units that are first theoretically designed and then implemented, so that they can modify the input to desired output. In digital electronics these are known as 'gates' that ought to manipulate the input for computations. In the same manner, Quantum Computers are being designed from quantum circuit or Quantum gates fundamentally. Turing machine is the basic theoretical model that gives design implications for the realization of a physical machine for computations. Deutsch proposed a model of universal or Quantum Turing Machine as a first theoretical prototype of Quantum Computers but it failed as it proved unwieldy in practical situations. So instead of this, the gate model was defined and implemented in the same manner as that of Digital electronic evolution. DiVincenzo proposed set of five requirements for physical implementation of the gate model of Quantum computers. They are listed as:

- A scalable physical system with well characterized qubits.
- The ability to initialize the state of the qubits to a simple fiducially state.
- Long relevant decoherence times, much longer than the gate operation time
- A "universal" set of quantum gates.
- A qubit-specific measurement capability.

The first DiVincenzo criterion calls for scalable, well characterized qubits, quantum analogue of a bit, it is simply a two-level quantum system, described in an arbitrary state can be described by

$$|\psi\rangle = \alpha |0\rangle + \beta |1\rangle$$

In the density matrix formulation, the state of a qubit can be expressed as follows;

$$\rho = I + r \cdot \sigma/2$$

Where r = (x, y, z) is the Bloch vector which describes a point in the Bloch sphere and σ = (σ x , σ y , σ z ) is a vector of the Pauli matrices. The Bloch sphere is a unit there are sphere which provides a convenient geometrical representation of a qubit's state space. It. a number of different candidates for physical realizations of qubits, e.g. single electrons, trapped ions, and superconducting circuit. The criterion for a "universal" set of quantum gates, this is a set of gates include be the set of the Hadamard gate, the π/8 phase rotation and CNOT gate. The third DiVincenzo criterion is generally considered to be the most difficult to meet experimentally. It involves finding a type of qubit that is sufficiently well isolated from noise sources in its environment to remain quantum coherent but is still readily controllable. They therefore require significantly larger physical systems. In this context, we introduce quantum adiabatic model for computation and is shown equivalent to gate model. It is believed to control the ineffective of gate model such as decoherence and noise factors.

## 1.5.1 Quantum Adiabatic Evolution as a Computational Tool

The main aspect of the quantum system is that it evolves via Schrödinger equation. The Hamiltonian is time dependent throughout the evolution. A quantum algorithm is run via this evolution and having the set initial state of the system. The nature of final state is governed by the "type of evolution", sometimes we need to evolve the system suddenly so as to reach the resulted output final state and vice versa. The time is chosen so as to encode the answer of the system. In designing the algorithm QAE has been conducted so that it may yield us the final state of the system. First the initial state is set to be the ground state and then the system is evolved slowly with time in such a manner so that it may always remain in the instantaneous state of the system. If the time taken to evolve is sufficient there will be least errors and then we may yield the final state. So the computational resources that shall be utilized via quantum adiabatic theorem will be least and hence this reduction may yield optimization. Hence it is utilized as a computational tool. [21]

# 1.5.2 Reduction of Satisfiability/Optimization Problem to Physical System using Quantum Adiabatic Evolution

Edward Farhi defined a new technique of understanding the optimization and Satisfiability problem in terms of a physical system and then reducing and evolving it using quantum adiabatic evolution. In case of Satisfiability problem is n bit instance having formula [22-24]

$$C_i = C_1 \cap C_2 \cap C_3 \ldots \ldots \cap C_n$$

Where $C_i$ is true or false depending on the values of all the bits of given instance. The algorithm is devised via quantum adiabatic evolution, first the initial state is specified and a time dependent Hamiltonian governs the state evolution as per the Schrödinger equation, taking form,

$$H(t) = H_{c_1}(t) + H_{c_2}(t) + \cdots H_{c_n}(t)$$

The H (t) is defined for time 0 to T and is evolved adiabatically.

# 1.6 Information meeting matter

In order to handle hefty data and computations in physics, certain numerical techniques and algorithms have been proposed that promise to be most suited models. The need for these techniques arises when conventional methods fail to provide any insight. The prominent areas of physics where we encounter these things might include condensed matter physics, scattering physics, particle experiments etc. The data so gotten is too big to be processed by conventional computers. The reason for this difficulty lies in the fact that in quantum-mechanical description of these processes, the Hilbert space grows exponentially large as soon as systems or particles involved in the process increases. This could even go beyond Avogadro's number of $10^{23}$ particles. Though we have a number of such statistical techniques to strike with, like Monte Carlo method, Renormalization Group, Density Functional Theory, Tensor Networks etc, but the suitable technique employed in strongly correlated system is that Dynamical Mean Field Theory(DMFT). Here we review the overall physics of strongly correlated electron system in a solid dealt with DMFT. In particular, we will discuss Green's function for such a system with regard to Hubbard or Anderson model which will serve as a guiding tool for determining the equation of motion and thus the time evolution of such system.

The behaviour of a huge collection of electrons as being quantum particles hopping off and unto the lattice with the massive immobile ions firmly embedded in it. The very characteristic features of such interacting systems dictate the electrical and magnetic properties of solids.

As evident from this preamble, the usual techniques of quantum mechanics in writing down the wave function of this huge collection of electrons and later to follow the time trajectory of this system is indeed a daunting task, keeping in view the various parameters involved thereof. So the conventional quantum mechanics seems obsolete to handle this physical system. People started formulating new recipes as approximations which aimed at reducing the complicacies of the problem to a significant level. Let's briefly follow timeline of these various approximations. Before that, we need to learn these concepts from the second quantization perspective.

## 1.6.1 Second Quantization Framework

Generally, solids are a thought of constituent electrons and ions behaving like a statistical ensemble where energy and interactions play a decisive role in determining its dynamics. Usually, the conduction electrons which are free to move within the lattice are considered to play significant role in defining the dynamics of the system while the inner core electrons are tightly bound to the ions and don't take part in it. How could such a system be governed by the moving electron and completely defining its comprehensive dynamics? Intuitively, we can imagine it as an electron at some lattice point when hops from one point to other, its amplitude or hamiltonian is already compromised by its inherent structure of this respective lattice, that is 'frozen electrons' influences inherently the hamiltonian and dynamics of the 'moving electrons' so as we can model whole of the dynamics by the later. The single particle quantum mechanics is constructed by considering a state vector in a Hilbert space H. The space for this single particle can't accommodate more than one particle. For many particle systems, we need to extend this space because the degrees of freedom increase and tend towards infinity with the increase in particle number. Still the complexity of the problem is not too complex for infinite non interacting particles that can be well modelled as tensor product of the wave functions of the individual particles as

$$\psi(r_1, r_2, \ldots . r_n) = \psi(r_1) \times \psi(r_2) \times \ldots \ldots \psi(r_n) \dots\dots\dots\dots\dots\dots\dots\dots\dots\dots\dots (1.1)$$

So the Schrödinger equation for such system given by

$$i\frac{h}{2\pi} \partial/\partial t \, \psi(r_1, r_2, \ldots . r_n) = H\, \psi(r_1, r_2, \ldots . r_n) \dots\dots\dots\dots\dots\dots\dots\dots (1.2)$$

What if we introduce the interaction between them? This is really a hard problem to deal with and makes it more complex. Hence wave function that corresponds to this system should take all of these parameters into account. If now we have a system that interacts but in a stochastic manner, that some of its particles interact and some don't, it is even more difficult insight to draw out a line between the dynamics of interacting and non interacting particles. As evident, solving Schrödinger equation for a system is extremely difficult. Hence, a reformulation is required. As well known, conventional quantum mechanics only quantizes the motion of the particles while treating the background electromagnetic field associated with charged particles classical. However, in the new scheme, the field is quantized in such a way that various excitation modes of a field correspond to the particles while rendering the original state vectors as field operators. In this new framework, an extended basis of Hilbert space called Fock space is used such that one takes care of a given number of particles occupy a particular state among the complete set single particle states. This scheme helps us to get rid of symmetrisation (antisymmetrization) requirement of total of bosons (fermionic) wave function. For example, in an N particle system, if $n_1$ particles are in state 1, $n_2$ particles in state 2, then the representational state corresponding to whole system would be like

$$|State> = |n_1, n_2, \ldots, n_j \ldots>$$

Where

$$\sum_j n_j = N$$

is the total number of particles. The above notation means there are $n_1$ particles are in state particles in state 1, $n_2$ are in state 2 and so on. Thus the occupation number operator $n_j$ has eigenstates $n_j$. Since spin consideration is very important in physics of correlated quantum particles, the value of $n_j$ is taken as

$$n_j = \begin{Bmatrix} 0,1 : For\ Fermions \\ 0,1,2,3,4\ldots : For\ Bosons \end{Bmatrix}$$

One important thing about the second quantization is that the state vectors turn into operators so that the algebra is a bit convenient. Hence the state vector in ladder algebra is linear superposition of ladder operators given by

$$\psi^\wedge(r) = \sum_i \emptyset_i(r) c_i$$

And

$$\psi^\wedge(r) = \sum_i \emptyset_i(r) c_i$$

From this formalism, it becomes simple to construct boson and fermionic system wave function and Hamiltonian. Here, as we are dealing with an electron system which is a fermionic system, the wave function character is undoubtedly antisymmetric. For such a system, the Hamiltonian reads as

$H = \sum_{ij} <i|H_0|j> c_i^* c_j^* + \sum_{ijkl} <ij|V|kl> c_i^* c_j^* c_k\ c_l$ ................................................ (**1.3**)

In a solid lattice, this corresponds to average energy of the electron hopping off and onto the lattice sites. Second term denotes the average interaction energy this electron experiences at the lattice site due to electron(s) there. Here $c^\dagger_i$ and $c_j$ are the electron creation and annihilation operator at the sites i. With these tools in hand, we are now in a position to describe the formalism of the electron dynamics in a solid which is responsible for studying the magnetic, insulating and superconducting properties of solids .

The earliest model viewed the electron as being nearly free moving inside the solid where the average potential due to all these ions is treated like perturbation. Electron motion and wave function analysis shows that band structure thus formed suffices to explain certain conducting properties of materials. Another model to approach the same problem is tight-binding model. Although both of these approximations lead to same results; however, free electron approach is most suited for metallic solids and tight binding describes insulators.

There are various generalizations to the above approximations for studying quantum many body systems. For example the Hubbard model for electron analyses in solids and thus the physical properties is a generalization of tight binding model with an extension to bosons (Bose-Hubbard model). The Hubbard interaction in the solids is modelled by a Hamiltonian which is usually expressed in terms of fermion creation and annihilation operators $c^\dagger$ and c as depicted. Let's first write down the interaction Hamiltonian for this interaction in view. It reads

$$H = \sum_{<ij>\sigma}(c_{j\sigma}^* c_{l\sigma} + c_{l\sigma}^* c_{j\sigma}) + U \sum_j n_{j\uparrow} n_{j\downarrow} - \mu \sum_j (n_{j\uparrow} + n_{j\downarrow}) \quad \text{...... (1.4)}$$

It is linked to the fact that electron destruction at the site j is equivalent to the creation of electron with spin s at site l. U is the interaction term only present if the electron meets another electron at the neighbouring site j which is typical local nature of our system correlation. The third term which involves the chemical potential m signifies the fact that the site filling is related to fermion statistical occupancy as dictated by Pauli's exclusion rule i.e., at a given time a site can be either zero occupancy, one or two with one up spin and one down spin. This Hamiltonian is the key to understand the collective stochastic (probabilistic) behaviour of this electron system i.e. the mechanism of electro dynamics on and off the site, the related conditions and the possible implications. It is often simple to analyse this system for one dimensional configuration. For greater spatial dimensions, including coordination number, time perspective, and quantum degrees of freedom, it becomes highly complicated to be dealt with.

This is the local attribute of self-energy. The recipe to gather information about the background lattice visa-a-visa its time evolution is provided by constructing a Greens function for the single site. This function bears a mapping to the bath and thus helps us to capture the whole dynamical picture of the lattice. One particle Green's function is actually the time-ordered average value of the electron creation and annihilation operators in second quantization formalism as given by

$$G(t, t') = -i < Tc(t)c^*(t') > \quad \text{............ (1.5)}$$

where T shows the time ordering of the electron dynamics here. That is the creation of electron at some time t and the subsequent annihilation at time $t^0$ is a probabilistic phenomenon with amplitude given the above relation

$$Green\ function\ = <state|\psi(x,t)\psi^*(x',t')|> \quad\quad (1.6)$$

This means the single site may or may not have an impurity at a given instant of time hinting at the lot of information being stored there. This subtlety refers to the quantum fluctuating nature at the site that will encode the whole time evolution dynamics of the lattice. It is worthwhile to mention here that once this Green's function is gotten, we try to construct it for the whole lattice configuration

Now the method to compute the bath Green's function is follows: Taking an initial value of self energy, we compute local Green's function. Hence bath Green's function can be obtained which is used later to compute the local action and then the partition function for this grand canonical ensemble and then again the local Green's function and then a new bath function. This iteration process goes on until we find bath function. This whole scheme outlined above could possibly provide us a great hint at big data analysis.

## 1.7 Problem Statement

Optimization algorithms are hard to device for big data due to huge volume and complex, non linear and dynamic interrelations with high dimensionality. It is almost impossible to traverse each data point in this space via iterative fashion or hill climbing and via comparison of each data points, finding global optimization. What if now if the space is almost approaching Avogadro's number! The configuration of data points in the space is dynamically changing with different number of minima. The algorithm gets often stuck in the local minima and hence impedes the global optimization. We utilize technique from quantum mechanics- quantum adiabatic theorem to solve this problem with big data. The system is reduced in terms of configuration with Hamiltonian associated; using greedy approach to find set of ground state Hamiltonian and then via tunnelling optimizes the algorithm to global optimization. The algorithm is a five stage structure including reduction phase, evolution phase, mapping phase, optimization phase and simulation phase. The time complexity is reduced to O (log n) depending upon input varying energy to pulse width ratio. It runs for almost every configuration of space till saturation of pulse width.

## 1.8 Research Questions discussed in Thesis

1. How to reduce Big Space of Data to some physical system?
2. How to conduct simulation of this space of big data?
3. How to map the parameters of the simulation for big data and quantum adiabatic evolution?
4. Transition has been from classical to quantum now what's next?
5. What are the possible applications of the reduced model?
6. What is the efficiency and time complexity of algorithm for global optimization and its comparison?

# 1.9 Overview of Thesis

We initiate a first step for designing a theory for big data based on its characteristics, conceptually driven by Quantum field theory at the interface of condensed matter physics. We formulate a new way of understanding Information being physical and a roadway where Information meets Quantum Condensed matter, expected to open new understandings in Machine Learning, Data Mining and Neural networks. The complexity of evolution of the data point in this 'trapped mess' is reduced to O (n log n). We have initialized a slight mathematical rigor for this space and neglected to go into mathematical derivations for different scenarios at this point. Ising, Spin Model and mean field has been already utilized for neural networks, Boltzmann machine and Machine learning. We start similar perspective for big data in the formalism of Hubbard model. With the new understanding of Information, the paper has strong future perspectives. Then we mapped the parameters and characteristics of big data with laser and simulated this system. First the information carrier was defined and it was simulated via rhodium atom, storing the quantum of information in the form of its excitation: a qubit and then the energy levels were studied and we mapped the parameters of dye laser: energy and pulse width with that of big data and quantum adiabatic evolution and then slowly varied the system by choosing the 'appropriate' gap as our ideal working area. After mapping we conducted our simulation for global optimization that decreased the time complexity to O (log n) unlike classical methods. There was a randomized behaviour seen in the dynamics of the qubits and it was concluded that qubits are not realizable for communication and there is a need to redefine the quanta of information that exhibit the properties that are not in qubits and w concluded that topology has the role to play. Topology remains invariant to perturbations and that there is a well defined mathematics for it. We at the end showed the application of our proposition in neural networks and random number generation via mapping and defined network in terms of quantum matter. Neural state mapped to fermionic states, neural superposition to fermionic braiding, Spacio-temporal integration to Feynman Schrödinger equation, synergetic order parameter to partition function and Hebb's learning rule to green's propagator. We gave insights towards the mathematical structure as an initiation for data sciences in terms of topology and category and algebraic topological theory.

# Chapter 2

# Literature Review

*"Happy Birthday Mrs Chown! Tell your son to stop trying to fill your head with science – for to fill your heart with love is enough", Richard P. Feynman*

## 2.1 Classical Methodologies

The main algorithms and techniques that are being used so far for big data are discussed as: [25]

Sharding and Consistent Hashing: Data is being divided between the machines, when we are dealing with big data and each division being a horizontal partition called as shard. This algorithm is based on hash function mapping each replicated data point to a point on an edge of a circle. These corresponding shards are assigned points on the same circle, responsible for each data point in the arc immediately clockwise adjacent to it. To make it more scalable and fault tolerant it is back up to shards farther from responsible arc enables to protect data loss. [26]

Bloom Filter: It is based on an array and hash function. When a key is passed through the search then its hash function is calculated and accordingly the corresponding index of array is set one. This means in big data analytics if we want to know if any of the key has been already searched then if the indices's are set to 1 then it means it has been searched else 0. There may be some false positives but then can be manipulated by altering the amount of hash functions and the length of the bit array. [27]

The Actor model: The local computations are performed by small subsystems called as actors and accordingly send messages to each other for synchronization. They don't share state, so we can have many actors without other requirements like mutexes, locks and other synchronization primitives. Prominent users of the actor framework are WhatsApp (using Erlang) and Twitter (using akka and scala). [28]

Map Reduce: This algorithm is based on two steps, first a Map function that maps data to (Key, Value) pairs and other being the reduce function; reducing a (Key, list of values) pairs to (key, result) pairs. These can be used for big data analytics for massive parallel computations. [29]

Sub linear algorithm: Another approach to dealing with big data is to deal only with parts of it, while keeping an eye on the error bounds. The data samples itself can be calculated and maintained before or done as a part in calculations. Some of the sampling methods include random sampling, Gibbs sampling and Metropolis-Hastings sampling. [30]

Single Instructions on multiple data: SIMD computations achieve parallelism by performing the same instruction on multiple data simultaneously. This can be made possible by hardware support and involves low level programming. A novel SIMD based system called MapD uses CPU/GPU hybrid computations that splits the query execution in such a way that it can take strengths of both. [31]

Storage Systems: With the advent of big data the storage became a challenge and with it other issues like backups have had to be maintained. More storage means high and sophisticated architectural designs and also fault tolerant platforms need to be emphasized. Brewer's Theorem: This theorem is also known as CAP theorem as it states that storage implementation could choose any two properties from Consistency, Availability and Partition tolerance, but not all three. [32]

The Big Data-verse Network is taken as read intensive system, where data sets are normally not deleted or altered, so one of the properties called consistency can be neglected specifically although we may rely on eventual consistency. Just as the case of Internet DNS system in which different servers may return different replies on same query but the overall consistency is maintained without adding much data overlays. Normalized schema avoid data duplication and preserve model consistency, but rely on table joins at every query time, Joins cannot be used over big data not only because of its big size but also due to its distributed nature. Big data storage systems are grouped under the term "NoSQL", but vary in almost every field: data model, performance, API, and so on. [33]

Key-Value Store: Project Voldemort: In this storage type, key value stores are large, distributed, persistent, fault tolerant hash tables. This pair is distributed among all nodes using consistent hashing over the key. The data associated with the key is opaque and cannot be altered but only retrieved. The original key-value store named as Dynamo was first developed by Amazon, now it's used by LinkedIn, eBay and Mendeley. [34]

Wide-Column Store: Cassandra: It's a column based NoSQL database, able to store huge data. It also stores Key Value pairs and each row can have almost 2 billion columns and can be dynamically changed as well. It was developed by Facebook, based on Google big table and Amazon's Dynamo. It uses hashing with bloom search for failed cases.

Relational DB Sharding: It has at least one of its tables split between different machines. It is early form of big data storage till now.

Document Store: MongoDB: It is NoSQL database using document model to store data in JSON format with key value pairs. It supports Map Reduce, text search and data Sharding. It's used today by New York Times, eBay and more. [35]

Graph Database: Neo4j: It stores data in the form of graph with vertex and directed edges and can handle big data via simpler implementations.

Polyglot Persistence: All Together Now: In this storage model all the above discussed models are implemented in big data analytics. It's actually knowing the part of CAP theorem and simultaneously choosing the type of storage system it can best suit for simpler implementation.

For example, the Connection project keeps large images, and their meta-data. A possible combination would be storing the meta-data in a wide-column store, such as Cassandra, and storing the image data itself in a key-value store, such as Project Voldemort, preserving the functionality, but allow smaller rows in Cassandra, boosting its cache performance, and possibly improving the system performance, even though two systems have to be queried. If, instead of Cassandra, we would have chosen MongoDB, such split would be a must; MongoDB documents cannot exceed 16MB [36]

The huge amount of Big data is not a big problem rather the curse of high dimensionality is what makes it more complex and many techniques have had been devised to reduce this high dimensional data into low dimension without loss in information. Methods like linear mapping methods, principle component analysis (PCA), manifold learning techniques such as Isomap, locally linear embedding, Hessian LLE, Laplacian eigenmaps and LTSA [A] Recently, a generative deep networks, called auto encoder [B], perform very well as non-linear dimensionality reduction. Random projection in dimensionality reduction also has been well-developed [37].

Hadoop: It is one of the most well established software platforms that support data intensive distributed applications via Map- Reduce paradigm. It mainly consist of three major things that include Hadoop kernel, Map/Reduce and Hadoop distributed file system. It is in base a programming model for execution and processing if large amount of Big Data sets being pioneered by Google and developed by Yahoo. Map Reduce uses divide and conquer and recursively breaks the complex problem into sub problems till it is solved directly. In terms of Hadoop infrastructure there are basically two types of nodes one being master node which divides the problem complex in nature into subproblems and assigns it to worker nodes so that they perform possible computation and then it is finally aggregated to yield final results. [38]

The other software and programming tools include Dryad for high performance and distributed execution engine, Apache Mahout for machine learning algorithms in business and proves to have good mature platform. Jasper-soft BI suit for business intelligence with cost effectiveness and self service, Pentaho business analytics for business analytics with robust, scalable and flexible to knowledge discovery. Skytree server for machine learning and advanced analytics, processing massive data with high speed, Tableau , karma sphere studio and analyst for Big Data workspace having collaborative and standard based unconstrained analytics and self service. [39-42]

Hadoop is designed for processing large amount of data in parallel by portioning mechanism performing batch processing making it multipurpose but not real time and high performance engine. For these applications stream processing for real time analytics was introduced. Streams of big data are high volume with high speed and complex relationships having high dimensionality and to tackle this many platforms such as SQL stream storm and stream cloud have been developed.

## 2.1.1 Challenges of Big Data Analytics

On one side it becomes manifest that the nature of big data is very complex and ever expanding on an accelerated rate but on other side of the analysis, more data brings in more contouring and defining of the concerned area. So mining this big data and prediction of the nature of the concerned domain is the main task ahead. While handling big data there are many difficulties inherited and propagated due to the nature and characteristics of Big Data like Data capturing, searching, sharing, analysis and visualization. It is like gold ore but yet to get refined! [44-46]

The efficiency and speed of CPU is increasing every 18 months but on contrary the disk I/O speed lags behind as the rotational speed is least improved on comparative basis. Due to this imbalance it becomes difficult to hand data and more specifically when information is huge but the information processing tools and methods are still in their formative phase. Due to this the architecture are unable to solve real time problems although many parallel processors are being designed with many techniques to avoid data, control and resource dependencies and deploying many methods like replication and shelving to make the architecture more scalable and efficient but it does not suffice when we are dealing with Big data. In relation to this, a related problem is that of inconsistency, incompleteness, timeliness, scalability and data security. The characteristic of big data of having Variety in analysis of unstructured and semi structured data seems to be a big challenge in dealing Big data. The value of big data in making out decision out of noisy channels and other factors seems also to be an impending factor in Big data analytics. Although many data pre-processing techniques have had been formulated like data cleaning, data reduction., data transformation and data integration but still it needs more computational resources and is not easy for increasing bulk of data.[48]

The other related challenge is the storage problem of this huge voluminous Big data that is being created in every sector including information-sensing mobile devices, aerial sensory technologies, remote sensing, software logs, cameras, microphones, radio-frequency identification readers, wireless sensor networks, and so on. There are 2:5 quintillion bytes of data created every day. Conventional storage media have been designed including solid state drive (SSD), Hard disk drives (HDD) and other technologies like Phase change memory (PCM). Other technologies like Direct-attached storage (DAS), network-attached storage (NAS), and storage area network (SAN) are the enterprise storage architectures that were used for storage of data but all these existing storage architectures have limitations when it comes to large-scale distributed systems.[49] Data transmission problem is also one of the limitations when it comes to big data as we have a limited bandwidth so for routing mechanisms we need to devise new methods so that there is no congestion in network and enhances scalability. A related challenge is the data curation that is storage and retrieval of data from this huge ever expanding and accelerated data and it is not easy to derive a value out of this huge pit. It's like the NP hard problem of finding a hay stick in a sack of hay sticks and the amount is increasing with changing environmental factors as well. Moreover it becomes difficult to visualize data in the form of graphs and tables as there are complex relationships and high dimensionality of data so it becomes difficult and impractical. [50]

## 2.1.2 Three Tier Models of Big Data

The voluminous big data with complex relations and higher dimensionality needs to be analyzed in an efficient hierarchical order, with at each level of this order, almost every computation and techniques are performed so that to deal with this data. Big data is being studied under three tier model that gives us an in depth insights about the working of how to deal with it. This model is divided into three tiers which are as: [52-54]

Tier I: Big Data Mining Platform: This provides a platform for data access and computation procedures. Big data it is ever expanding and complex so it needs high computational resources so that we can be able to deal with it efficiently. For data mining of big data it needs to get input to some processor and needs to be loaded in some memory space but it is nearly infeasible to load and process this big data. Cloud computing and distributed systems have had provided an efficient platform for this very tier but to move data from one system to other is often expensive and data privacy issues are also to be addressed. [55]

Tier II: Big Data Semantics and Application Knowledge: This tier of big data deals with two main issues, one being the privacy and information sharing and other is Application Knowledge. There are two main aspects of data security. One is to restrict the access to the user's data such as banking certificates and pin and other is anonymize the user in this big data so that it becomes difficult to pin point the specified user. First is carried out by sending securely the certifications and other is carried out by adding randomness to user data. The common approaches include using suppression, generalization, perturbation, and permutation to generate an altered version of the data, which is, in fact, some uncertain data.

The application knowledge provides essential information for designing big data mining algorithms and systems. This can help in accomplishing the specified objectives of any organization. For example we have a market and it evolves and changes every second in price and other related aspects. There may be many metrics and deciding parameters that can control this change. This tier of domain discovery helps in prediction of the status of the market in next few minutes and along with knowing the rule controlling these deciding metrics

Tier III: Big Data Mining Algorithms: In this tier there are mainly three aspects studies around one single operation of mining out the data efficiently:

Local Learning and Model Fusion for Multiple Information Sources: In big data analytics the system is distributed and the operations are performed locally which gives a biased output or decision. Moreover we can't correlate whole of the data as it is cost demanding and nearly infeasible. So we try to optimize it by locally searching and globally correlating the data set, hence yielding approximately correct decisions.[56]

Mining from Sparse, Uncertain, and Incomplete Data: Spare, uncertain, and incomplete data are defining features for Big Data applications. It is sparse so it has only few points to draw a conclusion. Uncertain data are a special type of data reality where each data field is no longer deterministic but is subject to some random/error distributions. Moreover the incompleteness in data makes it more volatile to bias.

Mining Complex and Dynamic Data: Big Data is complex in nature with interrelationships so data mining algorithm needs to analyze this issue and with it, its dynamism to change makes it weirder to be dealt. Algorithm forms the highest level of this tier to draw out conclusion efficiently. [57]

## 2.2 Global Optimization Algorithm

Global optimization algorithms are optimization algorithms that employ measures that prevent convergence to local optima and increase the probability of finding global optima. There are many features or hierarchical steps that are being implemented on the GOP so that we can get our best possible solution of the desired problem. The problem of optimization is that of a search problem of finding an element of minimum magnitude among the data set. Traditionally each data point is compared with its previous point and the best one with minimum value is selected and iteratively all the data points are being fetched and compared and results are drawn out with global minimum data point. For an Optimization problem the following points are to be valued:

- The success of optimization depends very much on the way the search is conducted.
- It also depends on the time (or the number of iterations) the optimizer allowed to use.
- Possibility of prevention to get stuck at some local optima.

The valued points are conclusionary analysis of the basic factors of optimization algorithms, based on which we can seek the efficiency of the technique involved in finding the global optima. The factors include Gradient Operator, Iterative steps and termination criteria. Modelling and Simulation also forms the main platform to visualize the algorithm better. A gradient of a scalar field f: $R \to R$ is a vector field which points into the direction of the greatest increase of the scalar field. It is denoted by $\nabla f$ or grad (f). The gradient can be sharp, a valley or of any shape and geometry, due to which the efficiency of the algorithm can get affected. It can make the algorithm stuck at some data point in its local optima. The number of iterative steps in comparing the points takes time and needs to be fast and hence it is also a factor that decides the efficiency of optimization. Less iteration means more efficiency. The termination criterion termination Criterion() is a function that has access to all the information of the optimization process, including the number of performed steps t, the objective values of the best individuals, and the time elapsed since the start of the process. With termination Criterion (), the optimizers determine when they have to halt. [47, 51]

While dealing with data, we need to define the formalism of data and the nature of their interactions in an abstract manner so that we can at least theoretically define the situation over which the optimization is to be performed. A model is an abstraction or approximation of a system that allows us to reason and to deduce properties of the system. Models are often simplifications or idealization of real-world issues. They help us to draw out the factors that have least influence on the conclusion and evaluation of the problem. Once the Model is theoretically defined it needs to be implemented on some platform so that we can have some results and that is done by simulation. A simulation is the computational realization of a model. On one hand if the model describes the abstract connections between the properties of a system on other hand simulation realizes these connections and implement it in real experimenting environment, making reasoning that if the model makes sense or not.

Global Optimization is a multidisciplinary research field and branch of applied mathematics and numerical analysis that deals with task of finding the absolutely best set satisfying various constraints. Its finding global minima among many local minimizes. These are quite difficult to solve exactly since many of them belong to NP-complete problems. According to the handbook of Global optimization edited by Horst and Pardalos, the global optimization technique can be classified into two main categories, exact and heuristic problem.

## 2.2.1 Exact methods

Branch and bound algorithm: Include a systematic enumeration of all candidate solutions, where large subsets are discarded, and examples include combinatorial optimization, concave minimization, reverse convex program etc. Enumerative Strategies: These methods include complete enumeration of all possible solutions. Examples include concave minimization models and generic dynamic programming in the context of combinatorial optimization. Homotopy and trajectory methods: These strategies have the objective of visiting all stationary points of the objective function. Naive approaches: Include grid search and pure random search methods. Stochastic search method: These procedures are based on random sampling, and their basic scheme can be improved by adding different enhancements sample clustering.[8-15]

## 2.2.2 Heuristic Method

Convex underestimation: This strategy estimates convexity characteristics of the objective function based on direct sampling. Genetic algorithms, evolution strategies: emulate specific genetic operation, applicable for discrete and continuous problems under mild structural requirement. Simulated Annealing: These techniques are based on the physical analogy of coding crystal structures that spontaneously arrive at stable configurations, characterized by globally or locally—— minimum potential energy. Tabu Search: This search forbids or penalizes search moves which takes the solution, in a few iterations, to the point in the solution space that have been previously visited. Tunnelling Strategies: attempts to find of local optima gradually modifying the objective function [16-20]

## 2.3 Quantum Big Data

This conquest of Space by Big Data has defined new challenges to the classical computational capacities of computers for solving problems at high speed, keeping in view the constraints of space, time and processing. This nature of Big Data has been a challenge in storing, processing, classifying and extracting a value from this huge cluster. Many solutions to this have been already proposed, from cloud computing where data 'ascends to skies' to designing high computing systems. Many mathematical tools have been also formulated for dimensional reduction of huge big data and efficient algorithms have been proposed for data mining and Machine learning. It is an intractable problem for classical computations to draw out meaning from the infinite clusters of big data. The paradigm has shifted so far from classical definitions of data towards newer fields inculcating the power of utilizing the tools from other disciplines, at the interface of physics and mathematics. Broadly categorizing, there are two main frameworks that had been successfully utilized for dealing with Big Data analytics. One bearing its robustness from its principle of superposition, entanglement and interference other is utilizing the properties of geometry that remain invariant to perturbations. [50] These two approaches addressing big data problem is Quantum Mechanics and Topology. These two fields are promising fields to construct some theoretical model to deal with Big data so that it can device algorithms that can decrease the time complexity for classification, sorting and regression irrespective of size and complexity of data. As for the first case (Quantum Mechanics) the techniques are divided into two scopes both being equivalent to each other although formalism and the nature of how to approach problem are quite different. One is the gate model of Quantum computations in which equivalent Quantum gate of classical gates are defined and input to these gates is a Qubit with states zero and one and all possible states in between, that is superimposed in nature. [61-70] This Qubit is evolved with time seeking output via unitary evolution. Quantum Support Vector Machine has been devised for classification of big data that decreased the time complexity of algorithm. Many algorithms for searching in a large data set also had been devised like Shor algorithm, Deutsch algorithm, Deutsch Joza algorithms and Grover algorithm. The other formalism is Quantum Adiabatic Evolution in which data or problem is mapped or reduced in terms of Hamiltonian and then slowly changing the energy of the system till results. It has been used to solve Satisfiability problem in which the evolution of the quantum state is governed by the time dependent Hamiltonian, interpolating between initial Hamiltonian whose ground state is easy to construct and the final Hamiltonian whose ground state encodes the Satisfiability argument. Similarly Combinatorial Optimization, constraint Satisfiability problem and NP hard problems are also being solved by this approach. The Quantum Adiabatic Algorithm is used as optimization method for finding global minimum of cost function with best choice of Hamiltonian. The Evolution in such formalism is as per Schrodinger equation. Quantum error detection codes leaving the code-space demand energy Penalties as a method for error suppression in Hamiltonian based quantum computing. Generally, the evolution takes place in four steps, an instance dependent, time dependent Hamiltonian H(t), initial state, evolution from 0 to t and measurement. Both of the approaches are equivalent, meaning that the problem that can be solved by a Quantum Computer designed from Quantum gates is same as to map and redefine the problem in terms of Hamiltonians, initial and final, making final as the result and taking initial to evolve slowly outputting the final Hamiltonians. Hence circuit models can be well mapped with Hamiltonian formalism. So Quantum Mechanics is expected to have a promising edge for handling big data due to speed up of processing and no space and time constraint. This has been utilized and evolved as separate areas of research such as in Quantum Machine Learning, Quantum Data Mining, Quantum Support Vector Machines, Quantum Neural Networks, Quantum Cryptography Quantum databases, Quantum Adiabatic Machine Learning, Quantum Algorithms and Quantum gates.[71-81] All of these approaches have increased the efficiency of handling with data and even big data. These all fields have had increased the efficiency due to quantum processes that are going on at the fundamental level of these computations. The processes like Superposition, entanglement, tunnelling and interference. Since there is redefinition of data and particle in quantum mechanical framework and are dealing with wave equations associated with it so the psychology of this basic fundamental unit is entirely different so gets manifested in the philosophy of its dynamics when subjected to some computations. This distinguished nature of data in this field makes it a promising candidate for dealing with big data. The other parallel field in the process of dealing with the space of data and interpreting structures in clusters is Topology, which has evolved to Topological field theory of Data.

## 2.4 Topological Field Theory of Data

Topological data analysis is a statistical method of finding structures in Data utilizing the ideas from Topology. Since the complexity and Volume of data is huge, TDA [58-60] using the notions of shapes, connectivity and invariance to perturbations, summarizes and visualizes complex data sets in huge clusters. The real essence of topological data analysis is to distinguish between noise and data and then through some parameters reduce the dimensions of the problem so that complexity of interconnectedness decreases and analysis of structures become clearer. This form of analysis is manifold learning of which principle component analysis is a special case. Mostly among topology, persistent homology has been a successful concept to address data analysis. It is a multi-scale approach to quantify features in data. Intuitively, we chose a parameter ŋ and a sample space of data. We see data and set for each set of (X, ŋ). The key feature is that some of the data points appear and disappear at each time for each value of ŋ. That is, each point has a birth time and death time, due to tearing of data from other and gluing of data point to other point. At ŋ=0 there may be n connected components, as ŋ tends to increase some of the connected components dies (merges) until one component remains as a structure. So, TDA will be a tool to summarize out the irrelevant stories to get out some real data. This analysis is robust in the sense that it is coordinate free that is independent of where this data point lies in the space, considering only property in terms of closeness and connectivity. It is invariant to deformations and a perturbation that is doesn't change the identity of the data set by stretching out, unless glued or teared off. It also entails a highly compressed description of geometry of data sets. The qualitative nature of this approach doesn't give importance to individual samples and its nature of being coordinate free makes it well suited for dealing with highly complex datasets. Inspired by a paper by Mario Rasetti on Topological mean field theory of data: a program towards a novel strategy for Data mining through Data language, defining a parameter 'n' subjected to various constraints and range, defining a hyper graph that takes vertex and edges as signals if the constraints are satisfied else noise. The system is reconstructed topologically with its simplistic complexes with its 'n' approximations, then making gauge transformations of data space leaving its dynamics unchanged as connecting field on a surface (manifolds). Many sub linear algorithms for handling massive data have been developed, Map reduction techniques, reducing the dimensionality of big data. Currently works are going on handling Big data via Big Data Sub linear Modelling through a Statistical Mechanical Coarse-Graining, aiming at modelling big data by compression and utilizing statistical mechanics for mapping Big data into a small subsystem, making data coarse graining such as mean field theory and renormalization groups. Other research going on Theory of Sub linear Sparse Structure Extraction from Big Data, identifying sparse that is small number of correlations reducing the volume of data to be reduced using regularization theory and statistical methods. Other works going on include big Data Clustering Theory. Much formalism for Boltzmann machine has been constructed on mean field theory with Hamiltonian formalism. Now every discipline of physics meets Big data in a way that the data generated is exceeding with an accelerated rates, so far every discipline hypothesis some algorithm in dealing with their Big data based on tools form their respective discipline but till now there is no 'reverse theory' to our best knowledge that can formulate a generalized model based on its characteristics to deal with Big data.

# Chapter 3

# Information meets Quantum Matter: Quantum Formulation of Big Data via Quantum Field Theory

*"When it comes to atoms, language can be used only as in poetry. The poet, too, is not nearly as concerned with describing facts as with creating images". Neil Bohr*

Ideally we define every system to be formed from aggregate structures that work together to form a formal platform. In order that we understand the dynamics and mechanics of the system as a whole then we analyze the system and it's aggregate and by using tools of mathematics define and understand the dynamism of the system. But in real applications, the system is not so simple in nature. There is fundamentally an inconsistency in each system to know if it forms an aggregate system or not or that if it is a structure or not. It is after this understanding we can have any know how about the system dynamism. There are at least two main conditions for a system to be aggregate so that we may simplify its understandings:

- Identification of the components in regards to their explanatory function, without consideration of the system as a whole.
- The wide behaviour can be explained by reference to the operation of comparatively few parts.

In real systems there is an unknown understanding of the fundamental entity that formulates the system. Like we say a data set, but how actually that data set is in terms of its nature? Classically it is in binary format that is physically stored on the plates of the capacitor in RAM with specific potential difference and capacitance. Let us have a close view, If we analyze each charge on the plate, they are interacting with each other charge on its plate and on the plate parallel to it with either positive or negative interactions. So the system fundamentally is interacting bits. What now? It is hard to determine at the fundamental level about the 'nature' of the fundamental particle. It is often said that it constitutes what is called as the intrinsic nature of system. But how do we define this 'intrinsic nature'. We don't implicitly define this paradigm rather we define it relatively in regards to the responses of the bit. Intuitively, if we keep a charged particle at some distance from a current carrying wire, it experiences a force when electricity is passing, else nothing. We conclude that this force is different from electric field and is due to motion of charge and designated it to be magnetic force. So it is an observatory analysis and still our question remains. What is this intrinsic nature absolutely? It is still a mystery and hence there is a tussle between holism and reductionism. The systems in its nature are actually irreducible as we don't know what the particle is in its reducible form.

## 3.1 Reduction

Language A is reducible or mapping reducible to language B, denoted $A \leq_m B$ if and only if there is a computable function f, such that for every w, we have:

$$w \in A \iff f(w) \in B \quad \text{.................................... (3.1)}$$

The function f is called the reduction of A to B. But what actually reducibility means? How does it matter that we can map one problem to other problem? The reduction instils few conditions that can be checked to understand if they are reducible. Theoretically, we utilize a reduced machine in order to know if w belongs A and an algorithm for B such that f(w) belongs to B, implies w belongs to A. It is important to understand that while reduction process it is significant to take the same machine with same computational power and that if we have $A \leq_m B$ does not imply that $B \leq_m A$. It states some formal definition of reductionism but let's make it little simpler and intuitive. If we have a problem A and it is a hard problem to solve and have almost no formalism to understand its dynamics then if suppose we have other problem B and that it is an easy one with a well defined formalism and mathematical formulation in a formalized way. What we do is that we try to reduce the problem A in terms of the formalism of B so that we can 'interpret' the dynamics of A. The following steps are better to understand the reductionism:

- Analysis of the two problem A and B and knowing the 'hardness' of the two problems
- Interpreting and mapping the problem A in terms of problem B
- Once the system is mapped it is easily interpretable and the dynamics of the system is well evaluated

But to reduce the hard problem in terms of the formalism of a well defined problem is not an easy task. A reductionism approaches via the phenomenon B to understand the system B. To reduce philosophically means to express the laws and rules of secondary science in terms of primary ones as a case of an example the reduction of the thermodynamics can be well understood by Newton's law and the statistical physics. In mathematical logic the reduction is associated with deduction. If we are given a premises P we deduce new statement S and actually it is not a new reality as it is already contained in P.

## 3.2 Gödel's Incompleteness Theorem and Reduction

Gödel incompleteness theorem gives an insight about the nature of a theory that can well describe the dynamics of neurons, keeping in consideration the complexity and incapacity of reductionism. [112] Though there have remained many controversies about his viewpoint- a tussle between optimists and pessimists but there are few philosophical attributes associated with the theorem that I believe can help us to interpret the nature of theorizing neural dynamics due to its complex nature. David Hilbert recognized that axioms tend to be self consistent if we cannot prove that a statement S and its negation ~S are both true theorems. It can be considered complete if for every statement S we can prove either S or ~S is a true theorem (in terms of language) This statement of Professor Hilbert When taken along with Kurt Gödel we can have an intuition of the theory or a mathematical structure completely defining the Big Data. Then it may be believed that the inaccessible law of reality as posed by the theorem may 'to some extend' be solved. Our epistemology shifts from arthematics (posed by Gödel) towards groups or more specifically category.  Gödel proved mathematics to be inexhaustible in that finite set of axioms cannot encompass the whole mathematical world. I believe that it is due to our inaccessible conduction to know what an 'intrinsic property' of matter really is! We mostly understand the responses of matter to different scenario and statistically or based on coherent and hierarchical structure of literature we make a step in explaining it. May be it is just another way towards reality! Mathematics may now leap from integration, summation and arthematics towards structures and diagrams like arrows in category theory.  Now in order that we define what theory can be devised for brain dynamics I feel the two theories that almost satisfies the above posed condition and discussion is Quantum field theory and Category theory. The manifestation tough at philosophical level seems equivalent but these are two adjoint theories satisfying the first posed condition and secondly that at functional level topology gets manifested satisfying the Hilbert view of completeness. So there are basically two questions to be answered about reduction:

- The elementary components belonging to the complex studied system as in case of Big data analytics
- How well the lower component of complex object reflects its nature and properties.

## 3.2.1 On formalization

Descartes' invention of analytical geometry has been a leap in conceptual shift for understanding laws and objects of nature. It stated the fact that any physical object law or concept can be mapped to the Cartesian framework for analysis and deductions. So if we are able to formalize the equation for the system and understand the rate of changes then it becomes more formalized in a manner that we can have a good say about the system.  Big Data is always characterised with Velocity, Velocity, Veracity and Complexity. It has been understood in computer science perspective and that the solutions devised to handle such a huge amount of data is of classical origins. [82-84]   It formulated solutions from designing of high computing hardware towards ascending the data to skies via Cloud computing. Now what if we are able to reduce the system in some known formalism? If we may be able to analyze the same problem associated with big data with a different angle and prism then new perspectives can be involved in developing a new paradigm with redefinitions and epistemological shift. What should be this new epistemology like? Here are few points that are worth noting: [86-88]

- There must be a well formalized mathematical framework for big data in specific and computer science. Physics has been a strong subject due to fact that it has a well defined mathematical structure that initiates as a language of interpreting nature. But in case of computer science it is immensely a deficit that we have no language. How can a discipline be comprehensive without a mathematical framework? There must be a mathematics that needs to be defined so that the 'reduction' may have coherent implications.

- Big data must have a well defined framework to which it will get reduced. If we have a well defined system to which we may reduce the big data then there is more actualization of the system and hence may give an impetus for newer technologies beyond classicality.

### 3.2.2 Generalization of Many Body Innovations

Suppose a system of N interacting elements with each possessing a perfect internal dynamics in the absence of the interaction. Technically it means that each particle has an internal Hamiltonian due to "being" of its existence. But in reality we cannot isolate a particle from the system and if we want to have an unreduced dynamics of the system it really is a hard problem. In Big data analytics there are infinite data points in the space and that they are interacting with each other at any instant of time with a free will. It is in essence the case of many body problems. Such formalism are unknown to canonical science but even cannot predict its solution except for the logic that anything can happen. These systems cannot be added due to continuous nature and neither can be integrated due to heterogeneity. In such case it is difficult to talk in terms of Schrodinger equation. The system existence equation can be presented in the universal form as:

$$E\psi(Q) = [\sum_{K=0}^{N} h(q_k) + \sum_{i>k}^{N} V_{kl}(q_k\ q_l)](Q) \quad \text{...................................................} (3.2)$$

Where $h_k(q_k)$ is the generalization Hamiltonian for the kth system component in free integrable state in the absence of interactions and the other term is interaction potential between kth and lth component. If we can define the dynamics of big data in the related formalism then it will be easier to understand the space of big data. But the question is how to reduce the space in such formalism? It seems beyond NP Hardness and beyond just Turing. [89]

## 3.3 Complex System

Complex Systems are ubiquitous; complex, multi-level, multi-scale systems are everywhere, comprising of many components that interact with each other to constitute the dynamics of the system as a whole in totality. The main important impediment in designing a theory for a complex system is its complex nature that is inherited due to the infinite particles that are associated with the system and also in particular the nature of interactions associated with the system. If it would have been the case that all the particles interacting all the time then still it would have been a simple scenario but in general the nature of complex system in regards with the interaction is that there is a 'free will' associated with the system. However the system interacts and there is no seemingly rule defining the interactions. Complex system is also associated with non linearity in a way that there is no 'descent' dynamics of the system, it is to state that seemingly we cannot associate any 'rule' that can define the dynamics of the system. There is no doubt that the data system is complex and data based with almost 4 billion people owing a cell phone. Every day, over 300 billion emails and 25 billion SMS are exchanged, 500 million pictures are uploaded on Facebook etc, adding 4 zettabyte to space with growth of 40% every year with a hope of reaching yottabyte, a number larger than Avogadro's number. Due to non linear dynamics, complex systems are attributed as dynamical systems, in a way that the data points are changing instantaneously with time. The complex dynamical systems are solved via differential equations. Example of such systems is Lorenz System, produce a mathematical phenomenon called as Chaos. In such systems a small change in input and initial conditions can make a drastic impact on the output. As a part of abstraction, the huge space of data or space of big data is a space contained with huge data points with interactions, comprising a complex dynamics is demonstrated as:

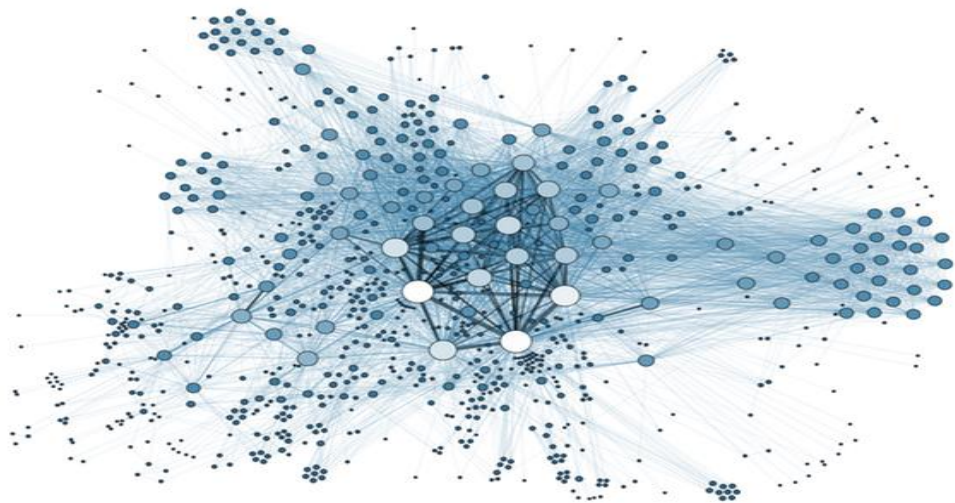

**Figure 1**: Big Data with Complex Interactions [137]

So there needs to be a specific formalism and a mathematical structure that needs to be developed for understanding big data. It is beyond the classical approach of formalizing new computations and technologies when we know there is need of understanding the problem from the very base.

### 3.3.1 Reduced Correspondence: A 'Sensitive Reduction'

The reduction is a very sensitive task for complex systems due to its unreduced nature as discussed above. For Big data to be reduced in some known formalism with mathematical structure then it must be followed by correspondence that is to relate the two reduced realities. The best way to understand this is via complexity of Black hole and information retrieval. Think of a spherical black hole in space contained with almost infinite information in vibrating strings. Now if bob wants to access this huge space of information, he needs to go into it but as it is clear that there is information loss if Bob is to 'jump' into it so how can he access this information? He 'scratches' the surface of the sphere and tries to peep into it and since the veil instils hidden information in the system so we may talk in terms of entropy. Maldecena came up with a correspondence that the dynamics inside the black hole is equivalent with the dynamics of conformal field theories at its surface. This is a beautiful correspondence! How does it help? It means Bob doesn't have to jump into black hole but to be at surface and see through it as being inside or at the surface are equivalent. In our case it is termed as dimensional reduction. Once reduced a correspondence needs to be defined.

## 3.3.2 Towards Understanding the Reduction of Big Data

One of the biggest challenges for Data analyst is to understand the nature and characteristics of Big Data and based on it formulates a theory that could start up as a prototype for theoretical construction of the big data sciences. Big Data is characterized with volume, velocity, veracity and complexity. Intuitively, it is a 'space of big data' with almost infinite data points, more data adding up at every instant of time with data sets dynamic and instantaneously changing with time, having high dimensionality and interrelationships between them. Most of the important aspects that one seeks from this mess of infinite points is to 'interpret' it to information and then to wisdom. In the framework of big data the prime goal initially is to have a prototype defining and visualizing how data is in space, how it interacts with other data points, how complex the dynamics change due to computational operations on it. This is a goal pertaining to construct a prototype for big data analytics and later it can be implemented for data mining, Machine learning and Neural Networks. The prototype can be modelled afterwards in relation with an automaton, defining the formal language for Big Data 'expressing' its hidden local and global dynamics. How can we define such a general model for big data? What are the basic conditions that this model must address? Some of the basic implications and qualitative aspects of the model should possess:

- The model on one hand must address the nature of each local data point in the space and at the same time be able to draw out global aspects of this big data.

- It must be able to predict from a local data point, the dynamics and evolution to interpret how global structure had been constructed based on this local data point.
- It should dimensionally reduce the space of big data and map it to some simpler framework for understanding the complex relationships of big data.

- It must be able to distinguish between what is data and what corresponds as noise.

- If possible there must be some insights of mathematics, at least qualitatively defining the intuitive picture about the parameters directly affecting evolution and dynamics of big data.

## 3.4 How Hard Big Data is?

Big Data is the term inherently defining the voluminous and complex data with high dimensionality having interrelationships and heterogeneous in nature being generated from distributed sources. This volume of data or 'space of data 'tends to increase in an accelerated fashion, adding up continuously more data points to this space and that too changing instantaneously with time. The scope of this problem to handle has even more been complicated with the introduction of Quantum Information theory and computations. Now Data that is stored as information in Spin momentum of the electron can have not only two discrete values rather it can contain a value that ranges in between, almost infinite, defined by $|\psi \rangle = \alpha |0\rangle + \beta |1\rangle$. Until measurement, we can't predict its information. The classical computation is obsolete and cannot handle Big Data efficiently. The definition and nature of Data is now shifted to a new platform with new epistemological shift and data redefinitions. The data that will possibly be input to the future Quantum Computer! Since the dynamics of Quantum Mechanics are entirely different so the data being input and processed is redefined. Now Spin that stores information is probabilistic and be in almost infinite states before measurement! Imagine this space of big data with almost infinite data points. Each data point stores information in the spin angular momentum just like an electron. This information can be a superposition of all possible states of being either 1 or 0. So the data point is a quantum particle just as electron and in this paradigm it is termed as Qubit. The space of big data is contained with almost infinite particles of same nature. There will not be deterministic interactions between the data points or Qubits in this space. It is not that all the data points will be interacting, so we need to draw a line between what is interacting and what is not. How does this system interact now? Imagine Bob as a data point storing some information in his device here qubit which is in the superimposed states. It's as if Bob is an electron and all the states are the virtual dimensions that he is defined with. Bob cannot remain isolated he sends and shares information with other data points in the form of images, text and videos, using many applications. So there is interaction between two data points and in the similar fashion there is interaction between all the data points. Now it is a complex system with high correlations but the complexity increases with increase in the type and nature of interaction. Imagine now Bob interacts with Alice but doesn't interact with Tom, and after some time Bob interacts with Tom but not Alice. It can be thought of sharing sending data in the form of images, texts and videos.

There can be some data points that don't interact at all. So this sort of interaction is even more complicated scenario than the former where all the data points are interacting. So, there must be some function that 'freezes' this system so that we can draw a line between what is interacting and what is not and also what is when interacting and what is not in that when! Some points in the space have same value or states too. If a data point at some instant stores data in form of 1 other time it can be 0 and may be in different superimposed state. This can be though as if the lattice is subjected to the external magnetic field or to a huge bath which instils perturbative energy to this lattice, making it changing its states dynamically. There is an addition of more data points in this space of big data almost at every time. Most of the data that gets added to this 'big data space' is mostly unstructured in nature that is to say it is heterogeneous and variable in nature with many formats, including text, document, image, video and more. We find it hard to structure this data and draw out meaningful information from so that this huge complex system can be well understood. Intuitively, to visualize this situation, if you have a dataset with 300 points, a conventional approach to analyzing all the topological features in that system would require "a computer the size of the universe," he says. That is, it would take 2300 (two to the 300$^{th}$ power) processing units — approximately the number of all the particles in the universe. In 1980, a work station with 32 Mbyte RAM could solve eleven interacting electrons and after 20 years the computation power has increased by 100 folds that allowed solving system with two more electrons. What about particles almost Avogadro's number! How does our brain make out information and handle it in such an infinite mess? How does nature do its computation? How does nature find the states of particles ranging to Avogadro number? So far the techniques that have been utilized, there has been an attempt either to dimensionally reduce Big data and qualitatively simplify the complexity to some extent. Many Quantum algorithms have been devised for classification, sorting and constructing universal Turing machines, Quantum neural networks, Quantum support vector machine to deal with this data. But explicitly to our best knowledge there is almost no theoretical model that can address Big data based on how it is and how hard it is as discussed.

For real drift in Machine Learning, Neural Networks and Data Mining it is essentially a big goal to construct a theoretical prototype or a structure that can envisage all the five conditions defined above, keeping in view the nature of big data. This can help us to develop a new field and even a disciple for directing the research for Big Data. Once we are able to understand how Bob acts, or how the electron feels as with itself and how it interacts owing to its complex nature, then we can easily know what to extract information from this structure. Psychology of electron defines the philosophy of whole cluster or mesh of infinite data!

On March 31 2017, Professor Andrew Millis delivered a lecture on Quantum Condensed matter physics meets Big Data and other sciences meets big data in the sense that the data being accumulated from experiments and other sources have exceeded our conventional standard towards Yottabyte. Many conferences had been held on 'From Quantum Matter to Quantum Information' seems to be a radical statement due to seemingly diverging aspects in the areas these two fields address. Big Data on one side concerned with information and dealing with huge complex mess of data and Condensed matter theory dealing with physical properties of condensed matter. There has been a leap from classical to quantum and on parallel to topology now what? How can we define the basic and fundamental prototype of machine simulating theoretically the above defined system? What could be the foundational?

What model can we reduce this complex system to? What next? We construct a structure or prototype at the interface based on nature of Big data. We propose that the best way and platform to understand and analyze this system is in terms of Quantum field theory and condensed matter physics.

## 3.5 Reduction of Big data to a Complex System

Reduction (correspondence) are the basic techniques that have been utilized for complex systems in which problem is reduced to some defined problem and then utilizing the correspondence of the two problems can yield efficient results. Informally, reduction is a technique in which two problems A and B where A is hard to solve and B with a solution. If somehow we can reduce the former problem in the later formalism then we can solve the former too. Reduction techniques has been implemented in solving Satisfiability, NP hard problems, Optimization problems via 'reducing' them in a physical system with Hamiltonians and then including physics in solving their complexities, ending with an efficient result. Seth Loyd used topological mathematics in dealing with Big data via reduction. The complexity of black holes and information retrieval is an impossible problem to even think of! Due to equivalence of Ads-CFT correspondence has defended the foundations of string theory, simply stating that the complexity in the volume of black hole in D+1 dimensions is equivalent to the studying CFT at the surface of the black hole in D dimension, hence reducing the dimensions and complexity as well With the paradigm shift of computation from a classical computer towards the quantum platform has led to the dramatic changes in the nature of data and how it behaves. The Space of Big data is quite different from the space of data and is associated with following challenges: [86-90]

- Data points are almost infinite with more data points adding to this 'space'.(Volume)
- Each data point is in superimposed state, so is the whole space of big data.
- There is an instantaneous changing in each data point (Velocity)
- There is a very complex interrelationship between the data points. (Complexity)
- The type of interaction between data points is changing from no interaction or interrelation to interaction, changing with time.

Reducing this formalism in terms of a physical scenario we can say this space of data is infinite, heterogeneous, instantaneously changing, complex, correlated and dynamic system. So there is a correspondence from the characteristics and nature of Big Data which maps it to a physical scenario that is a complex system. Volume mapped as infinite, Velocity as Instantaneously changing and dynamic, Veracity as noise with environment, interrelations as interaction and correlations. Hence Big Data is reduced as a Complex system.

# 3.6 Towards a Model

So as to 'reduce' our proposed problem of big data to some known formalism that can meet the challenges that are associated with it. For reducing the problem we need to analyze what type of model we want to propose that can be well suited to the above defined 'redefinitions' Here are some notable points that our model to which we will reduce our problem are:

- It must be compatible with particles with infinite dimensions and number
- It must be compatible with interaction of a particle with other even if the other particles are infinite.
- It must consider all the interrelations with the data point.
- It must be well defined for correlated systems.
- Dynamic in nature.
- Considering non interacting data points.
- A parameter that can map local information stored in data point to the overall points
- Address Superimposed Data points.

These few notable points clearly addresses the nature of our space of big data being infinite in number as if in Hilbert Space of infinite dimensions, highly correlated, dynamic in nature. These Conditions cannot be handled by conventional Quantum Mechanics rather we need a tool that can address all the above conditions. We make rescue by introduction of Quantum Condensed Matter utilizing the essence of Quantum mean field theory. It is where matter meets information.[91]

## 3.6.1 Information is Physical

The information stored in Quantum computation is Spin angular momentum, having infinite number of data points, [89] with each point having some information stored as spin momentum having values +1, -1 and in between. It is a Qubit. The data point will be now having its well defined 'influence' which is defined as the Hamiltonian. Each data points having Hamiltonians in this space and each data point interacts with the infinite number of other data points that surround it, which are the interrelations among data points. Intuitively, it is a big space of this data spread over virtual space that can be a cloud or infinite memory (qRAM) with same characteristics. Each data point stores some data in the form of superposed states that inculcate some energy in it, just like a capacitor storing some data in RAM is actually in series with the transistor, storing data in the form of potential on plates, So data in RAM has some energy associated with each data stored. Quantum mechanics stores data in the form of spin angular momentum for some amount of time so it too has some energy associated with some data point stored in an analogous fashion. Pertinently, each data point has a Hamiltonian associated with it. This Qubit can be stored in different ways like in spin of electron, polarization of photons. The data point stored in a memory practically has not infinite degrees of freedom as in case of Qubits rather in practical scenario, memory has a spacial confinement due to which data can be stored in the defined and finite degree of freedom, fermion computation provides us with the best framework. So the data point is actually an electron that stores information in its angular momentum. Since the dynamics and information of each data point cannot be evaluated independently due to interactions and correlations. We defined a rule essential for getting interrelated or interacted that is interaction is only possible when there is communication between data sets so a fermion has to hop from its point to other point. The data point if Qubit or fermion is located in some physical memory so it cannot have infinite dimensions due to spatial confinement. As an illustration, two electrons with different spin direction occupying same narrow d or f orbital in real materials is a correlated system, an analogy for understanding a qubit or a fermion in a spacial confined memory. If $\Omega$ is the average time spent by an electron on a lattice and 'a' be the lattice spacing, W being bandwidth then: $\Omega = h/W$ It clearly defines the effect of spacial confinement and degree of freedom on the makes an electron to spend less time in orbital. Narrow the orbital, larger an electron resides on an atom and there by feels the presence of other atom. The spacial confinement enhances the effect of coulombs interaction between electrons making it strongly correlated. These systems are strongly responding to the external parameters like spin susceptibility. In reality the parameter and the interplay between spin, charge and orbital degrees of freedom of correlated system and with lattice degree of freedom leads to various effects. In such systems we can't analyze the dynamics of a single electron completely without considering the effects of its interactions with other electrons and moreover the spacial confinement and degree of freedom of the lattice provides appreciable effects on the specified electrons. In similar fashion, the fermion or Qubit stored in a memory experiences a sort of spacial confinement, nature of information carrier and degree of freedom of the lattice. [85]

### 3.6.2 Mapping Qubit to Fermion

A Qubit with almost infinite virtual dimensions depicting a data point can be well mapped as a fermion. The n Qubit state can be envisaged as a n fermion states with 2n modes. This is because one could pair those 2n modes to get n pair and then use one mode of each pair to fill with a fermion The Qubit is stored in a memory having spacial confinement, Imagine two electrons with different spin directions occupying d or f orbital, the dynamics of the interacting electrons is interplay between spin, charge and orbital degrees of freedom with lattice degrees of freedom. In such case we can't analyze the dynamics of single electron completely without considering the effects of its interactions and spacial confinement and degree of freedom of lattice too affects the system. In similar fashion is the case of Qubit in a memory cell in an interactive system. A fermionic computation well takes care of all these parameters and qubit and fermions can be too mapped via Jordan-Wigner transformation.

[91] This transformation addresses the superimposed states of data and decreases the complexity.

## 3.7 Construction of a Prototype for Big Data

Quantum data points are distributed although in a complex manner without any structure but symmetry forces us to think that there must be an inherent rule that makes this huge mess to have some organized cluster. [92-93] when we deal with this space of big data we won't input whole of the input space to the Quantum Computer and then output any value, rather a subset of this huge data space is input to the machine. This subset will be intuitively a set of huge cluster in itself with same properties as that of its universal set (Space of Big Data). Let us suppose for simplicity that each data point as a vertex denoting the site in which information is stored in its spin momentum and the edges represent the correlations with other data points. These clusters of data form a lattice type model and for simplicity we take it to be symmetrical. From a data analytical point of view, it is a multi dimensional array of data points that are analyzed in such a manner so as to also include the interrelationships of other data points in this array to the data point. The head of the pointer of the array changes after a data point is visualized and operated on to the next point and same procedure is repeated. The moment of the pointer from one cell to other is hopping of an electron from one site of this lattice to other, having some kinetic term and the interrelationships denoting the interaction term. This multidimensional array is equivalent to the lattice model with each cell containing a data point having spin up or down stored as information. The spin can have a superimposed state in between the two values in the same manner as defined by the lattice model. At this point, we can restrict ourselves with data stored in a cell of array or the data point on one of the lattice having only two value so that we can reach some simpler conclusions and at the end analytically determining this superimposed states and it is neither a compromise as we are dealing with fermions that are well mapped to a Qubit. Assuming for simplicity that these Qubits are arranged in a lattice like structure where each node or point of the lattice signifies the Qubit storing a superimposed information as spin angular momentum. The data points have a 'specific interactions' that is some interact some don't and some interact in other point and some don't interact at that point. We can't assume each data point to be interacting every time if we do so then it can't model the Big Data characteristics. There must be some 'rule' that makes a data point to interact with other data point. In order Bob wants to interact with Alice that is wants to share information with her, he sends information say in Qubit or fermion to her, so this defines a rule for interaction. So interaction between data points is only possible when the electrons meet at the lattice. This defines the 'rule' for interactions. Now pertinently, we can have now any type of interactions possible. Some non interacting meaning the data point is static at its own lattice site, Bob just storing information not sending. Some are interacting in some time and some in other, depicting when fermion is sent at other site to Alice for interrelation. At initial we assume that all of the data points are non interacting, as in data analytics it is not that all the data points will be interacting, so, there must be some function that can distinguish between them and draw a line between what is interacting and what is not. This initial scenario is the simple state of this infinite space of data where only particles are infinite but not interacting. The lattice consists of N data points, where N tends to infinity with strong correlations with each other, some of data points being 'specific interacting'. It is as if in contact with a heat bath that provides it energy so that the spins at each of the sites flip instantaneously or we can provide a magnetic field to it so as to denote the operations that are being implemented on every site, making the system instantaneously changing and dynamic. Language of Hamiltonians: The Space of Big data in some physical space like memory or 'infinite cloud' have Hamiltonian associated with their data stored. This lattice of Big data is huge to be analyzed or input to Quantum Computers so practically let us take a 'slice' of a lattice from this infinite set but having all the characteristics and nature of big data. That is to say those N points are tending to infinity, more data adding, dynamic and correlated. A question may arise, how can we understand the dynamics of big messy data by studying the 'slice'? How can such a subset be used as a framework for big space of data? We take help of evolution of system with time. The function or real parameter defining the change in energy of the system at each interval is called action. The difference of which can tell us how state of a system has changed. We start from this small lattice slice and evolve it in time till it evolves into full space of big data. And this will be defined by action parameter. [93-97] Since for 'specific interactions' we defined a rule essential for getting interrelated or interacted, that is interaction is only possible when there is communication between data sets so a fermion has to hop from its point to other point. So for now the Hamiltonian there are broadly four energy parameters:

- Interaction only when Alice sends fermion to Bob depicting the Kinetic term in the Hamiltonian.
- Interaction with data points in space denoted as Interaction term in the Hamiltonians
- Potential Change to continuously changing particles, dynamic system denoted as Potential term.
- Self energy of data in Space.(Information energy or Information content)

The candidate that can address all the defined aspect is Quantum condensed matter with Quantum Field theory. Quantum field theory is an extension of Quantum mechanics but the former is more generalized theory for particles that are large, too large to be contained in Hilbert space, so it gives particles an extended Hilbert space known as Folk space. What if there are many different particles in the same state, what if N is very large, what if the particle number is not fixed in space, it instantaneously changing, with infinite degree of freedom. What if the dynamics of single particles cannot be evaluated independently, rather this very particle is interacted and correlated with other particles and that too these particles being interacting are infinite! It can't be addressed by Quantum mechanics to a appreciable extend, rather it as depicted by 'field' in QFT has a different values in space time and can be better for understanding the problems defined. The Schrodinger wave equation is differential equation acting in the argument of the Schrodinger wave function and can never change the argument so cannot deal with changing infinite particle number. It is vague to think about the wave equation of particles almost Avogadro number. All of these 'ifs' are characteristics of Big data. These define nature of big data and to have solution of dealing this problem QFT is the rescue. In QFT since there is a continuous changing and dynamicity so QFT makes use of language in terms of creation and annihilation operators. [98-100] while defining the energies that are associated with data point in QFT paradigm, it needs to formulate in the language that is best suited to it. Some notable points are listed as:

- The existence of antiparticles and the ubiquity of creation and annihilation operators.
- Transition between states.

So QFT is the best candidate to construct a conceptual prototype for understanding and addressing almost all the complexities associated with big data analytics. In order to structure the model condensed matter physics provides a probe to model the data. From the Hamiltonian and concept point of view Hubbard Model is best for modelling and constructing a prototype for big data analysis. The spin can have a superimposed state in between the two values in the same manner as defined by the lattice model. Solution to this lattice problem is an intractable case as it is in theoretical physics. Going to each point of the data set and analyzing its Hamiltonian and its interaction with other data points becomes a hard problem to be solved. When the data set is from the space of infinite data then the nature of problem becomes more complex. This problem requires infinite resources and hence the classical computation cannot frame its solution. Memory resources, CPU clock speed and other parameters are not so fast to make this problem solved. We propose that we can use the mean field theory approximations to get a generalized solution. Mean field approximations for Boltzmann machines have already been proposed. The data point in this lattice that is to be studied is influenced by the Hamiltonian of other data sets and that too are infinite. We treat the influence of other particles as a sort of magnetic field around the data point. This magnetic field is the summation of the average effect of all the data points. In this scenario analyze the effect of this average, treated as bath on the data point. So the lattice model is reduced into a particle in the bath with average effective field, and this problem is solvable. Since the system is dynamic in nature and instantaneously changing so this model is best explained with the dynamic mean field theory. Since the system that is space of Big data is infinite set of data points, instantaneously changing with time, dynamic, correlations so it is best reduced to a Hubbard model whose dynamics is defined to the best by dynamic mean field theory?

## 3.8 Working of Model

A Subset of infinite space of data, modelled in the language of Hamiltonian, having hopping Kinetic energy, Interaction energy and potential term. These associated energies define the dynamics and structure of how the constructed model will work to derive value out of infinite complex data. At t=0 Let the system be non interacting in nature and fix a point as a reference for analysis (Say Bob). As time passes, for interaction to take place it must hop from its point to other. The Qubit or Fermion whatever is the case hops to other site for interaction. As we have already made our point about QFT that it utilizes the notion in terms of particle-antiparticle so we need to associate our energies in same order for our datasets. Since for interaction, electron hops from one site to other, so there is Kinetic energy term as a result of this hopping that can be well described in terms of amplitude or frequency domain after

Fourier transform but it needs to be defined in terms of creation and annihilation formalism. That is c and c* which describe creation of a data point at other site say Alice and its annihilation in former site say Bob. It is not that creation and annihilation operates forces the structure of Big data to hold nothing rather creation is at one site and annihilation at other so it retains information in its structure. There are possibly many ways for the electron to hop from site one to other which can be well understood by Feynman's propagators. But we average out the amplitude of hoping for our simplicity as at this point of time we are only concerned about a prototype model.
Mathematically,

Kinetic term = SUM (Average expected value of electron hopping from i to j) in terms of creation and annihilation operators When the fermions interact with each other at the site then there is an interaction term associated with it that should be in creation annihilation formalism.

Interaction term= SUM (Interaction at one site between two fermions) in terms of creation annihilation operators

Now this complex system is dynamical in nature so the number of datasets is adding up due to velocity of big data, This may add potential to the Hamiltonian. It should take care of the stochastic nature of dynamics at this particular site visa viz Pauli principle. It makes a 'rule' to define how data that is coming in should get aligned. There needs to be a sort of distinction between a fermion and a Qubit although both can be mapped to each other and well corresponded by JWT transformations. A Qubit can be in possible many degrees of freedom at a time but in terms of occupancy in the memory will be strictly followed by Pauli principle. It can be have a location filled, half filled or empty. We store our information in matter based on this principle. This is expressed by a potential parameter and it decides 'fate' of new coming data set. It can be best felt in terms of fermions.

Potential term = SUM (Rule for loading into memory as per Pauli Principle) This model modelled by Hubbard model Hamiltonian. For this subset of N particles in Folk space the Hamiltonian is of the formalism

$$H = \text{Kinetic term} + \text{Interaction term} + \text{Potential term}$$

Each data point has some wave function associated with it and now if we have infinite data points it becomes a complex problem. The single particle quantum mechanics is constructed by considering a state vector |ψ⟩ in a Hilbert space H. The space for this single particle can't accommodate more than one particle. For many particle systems, we need to extend this space because the degrees of freedom increase and tend towards infinity with the increase in particle number. Still the complexity of the problem is not too complex for infinite non interacting data points. What if we introduce the interaction between them? This is really a hard problem to deal with and makes it more complex. Hence wave function that corresponds to this system should take all of these parameters into account. If now we have a system that interacts but in a stochastic manner, that some of its datasets interact and some don't, it is even more difficult insight to draw out a line between the dynamics of interacting and non interacting particles. In our case we can't write or talk in terms of formulating Schrödinger wave equation rather there needs to be a function that 'freezes' the lattice and can formulate the dynamics and time evolution of this complex system. Once it is frozen we can see how every data point is interacting and when. Now, why would those in statistics and machine learning pick the Green's function instead of the corresponding energy functional? In physics the system being modelled often has symmetries and conservation laws that are more easily seen in the form of Hamiltonians and Lagrangian (for example, in the case of Noether's theorem). However, for messy, complex data that does not possess any clear symmetry or conservation properties, it might make more sense to guess the Green's function instead with a consideration to the possibly inhomogeneous local properties that will have to be determined by large amounts of data. Thinking about this different choice in modelling strategy can provide a lot of insight to physicists. We evolve this lattice and its dynamics is defined by the green's function. It depicts probability of data point to hop from neighbouring site to other, and it is defined in terms of creation annihilation operator. Here Bob is set as a referenced point with its infinite interactions with other data points. With time we look how many data sets are being created and annihilated in regards with the complex systems. After some imaginary time we can have full dynamics of this complex system.

Green Function = Creation annihilation at some site evolved in some time

The Space of Big Data is well modelled in Bose –Hubbard Hamiltonian described as:

$$H = -t \sum_{<j,l>\sigma}^{n}(c_{j\sigma}^{+} c_{i\sigma} + c_{l\sigma}^{+}c_{j\sigma}) + U \sum_{j}^{m} n_{j+}n_{j-} - \mu \sum_{j}^{n}(n_{j+} + n_{j-}) \quad \text{..........................} \textbf{(3.3)}$$

Once modelled, we need to analyze the dynamics of each fermion and its interaction that can't be defined by wave function and Schrödinger equation rather a 'freezing function' is needed and that in our case is Green function. Why we need Green's function instead of corresponding energy functional? The systems modelled have symmetries and conservation laws and are more easily defined in terms of Hamiltonians and Lagrangian. But for messy and complex systems, it might be more sensible to guess Green's function instead of local properties of large data.

The evolution of this space of big data is modelled in Green's operator. Considering that this system of data sets evolves from an initial configuration($x_0$; $t_0$) to a final configuration($x$;$t$) given as,

$$|x', t'\rangle_{state} = \psi^{+} (x', t') | State\rangle \text{..........................} \textbf{(3.4)}$$

$$|x, t\rangle_{state} = \psi^{+} (x, t) | State\rangle \text{..............................} \textbf{(3.5)}$$

Then the Green's function reads as:

$$G(x, t: x', t') \rightsquigarrow \langle \text{State}| \psi^+(x, t) \psi^+(x', t') |\text{State}\rangle \quad \text{............................................. (3.6)}$$

This is interpreted as evolving a system with a data point at position x0 and timet0 to a location x where we add another data point at time t. Then the Green's function is the scalar product of these two states and tells us the probability of occurrence of this transition. Since energy changes are the driving factors of this data evolution, the system has undoubtedly an associated action given in a generic form $S = \int L \, dt$ Construction of this action S helps us to write a partition function for this system as it behaves like a grand canonical ensemble and eventually Green's function for the system. The Whole dynamics is analyzed by the change in the interaction of the data points that can be described in terms of action:

The action of the neural dynamics can be evaluated from $S$: for field $w(x,t)$, we have

$$S(w) = \int L(z, t, w, \partial_{zt}) \, \partial_x \partial_t$$

Where L = T- V is the Lagrangian, difference between the potential and the kinetic term. The probability distribution of w (x, t) is

$$p(w) = e^{-S(w)}/z$$

The dynamics can be formulated in terms of Klien-Gordon equation, pertinently defining the correlation between two data points at w ($z_1, t_1$) and w ($z_2, t_2$)

$$\langle w(z_1 \, t_1) w(z_2 \, t_2) \rangle = \frac{1}{z} \int w(z_1 \, t_2) \, w(z_1 \, t_1) e^{-S(w)} Dw$$

The way Big data networks can be related to Feynman propagators and diagrams so that correlation between two data points in space and time.

$$\langle w(z_1 \, t_1) w(z_2 \, t_2) \rangle = G(z_1 \, t_1 \, z_2 \, t_2) \quad \text{........................................................ (3.7)}$$

Where G is the Green's Propagator and encodes all information about the system.

## 3.9 A Surprising Dimensional Reduction

Now we want to make a dimensional reduction of our data lattice so that we can qualitatively understand the complexity of the data. This can be well done by mean field approach. Tanaka formulated mean field model for Boltzmann Machine. Mean field theory reduces a many body problem to a one body problem with an effective external field and dimensionality plays a central role in deciding determining whether mean field works for a particular problem or not. In technical terms the Lattice model intractable in nature is reduced to Anderson Impurity Model which is solvable. As in case of Weiss mean field theory, the observable is the magnetization in a similar fashion the observables in DMFT are the Local Green's functions. So now in order to calculate the greens function for this impurity model to know about the dynamics of the system, The complex structure of Big data structurally in coherence with Hubbard Model and conceptually driven by QFT is reduced to a data point surrounded by all the interactions from other data point same as that in case of Anderson Model. It is a dimensional reduction so plausible! Now Bob sits and waits for all types of interactions whether some interact in some time or other in other time frame. No problem now! Almost all the complexities are reduced to a simplified problem. Now we have a little thing to be addressed. The medium of this bath is dynamical in the sense it is changing with time. It will act as a source for equivalence between two models that is Hubbard and Anderson. At each point of this bath has a varying 'self energies' and all interact with Bob in their own way. The Hamiltonian of this reduced impurity model is

$$Ha = H_{atom} + H_{bath} + H_{coupling}$$

Since now there are almost infinite states the data points can be in, so it can be well defined by the partition function, At the end how the system has evolved in some imaginary time domain can be visualized by greens function, partition function in terms of creation and annihilation operators. It is called action [101-111]. Bob has some information stored and there can be now many possible states in which he can be, sending to Alice also in different states, via creation and annihilation operators for some time evolving defined by greens function. This is action for this system of two lattices.

Action = (Partition function, Greens function in terms of creation annihilation operators)

## 3.9.1 Taming the infinities

The intractable problem effective mean field of all the interactions of all data point and its interaction to a single data set makes it solvable. By changing the Self information, a potential due to information content in data point we can make the greens function equal to the impurity greens function. What does it mean now? Greens function is evolution of the system and it defines how the system has evolved in time. When the greens function of a single site data set gets equal to the green function of lattice it means that they both have evolved in a similar fashion, so the dynamics of this intractable mess is well evaluated by solvable impurity model. What about the infinite mess? DMFT gets exact when the dimensions tend to infinity or coordinate number tends to infinity. Mean field theory reduces a many body problem to a one body problem with an effective external field and dimensionality plays a central role in deciding determining whether mean field works for a particular problem or not. In high dimensionality when the dimensions 'd' or the coordinate number 'Z' goes to infinity then the MFT becomes accurate and exact for the specified problem. That is d→ infinity or Z tends to infinity then MFT is exact reduction for the Big data analysis. Dynamic Mean field theory calculates the dynamics of many body systems like Hubbard Model and reduces it in terms of Anderson impurity model with a spin interacting with effective field of all of the spins. In infinite dimensions the self energy of impurity gets exactly equal to the lattice self energy, it means the information contained in the bath as a reduced effective information gets equal to the lattice information, perfectly simulating the models. The DMFT becomes exact in the limit of infinite lattice coordination (infinite number of spatial dimensions).Local Green's Function -----→ Self information -----→ Impurity Green's Function (Loop till it converges) The dynamics of this model is initialized by a impurity green function and it is changed and the parameters on which it based is changed in such a manner so as to converge at lattice green function. The parameter is Self energy. Now infinity is tamed in a manner that our proposition gets exact in infinities!

Diagrammatically it may be depicted as:

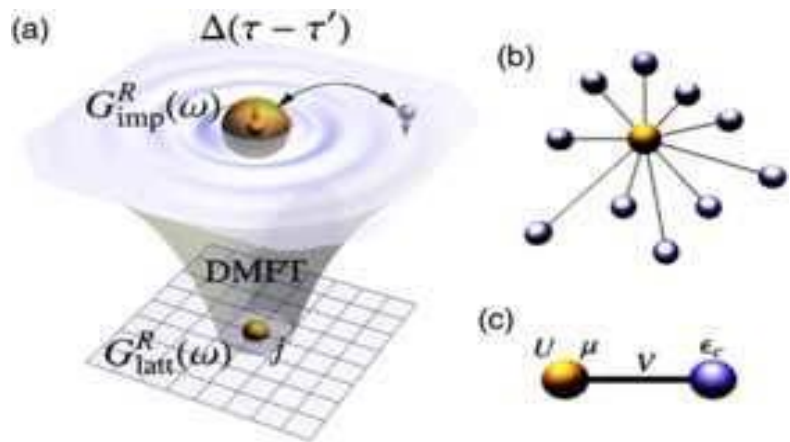

**Figure 2:** The evolution of system in DMFT.(a) Dynamics of big space reduced to a data point in effective field via Green's function(b) A data point with complex relationship and interaction with other data points trapped data point(c) Origin of various energies for the impurity site.[136]

## 2.10 Algorithmic Analysis: The 0(n log n) Complexity!

**Reduce** ($G_{lat}$ :$G_{imp}$: $\Delta$: $\sum$)
**Store** $G_{lat}$
**Initialize** $\sum= 0$
**Incremen**t $\sum=\sum+1$; α a constraint
**While**  ‖$\sum$- $\sum^{old}$‖ > α,  do
**Calculate** $H_{imp}$ from $\sum$
**Calculate** bath from $\Delta$
**CONDITION**
z→ infinity: d→ infinity

**Then**
$\sum(k:i\omega_n)$~ $\sum_{imp}(i\omega_n)$
**Compute** $G_{imp}$ and $\sum$
**If** $G_{lat} = G_{imp}$
(Return value)
**else**
**update Self Energy** $\sum$, $\sum^{old}$
**update** $\Delta$
**end while**

The time complexity for computing the impurity Green's function from self energy of bath takes O(n) time, and its comparison with lattice Green's function after each iteration takes O(logn). So the total running time complexity of the algorithm becomes O (nlogn). Volume, Velocity and Complexity of Big data have been addressed, utilizing Quantum matter physics.

# Chapter 4

# Quantum Simulation: Mapping Laser Parameters to Quantum Adiabatic Evolution

*"The most important application of quantum computing in the future is likely to be a computer simulation of quantum systems, because that's an application where we know for sure that quantum systems in general cannot be efficiently simulated on a classical computer"-D.Deutsch*

As Feynman has well quoted that "The rule of simulation that I would like to have is that the number of computer elements required to simulate a large physical system is only to be proportional to the space-time volume of the physical system. I don't want to have an explosion. It in future and develop and initiate work for a mathematical structure of computer science in general and big data in specific. Feynman has already pointed out that it is really hard for the classical computer to simulate nature and in other words we may say that it is hard to simulate the quantum mechanics on a classical computer. So the question arises that if we don't have a quantum computer then how can we understand the possible efficiency of utilizing quantum mechanics as the source? It has already been manifest that simulating the quantum system is a challenging task and more specifically for large and open systems it is much more tedious task. In order to store the states of the qubit that theoretically tends o be infinite on a Bloch sphere but we need memory of um bit almost infinite capacity and it is practically infeasible!

The state of the quantum bit is described by a number of parameters and to give a complete computational understanding of the quantum information increases the resources by exponentially leaps. In addition to this the temporal evolution of the system requires also many operations and hence increases the complexity of the problem. The system seems to be modelled with some different tools and techniques and it cannot be directly implemented on a quantum computer as it is not practically fully invented! So what can be the alternative? It is simply to provide the environment for the quantum particle so as to understand its dynamics and it can be carried out by simulation via approximation methods like quantum Monte Carlo, but depending upon the system and the specified problems there are not always good approximation techniques and hence face some limitations. Quantum simulation has been extensively utilized for understanding the efficiency of the quantum computations. With advancements in the manipulation of quantum systems such as atom in optical lattice, tapped ions, nuclear spins can lead a good path towards laser being efficient for quantum simulations. Still the field has not been compensated though there is a great advancement in the quantum simulators and defining the different aspects of simulating the system. The comprehensive introduction to this subject can benefit the researchers to a great extent.

# 4.1 How hard is the Simulation of Big data?

The probability of utilizing the quantum simulator depends on various parameters and nature of the system and after the system has been analyzed with utmost precision, only then can we have a say on the aspects of simulations that can be utilized foe the simulation. As for the big data we have already discussed the complexity of its dynamics, as infinite space of data with instantaneously changing states of the data points added from the heterogeneous and autonomous sources and having complex interaction at their free will. Even if we run the simulator on the physical system then the memory, processor and the other resources required for classification, optimization and sorting may increase exponentially. Additionally, the system is temporal changing and hence at each interval of time there is a different state of the system, making it more complex to handle. In order to know the complexity of the system under consideration we make a situation to discuss how hard it is for big data to simulate on the classical computer and as well on the quantum simulators [133]

Let us consider a general simulation problem of finding the state of a quantum system described in terms of a wave function ψ at some instant of time t and computing some parameter of interest. The solution of the Schrödinger equation:

$$ih\, \partial/\partial t\, |\psi> = H|\psi> \quad \quad \quad (4.1)$$

is given by $|\psi(t)> = \exp\{-ihHt\}|\psi(0)>$. In order that ψ commute we need to discretize the problem so that it may be encoded in the memory as the amount of memory increases exponentially with the state of the system. As an example, if we want to represent N spin particles requires 2N numbers without including the motional degree of freedom. So calculating the time evolution of this system requires 2N * 2N matrix. If we take N=40 implies that storing 240 for ψ requires 4 TB for representing the spin states of just 40 particles. Just for the intuition, The US Library of Congress has almost 160 TB of data, doubling the number of spins would require 5×1012 TB which is about ten thousand times more than the amount of information stored by humankind in 2007. Many simulations have been developed for understanding and analyzing the efficiency of the quantum computations like classical stochastic methods namely Monte Carlo algorithms for many body systems that goes well for simpler complex systems but for frustrated and chaotic systems the simulation needs to be revisited. Other methods for solving many body physics are density functional theory, mean field theory, many-body perturbation theory or green function based methods. This intuitive mathematics defines the complexity of the system. By quantum simulator means a controllable quantum platform to simulate or emulate the other quantum systems. In general there is an initial state of the system and we wish to reach it to a final state via unitary transformation, In simulation we prepare the initial state and define the unitary operation to change this initial state so as to reach the final state and if there exists a 'mapping' of the system with that of the simulator then we can say that simulation of such system is possible. The state is associated with a well defined Hamiltonian that we evolve via quantum gates or via quantum adiabatic evolution or through some other process so that we have some ample amount of possibility to reach the final state and reside on the final desired Hamiltonian. It is not that every time it may be possible as Unitarity of the system is not preserved always. Now for big data the aspects are worth noting to discuss here are as under: [114]

- The huge space of big data is a main impediment that makes it hard for simulation. If we have a controlled system that can be designated by a few particles, even if the interactions are hard but since the number is less and that makes the system easily mapped to simulation. The space of big data is huge and this makes the simulation hard. It becomes hard to define the individuality of each electron and its dynamics in this huge space.

- The space is instantaneously changing in nature that is the system is dynamic in nature. Even if we have huge system of particles is a hard problem but now what if more data is adding up in this space and at an accelerated rate. Then the system complexity is increased and makes it harder to simulate the system of big data.

- Big data is a non linear system meaning that the dynamics of the data doesn't change in accordance to some rule or 'discipline' so is a frustrated and chaotic system. How can we map a system that is unpredictable! It makes the simulation hard to be mapped.

- The interaction between the data is also a big problem. If we have system of space of big data that doesn't interact at all then it is a simple case but that is not practically feasible data points do have to interact with each other. Even if we have all data points interacting at every instant of time, still makes the system simpler due to predictable interactions. But what if the interactions have a free will! It means any data point in this infinite, dynamic and non linear mess can interact with other data point at every instant of time. It makes the simulation of big data hardest.

- Big data is still a research area of interest and not yet traversed in terms of reducing it into a complex system, so it becomes harder to make an exact simulation tool and when it is formulated in quantum language then it becomes hard to talk simultaneously in data language and quantum formalism.

Though for frustrated and chaotic system there are many simulators available but then for a dynamic system parameters need to be changed in different formalism, so there is a need of simulating the problem that can map almost all the parameters of the system. In our thesis we have utilized Laser simulations so as to model the big data via laser rate equations and to map its parameters with space of big data.

## 4.2 Dye Laser

The term LASER means **L**ight **A**mplification by **S**timulated **E**mission of **R**adiation. The central issue that emerged and that can be considered as a precursor for initiating the Lasing action is the light matter interactions, in a way to understand what does light or a photon do when it is incident to the matter. In our previous chapter we have defined a new epistemology for big data in terms of Hamiltonian formalism and that it was an attempt to consider information as physical entity, meeting matter. So now we want to simulate the system with the dye laser and see how it affects this system. The goal is to find the global optimization in big data via dye laser simulations. The dye utilized in this laser is that of rhodium and that forms the central molecule over which we shall be experimenting and simulating. In general the working of laser is based on the stimulated emission. When a phone with specific energy and frequency is incident on the atom in ground state or level 1 then the electron will only excite to the other level only when the energy supplied or incident is equal to the energy difference of the two levels. Once the electron is excited then we may say that it has absorbed some quanta of energy and hence increased the self energy of the electron. So the energy of this excited electron is the self energy of the electron and the added potential due to the incident photon. Now the electron in the level 2 or the excited state is made to absorb another photon incident to it. Now there lies a clever logic! If we incident a quanta of energy that is equal to the energy difference between level 2 and level 3 then it may make another excitation but here is a game to play! We incident on it the photon of exactly same energy! Now how should the electron behave? Due to quantization principle the electron is not and never capable of making an excitation but at the same time it must do 'something' with the electron. What nature chooses for this electron? How does it respond to this change? As implied that there are two energies associated with the electron- one is the self and it cannot be altered and even if altered can change the nature of the electron into some other fundamental particle, an alpha rays or whatever. Instinctively, the energy that is absorbed remains as a cloud around the electron and then the photon of the same frequency incident on this electron makes a resonance and hence 'cools' the electron by taking the absorbed energy out, amplifying the lasing action. Though due to stability parameter the electron may come down itself but the action is spontaneous and in the laser case we make it to come down to lower energy state before the specified time and hence is a stimulated emission. This is the laser action and its dynamics of action. [115]

### 4.2.1 Parameters of Laser

There are mainly three main essential aspects in carrying out the simulation for our system. These are discussed as:

- In dye laser we utilize some dye and here we utilize rhodium and it is the real site of understanding the dynamics of the data point. Once we are able to map the parameters of laser with that of data analytics then we may be able to understand the data point as a case of rhodium atom. Though rhodium is not any form of information but information is physical and at the elementary and fundamental level is nothing than excited or quasi electrons. The rhodium atom has a structure that is well defined and can be encoded for information storage, in the geometric alignment and the excitations. The rhodium has atomic number of 45 and as one electron in the outer most shell. We understand the dynamics of this electron as a data point.

- The other parameter in laser is the energy to input to the laser and it is given in the form of photons. The photons are incident to the rhodium and the mechanisms and dynamics of the processes start. But it is good to know that we may be able to change the energy and hence the power of the laser by changing the number of photons and the intensity of photons.

- The last important parameter is the pulse width that is the site where dye is placed. It contains rhodium and greatly affects the degree of freedom of the rhodium atom. We also see in due course that what happens if we change it and how it may benefit the simulations.

## 4.2.2 Mapping Laser to Data System

It is the most important and significant move to understand how we may map the system of our consideration in terms of the lasing action. Mapping the real system and emulating and simulating it via mapping are the real essence of the simulations. Once we know how the system can be evolved and how to encode the problem then it becomes easier for us to understand the dynamics of mapping. There are essentially three main aspects that need to be analyzed and understood for carrying out the simulations and following are they discussed as:

1. Encoding the problem is one of the challenging tasking i simulation. We need to choose a Hamiltonian that is degenerate and is simpler.
2. The Unitary evolution that makes the system evolve as per specified constraints
3. The final Hamiltonian that we must reach after evolution of the encoded initial Hamiltonian.

In our case we have a relation in a manner that we know the initial point and the final point of the system that is to say that we already know the Hamiltonians. It makes our simulation easier. The main point to understand is that we need to understand the evolution of the system and the constraints are we need to analyze is that the evolution of the system must be slowly varying that is the system must be adiabatic. The system initial Hamiltonian is encoded in the excited state and in our case it is level 3 and the final Hamiltonian is the ground state in a manner that we want to reach the global optimization. So it is favouring our problem. Now the evolution needs to be adiabatic evolving. The essence of the idea is that in understanding the quantum adiabatic evolution of the system we need to address the atomic and molecular ensemble. Many aspects of quantum adiabatic evolution has already been carried on Hadamard operation with phase modulated laser pulse and the selective population transfer has already been implemented in three level system experimentally [116]. The working area in our case is between the level 1 and the level 3 and in this area the main aspect is the adiabatic evolution of this system. Few points are worth discussing here:

- Since when the energy is given to the laser at every instant is constant say $V_1$ so the electron always remains in this working area and the principle of quantization will never let electron to jump into the other level or excited state. This condition makes the system controlled and hence there is a good correspondence and mapping of laser with the quantum adiabatic evolution.
- Now there is other problem that seems to be an impediment in mapping the simulation, if we incident the photon on the electron then it is a sudden jump and the de-excitation are also a spontaneous process so how can we map the two diverging aspects. The answer lies in the fact that laser is unlike normal excitations, it controls the movement of electron and can make it come down or de excite whenever it needs with the help of resonance. So we call it as a stimulated process. So this stimulated process makes the system controlled and hence can be adiabatically evolved as per the constraints.

- If we examine the space between the two levels or our working area then there are almost minute levels that occupy the intermediate levels. So it is not that electron suddenly jumps from one level to other, rather it is in itself a slow process. But at the same time the electron should not be in these intermediate levels and may add up more noise factor to the electron and for this case we make pumping potential or power or energy of the photons in such a manner that this case never comes.

- In order to encode the problem we need to know the nature of the final Hamiltonian to with we have to reach and in our case since it an optimization problem so it surely is the ground state and the initial Hamiltonian must be easy to prepare and in our case it is level 3. It makes our case more related to laser simulation.

So the above points clearly manifests that there is a map between quantum adiabatic evolution and the dye laser.

## 4.3 Reduction to Hamiltonian formalism

Each data point in this space has a defined configuration that instils an energy measure in it. This system can be reduced in terms of Hamiltonians. When information content is stored in RAM, at its physical level it is a capacitor in series with a transistor. Information gets stored on the plates of the capacitor in the form of voltage bearing some capacitance. This distribution of charge on each plate has some energy content and hence Hamiltonian. Now when a data point is stored in a memory cell, whatever type of data it is, a Qubit or fermions. There is a spatial confinement on each data set and it influences the dynamics of data points to a greater extent. This scenario is same as that of an atom trapped in energy well and needs to come out of it. The data set can be in any local minima and needs to tunnel to global minima, but there are many aspects that affect the dynamics of this very data point. Spatial confinement, coupling energy, external energy and the nature of data point are main parameters to be evaluated. To illustrate it we take an example of Potassium Dihydrogen Phosphate (KDP). In this molecule, each lattice point is contained by a double- well potential created by oxygen atom and the hydrogen or proton residing within it in any of the two wells. At no fluctuations, the Hamiltonian for the system in this spin picture is identical to the Ising Hamiltonian. However, proton being a Quantum particle there is always finite probability for it to tunnel through finite barrier between two wells. The tunnelling phenomenon is directly corresponded with the potential given on exterior and it can be magnetic field lines or even photon intensity, giving it a threshold potential to tunnel the barrier so as to reach the global optimization. So the Hamiltonian in terms of Transverse Ising Model is represented as:

$$H = -\sum_{\langle i,j \rangle} J_{ij} \sigma_i^z \sigma_j^z - \Gamma \sum_i \sigma_i^x \quad \text{...............................} (4.2)$$

Here, $J_{ij}$ is the coupling between the spins at site i and j, where $\sigma^{\alpha}$'s ($\alpha$ =x, y, z) are the Pauli spins satisfying the commutation relations. Ising model define distribution of data in space and also the dynamics and parameters that are associated with it, interpreted in terms of creation and annihilation an operator, that is, a data point hops from one site to other site and thus annihilates at one site and is being created at other site. Moreover, there is a perturbative energy that instils potential to the data sets to flip the spins of the lattice. It is being inculcated via external transverse magnetic field.

For simulating this theoretical aspect of Ising model and parameters affecting tunnelling, we have utilized photons as the source of flipping the spins of the Ising structure. In transverse Ising Model, external magnetic field analysed tunnelling process and here we used photon intensity. Our model utilizes Rhodium dye in a similar fashion as that of KDP. The molecular structure resembles in a way that it too has vents due to tri-oxygen atoms, in which protons reside. This molecule is subjected to the external photons that are being varied and its effect is seen on the molecule in terms of the output. We compare the input varying energy and output and analyse intensity of photons gets increasing as bandwidth of the rhodium atom decreases and hence there is more possibility of tunnelling. The overall mechanism of the method can be summed up in the following points:

1. Quantum Tunnelling is responsible for efficient algorithms of optimization. The energy required to pass the hill is reduced via tunnelling. It evolves as per the Schrodinger equation as:

$$i\hbar \frac{\partial \psi}{\partial t} = -\frac{\hbar^2}{2m} \frac{\partial^2 \psi}{\partial x^2} + V(x)\psi \quad \text{...........................................} (4.3)$$

2. Reducing to Ising system we varied laser energy instead of the external magnetic field. The dependence of this reduced system of this is based on energy input and the width of the rhodium molecule under consideration.

3. Pertinently, there is a mapping and correspondence between Hamiltonian and width of the barrier and the laser parameters like energy and the pulse width.

4. We change the input energy by varying it and analyse its effect on the output and the pulse width so that we can infer something about the quantum tunnelling. Simulation yields graphs and then we can understand the real cause of the efficiency.

## 4.4 Revisiting Qubits

Information is never an abstract and intangible form; rather it is has a physical representation. It can be stored in Spin angular momentum of an electron, superposed states of a quantum bit, and a hole in a punched card or some other equivalent. At the very physical level information stored in a digital circuit is binary code that has physical interpretations in terms of voltages Qubits are abstract mathematical objects that having two possible states $|0\rangle\ and\ |1\rangle$, denoting almost all the possible states between 0 and 1 . It is a linear combination of states known as superposition. The special states $|0\rangle$ and $|1\rangle$ are known as computational basis states, and form an orthonormal basis for this vector space. The efficiency of qubit lies in its superposition of being in almost infinite states reducing the resources. There are few aspects about qubit that makes it 'bad' for information and they are summed as:

1. Qubits are in definition abstract entities that are realized from linear algebra and develop the whole discipline via mathematical rigor. Though qubits can be prepared from two photons or cavity QED or even two electrons but they are far from a physical realization for communication and informatics.

2. The main efficiency lies in superposition, storing almost infinite information in qubits but at the time of measurement it loses all the states and collapses its wave function to either 0 or 1 behaving like a classical bit. The information that had been stored in other linearly combined states get wasted thus almost 90% of the information is lost. This creating an illusion of information storage.

3. Quantum Information cannot copied as per No Clone theorem , though making it safer for communication theoretically but once information is lost then there is almost no way to retrieve it.

4. The life time of Qubit has extended from 10 microseconds to 23 microseconds so that it remains in superposition but how does it help? Still we need time for communication and information exchange, it means Bob send information to Alice and it remains in quantum computer of Bob before it is lost!

5. Qubits are prone to noisy interferences like magnetic and electric fields, causing splitting to fluctuate stochastically in time, dephasing the qubit.

6. The process of loss of coherence or 'quantumness' of qubit via interaction with environment, known as decoherence is practically making qubit almost infeasible for computation as it takes all the abilities of a qubit to make an efficient leap

7. The design of efficient architecture for providing an error free and isolated environment for a qubit so that it can maximize its life time and retain its coherence is expensive

8. Not feasible to real open systems.

**Table 1:**

| TIME EVOLUTION | PULSE WIDTH/ DYNAMICS OF A QUBIT USING RHODIUM SIMUATION |
|---|---|
| 0.2 | 3.3145 |
| 0.5 | 3.3149 |
| 0.8 | 3.3151 |
| 1.3 | 3.31499 |
| 1.7 | 3.315 |
| 2.2 | 3.31319 |
| 2.7 | 3.3129 |
| 3.3 | 3.31319 |
| 3.6 | 3.31198 |
| 4.1 | 3.31281 |
| 4.3 | 3.31529 |
| 4.5 | 3.312 |
| 4.6 | 3.312 |
| 4.7 | 3.3135 |
| 4.8 | 3.312 |
| 4.9 | 3.3124 |
| 5.0 | 3.315 |

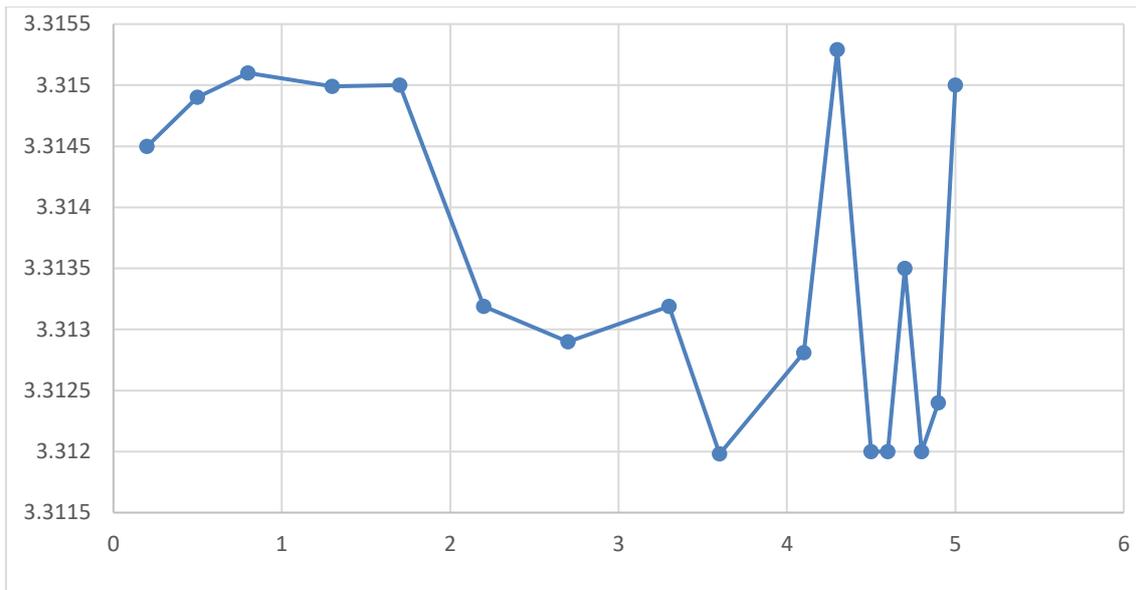

**Graph 1**: Simulated Result of the variation of energy and pulse width with time and effect on 'Quantumness'
(Time Vs change in quantumness via pulse width)

## 4.5 Methodology

We utilize simulation via laser and map the parameters of laser with that of quantum evolution of data point. These have been a progress in simulating topological states of matter with cold atoms in optical lattices the statistical properties of Majorana zero modes can also be simulated in cavity arrays and photonic quantum simulators. We simulate the system via dye laser simulations utilizing rhodium atom as a test charge to visualize the extent of 'quantumness' being exhibited by the quantum system. The main perspectives of the idea are summed as:

1. The parameters of laser are first mapped to simulate the given problem. For this case there are basically two main parameters involved in dynamics of the test data point. One aspect being energy being given as photons and other parameter being the pulse width. Varying these two parameters change the dynamics of the laser system and we map energy as input variation to yield optimization and pulse width in terms of the voltage or potential difference between the two vents created by di oxygen molecule of rhodium.

2. Many aspects could be studied like tunnelling probability based on the varying parameters of the laser as done in our previous research; here we analyze the data point evolution in the cavity of the laser. It is graphically depicted by pulse width-time graph.

3. The transitions of quantum evolution is analyzed and need for transitions from qubit to topological qubit, utilizing topology is analyzed to yield results.

## 4.6 Simulation

The simulation is being carried out on many aspects but three are worth significant: energy/power of the photons, pulse width and rhodium atom. The energy is incident on the rhodium atom that is contained in the cavity of laser having some pulse width; both of the parameters influence the dynamics of the rhodium atom. The process is as:

- The power/ energy are mapped to the external perturbation that is given to the system and it is a variable quantity in our case. It may be increased and decreased depending on the impact on the rhodium atom

- The pulse width is the space that is contained by the data point; it is just like a memory that offers spacial confinement to the degree of freedom of the data point.

- The rhodium atom that is a data point and we may encode our information in the excitations of the electron and hence information can be encoded in it.

There is a confined space given to the rhodium atom and there is a continuously incident photons incident, the atom has quantized shells having electrons. When the photon is incident on the atom it excite the electron from level 1 to level 2 but this level is not too stable so it has to be excited to the level 3 that is comparatively stable state. When population gets accumulated in time then it tends the electrons to de-excite on demand without letting the atom to rest in some quasi state, this process is repeated till we reach optimization

We performed a FORTRAN simulation of the above method using dye laser and rhodium as atom for analysis. Then the results are interpreted that gives an understanding of the evolution of the test data point for understanding the quantum nature of the data. We formulate the relationship between the dynamics of the apparatus using Euler's formula that forms a link between the previous states of the system with that of the next state. We used rate equation of the dye laser to understand the dynamics and performed the related experiment. The equation describing the change in the population inversion density 'n' and photon density $\phi$ can be written as:

$$\frac{dn}{dt} = -rc\sigma_{se}\phi \qquad \text{———————(4.4)}$$

$$\frac{d\phi}{dt} = \left(c\sigma_{se}\left(\frac{l}{l'}\right)n\phi\right) - \frac{\varepsilon}{t_r}\phi \qquad \text{———————(4.5)}$$

$\sigma_{se} \rightarrow$ Stimulated emission cross section
$c \rightarrow$ Speed of Light
$l \rightarrow$ Length of gain medium

$l' \rightarrow$ Length of Laser cavity

$\varepsilon \rightarrow$ Fractional loss per round trip

$t_r \rightarrow$ Round trip time $= \frac{2l'}{c}$

$n \rightarrow$ Difference between population in the upper multiplet ($n_2$) and lower multiplet ($n_1$)

$\sigma_{se} = \frac{\sigma_{eff}}{f_a} \rightarrow f_a = 0.41$ at 300K for Nd: YAG (Boltzmann factor) –gives the fraction of population inversion within $4F_{3/2}$ upper laser level. Care is also taken to adjust the initial inversion to account for the effect that it has on rate equations

$r \rightarrow$ Inversion reduction factor

$\frac{\varepsilon}{t_r} \rightarrow$ Loss

In Euler form,

$$n_{i+1} = n_i - (2n_i r_{se} c \phi_i)\Delta t \quad \text{...................................} (4.6)$$

$$\phi_{i+1} = \phi_i + (n_i r_{se} c \phi_i) - (W_L \phi_i)\Delta t \quad \text{.................................} (4.7)$$

The two dynamic equations are coupled and must be solved in iterative manner that allows successive values of population inversion density and photon density to be used to calculate the next value of each. The initial number of photons in cavity can be approximated from the energy density of the EM field inside the laser cavity from planks black body radiation theory and the number density of excited molecules from the ratio of initial stored energy in laser medium to the pump volume. $\Delta t$ must be small relative to projected pulse width.

The pattern of the change of energy in this regards is very gradual and it shows the extent to which the qubit gets randomized. With just slow change of the energy then there is a randomized behaviour of the qubit. The data points of the change in energy are as:

**Table 2:**

| TIME EVOLUTION | ENERGY |
|---|---|
| 0.2 | 0.50832 |
| 1.3 | 0.82611 |
| 1.7 | 1.08041 |
| 2.2 | 1.39838 |
| 2.7 | 1.71645 |
| 3.3 | 2.09831 |
| 3.6 | 2.28929 |
| 4.1 | 2.607686 |
| 4.3 | 2.73506 |
| 4.5 | 2.86247 |
| 4.6 | 2.926213 |
| 4.7 | 2.98988 |
| 4.8 | 3.053636 |
| 4.9 | 3.11733 |
| 5.0 | 3.18106 |

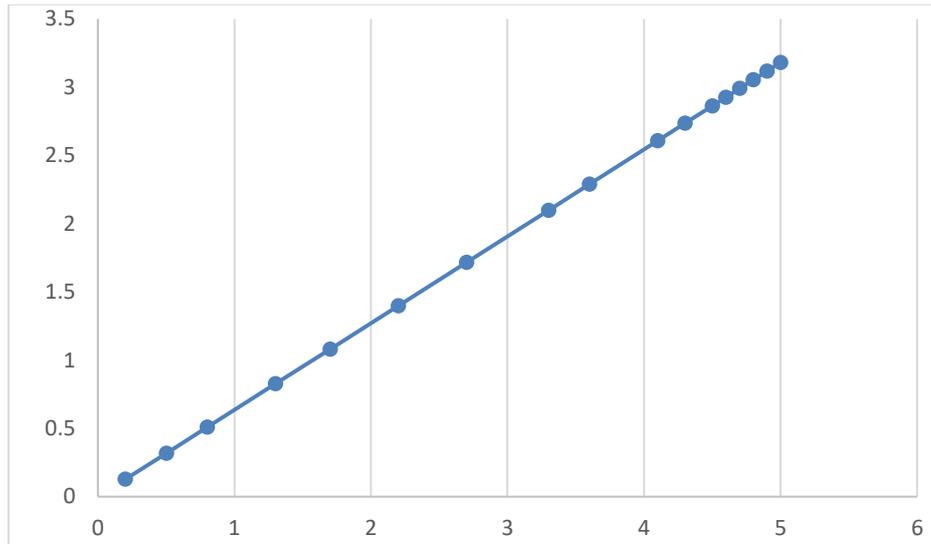

**Graph 2**: Time Energy Changing Adiabatically (Energy-Time graph)

**Tables 3 and Graphs3**

| $\Delta t$ (x10$^{-12}$)sec  Dynamic Parameter of Big Data | Peak Power(x10$^{-9}$)W  Evolution of data point: Tunnelling Optimization | Pulse Width(x10$^{-9}$)sec  Potential: Efficiency Parameter | Input Energy($\mu J$)  Input variations to yield Optimization | Peak Power(x10$^{-9}$)WEvolution of data point: Tunnelling Optimization | Pulse Width(x10$^{-9}$)sec  Potential: Efficiency Parameter |
|---:|---:|---:|---:|---:|---:|
| 5 | 2.04 | 3.43 | 140 | 0.0051 | 3.5 |
| 2.5 | 1.58 | 3.2 | 150 | 0.0064 | 2.4 |
| 1 | 0.635 | 3.7 | 200 | 0.0417 | 2.25 |
| 0.1 | 0.0636 | 3.2 | 250 | 0.086 | 0.86 |
| 0.09 | 0.0572 | 3.31 | 500 | 0.378 | 0.466 |
| 0.08 | 0.0509 | 3.19 | 1000 | 1.077 | 0.313 |
| 0.07 | 0.0445 | 3.29 | 1500 | 1.832 | 0.245 |
| 0.06 | 0.0382 | 3.27 | 2000 | 2.608 | 0.221 |
| 0.05 | 0.0318 | 3.31 | 2500 | 3.396 | 0.232 |
| 0.04 | 0.0255 | 3.3 | 3000 | 4.191 | 0.2145 |
| 0.03 | 0.0191 | 3.15 | 3500 | 4.992 | 0.2059 |
| 0.02 | 0.019 | 3.15 | 4000 | 5.796 | 0.2 |
| 0.01 | 0.019 | 3.15 | | | |

**Table 3:** Determination of Duration Δt & Variation in input energy (Ein) obtained from FORTRAN simulation.

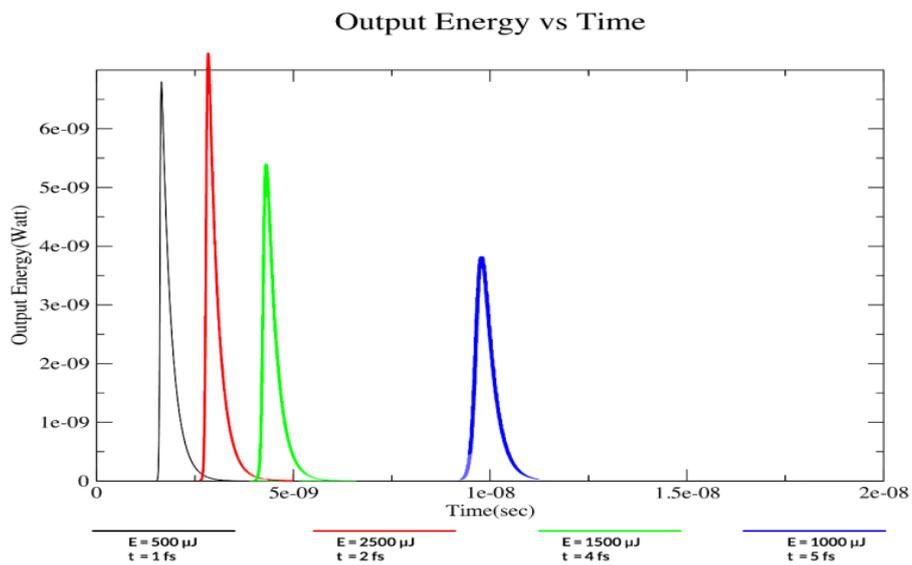

**Graph 3:** Energy-time graph and its variation with pulse width as per FOTRAN Simulation.(Energy to Pulse width for Tunnelling probability)

From the simulated results and graphs there are many conclusions to be drawn. As an interpretation of the results, as the input energy of the system is increased, there is increase in the pulse width. But the interpretation needs to be in a very little time frame so that it becomes so close so as to analyse the quantum nature of the data point and its evolution in this very little time frame. If we plot pulse width that forms the environment for this data point in which it resides with time and that if time is not frame to infinitesimal values then there seems no fluctuations or evolution of this quantum particles. The graph seems to be a sinusoidal in nature in this case, signifying a transit from being quantum to classical and then repetitive discourses. As per the pulse width, the environment in which the data point resides increases to some values to about 500μJ then there is no change in the pulse width and remains constant with varying energy. It concludes the optimum space for the data point. So in this area of analysis, as the input energy is increased, then the probability of tunnelling also increases so is the quantum nature of this data point – a qubit. We analyze the space in which the width too gets increased and we plot time with minutest progressions so as to get understanding in this area of analysis. Energy increases the width so there is increase in quantumness of the particle and then at some point there is decrease of the quantumness and it decreases it shows classical behaviour and additionally there is almost no scope of tunnelling till the probability vanishes to zero and at the same time the width of the potential gets constant and doesn't increase even if energy is increased. At this point of time, the qubit starts decoherence and there is loss of quantum nature of the algorithm and hence can affect the efficiency after some point of time. So it shows a randomized behaviour after this point. The time interval is from 0-1 fs for this analysis. As a corollary of this simulation, it means that quantum algorithms have a fixed time interval to show its efficiency and after this time it behaves classically and this is well predicted from evaluated life time of qubit being 23 microseconds. Our analysis confirms this results, moreover we evaluated the time for decoherence process and hence there needs to be a new concept for data storage and it is the genesis of new information perspective called as topological Qubits.

## 4.7 Topological Qubits

Topological quantum computation is associated with quasi particles called anyons that is a generalization of statistics of fermions and Bosons. Their evolution is governed by inherent property- topology that abstracts it from local geometrical details and perturbations. It provides an error free and invariance to slight perturbations to the system, generalizing a transition from Hilbert space to decoherence free subspace. It is an information unit that is described at the interfaces of condensed matter physics, statistical physics and quantum computations. Topology is the study of global manifolds that are insensitive to local smooth deformations. The idea was first formulated by Kitaev realizing information being viewed as spin lattice model, where the elementary excitations are anyons. There is an exchange of path in this form of information and there is no effect of the local changes as when two identical particles are exchanged then there is almost any concern about the local perturbations. This makes this system error free and there is no influence of local noise as information is stored in the intrinsic property-topology. There is a strong interaction between the information so intrinsic topological class is being studied for understanding topological qubits. It exhibits almost all properties as that of qubit like superposition via braiding and topological entanglement entropy Topological states are physically easily implementable and few physical apparatus that are utilized and support anyons are (i) strongly correlated electron gases under strong magnetic fields supporting fractional quantum hall states (ii) collective states of strongly interacting spins giving rise to spin liquid states (iii) topological superconductors and their engineered variant. Among the topological excitations most interesting is Majorana models obeying non abelian anyonic statistics the simple model is Kitaev chain of 1D spin less p wave semiconductor. Intuitively, each on site fermion split into two Majorana modes and then by appropriately tuning the model the Majorana models can dangling without pairing with the other nearby Majorana modes to form usual fermion and for that they are separated forming a Topological qubit. From the quantum information point, the topological qubit is like EPR like encoding information in a non local manner, making it invariant to noise, decoherence and perturbation, realizing best for open systems.

From the simulated results (Table ) and graphs (Graph ), it is manifest(Kay & Waldman, 1965), (Barry Coyle, et al., 1995)that as the energy of the photon is increased then there is an increase in probability of tunnelling and hence optimizes the algorithm to global minima and then as it starts increasing there is a decrease in probability of tunnelling, implying the optimum energy that needs to be given for tunnelling process and hence efficiency. This tunnelling probability T(P) and its rate R can be shown as:

$$T(P) = e^{-2a\sqrt{\frac{2m}{h^2}(V_0 - E)}} \quad\quad\quad\quad\quad (\textbf{4.8})$$

This optimum point is an interpretation of beginning of the loss of 'quantumness' and hence efficiency. Then after some point the behaviour gets randomized. It is due to decoherence with the environment and hence quantum nature is lost, evolving to classicality. This is a main drawback and point that reverses the efficiency of quantum computations due to decoherence and the noise factor associated with it. So there is a very short life span of qubits that makes it expensive as well as impractical for implementations. As a conclusive remark we expect that the future Quantum Computer needs to be designed in an isolated manner from the environment and that the qubits need to be reformulated in some form of information carrier that is error free and still retains quantum nature. This we expect will be performed via Fermionic computations using topology as its basic source of information. Storing information in this property, being invariant to perturbations can increase the efficiency of big data analytics. Additionally, we can use the concept of adiabatic for evolving the data towards optimization via slowly varying the Hamiltonians, yielding efficient results.[117-120]

# Chapter 5

# Quantum Adiabatic Evolution for Global Optimization in Big Data

*"Indeed to take full advantage of the memory space available, the ultimate laptop must turn all its matter into energy"- Seth Lloyd "*

## 5.1 Quantum Adiabatic Evolution

The term easily gives a complete understanding of the evolution in a manner that it is slow. It seems as if we are applying small perturbations to the system and the system retains and conserves its quantum numbers. Now what type of perturbations are they? It seems as if 'nothing' is happening to the system. The evolution is slow so that the system remains in the instantaneous Hamiltonian without any transition. The energy that we may instil to the system is slow and can be given via magnetic field or an electric potential but the main concept behind it is that the change is slow and energy is given at a very slow rate with sensitive flow.

### 5.1.1 Understanding Adiabatic Evolution via Laser: A mapping for Simulation

There is a direct relationship between the quantum adiabatic evolution and the spectral information in a way that it depends on the transitions of the information carrier like electron or qubit from one energy level to other. The main aspect of understanding the problem, a problem of optimization in specific and any other data science problem in terms of a physical system. The mechanism or the technique is referred to reductionism. In our thesis as already justified that we are utilizing the dye laser for simulation with rhodium atom as the main site of activity. Information can be easily encoded in the rhodium atom, depicting and manifesting the physical nature of information. In some case we may encode the information in the electronic configuration or in some aspects can store information in the geometric alignment of the molecular structure. We may also define a qubit out of it, defining the transition of electron as $|1>$ , as it can be in the ground state or first excited state and in between at the same time designated as the meta stable states. In normal cases there is a transition from one shell to another whenever any energy source or quanta of photon is incident on the shell but as a spontaneous process it is inherit in nature that it comes back to the stable state. Nature loves stability and symmetry! It has already been implemented in many Satisfiability and optimization problems by Edward Farhi [] in this mechanism, an optimization or other problem was reduced in terms of Hamiltonians and then evolved with time till we have results. The theorem is always stated for Hamiltonians that have a separated Eigen values from the rest of the spectrum. There is a well defined mapping from quantum adiabatic evolution to laser action.

1. There is a well defined relation between the gap condition of the adiabatic evolution and that of energy levels of the laser. Since the evolution is gap oriented and every other parameter is set so as to optimize this working area, it can be well perceived from the laser prism. There are many working areas of laser like that of two, three and four level systems and based on the parameters of laser we can understand the 'gap condition' of quantum adiabatic evolution.

2. The evolution is characterised by its manner or formalism of evolution. It is important that the evolution must be slow without changing the Eigen values of the Hamiltonian, it is well justified in laser. There are two main phenomena or principles that are basic to laser, making it coherent with adiabaticity. First is the principle of quantization that makes our data set forced to be in the working region, even if we supply any quanta of photons to it. The main thing to keep under consideration is that the energy must always be equal to the difference of gap (working region). The other aspect favouring this action in adiabatic formulation is the stimulated emission of laser. Unlike spontaneous action we force electron or our encoded data point 'dance ' or move as per our need by simply changing the parameters. These two aspects of laser make it fully satisfactory to simulate the quantum adiabatic system.

3. There are other simulations that may emulate the system well than laser like Monte Carlo methods of simulation, but there is no quantumness it in. Since our data is a qubit and we need to have some perspectives of its dynamics in our system. It is well encoded in laser simulations. The two level excitation is nothing than a qubit, because it can be in the either of the two states or in between the region characterised as a metastable states in lasing action.

4. There is a well defined correlation of the mathematical structure of the two mapping cases. Though one seems to be in differential equation formalism and other in the form of linear algebra with Hamiltonians but at the simple level we are just giving energy to the system in the form of photons and that controllably affects the dynamics of our data point is same as talking in Hamiltonian formalism and gap conditions.

5. The most important parameter in quantum adiabatic evolution is the time factor that theoretically should be infinite so that the data points gets enough time for its evolution. In our case we are using a three level laser. There is a main problem in our dye laser that may directly nullify the adiabaticity of the system and that is a fact that the data point or electrons cannot remain in the level 2 for much time as it is not a stable level. So if after excitation we are not able to 'dictate' our conditions to this data point then there is no mapping to adiabatic evolution. Fortunately, the electrons are settled to a new state that is having less energy as that of the level 2, designated as Meta stable state (level 3). In dye laser the electrons in level three can be easily controlled and hence the time factor of the evolution is satisfied.

6. There is a good map and correspondence between the optimization problem and that of the laser. In the former we need to have a data point with minimum or maximum value and in terms of later it is just to find the data with least Hamiltonian.

## 5.2 Data set and its configuration in space

In dye laser we have already understood the parameters and how they are varied and the effect that it manifests in the lasing action. First of all the information can be well encoded in an electron or excitations as inferred by quantum mechanics, so rhodium atom with electrons is also the physical manifestation of the information carrier. The laser is filled with the rhodium dye and it contains millions of atoms and in each atom there is huge number of electron. As to clear the scenario, we have used about 10g of dye and it contains almost infinite number of atoms as per our need of big data. The molecular mass of rhodium is 102.9g/mol so the number of atoms in 10 g of rhodium is $102.9 * 6.023 * 10^{23}$ which is beyond Avogadro number! Almost 100 times! It is beyond the current understanding of big data.  Now as stated earlier that information is encoded in the electrons and its excitations termed as qubits. When the photon of some frequency is incident on the rhodium atom and this parameter is in our hand then there is excitation in the electrons towards level 2 and as already stated that this level is not good and efficient for the quantum adiabatic evolution so it come to a meta stable state called level 3 and now the electrons are pumped to this region of space with varying energy parameter. Out of this infinite mess let the data in this space be of Avogadro number that is huge for the current understanding of big data. Now let us analyze how data is understood in this region. The level 3 or the metastable state contains the infinite mess of electrons and that to us is the big space of data [121-125]. This big space of data can be only assigned big data only when it satisfies the conditions already posed for the definition of the big data. Big data is characterised by volume, velocity, veracity and complexity. That we have already reduced as a complex non linear and dynamic system. Level 3 is the simulation of the space of big data along with the changing parameters. Following are the points of understanding this space as the space of big data:

1. There is a well correspondence between the excitation electrons and that of qubit so we can consider that the information is encoded in it, making electrons in our case as the data points. Though for encryption it needs more understanding about what text is written but for optimization who cares for it!

2. Big data is characterised by the volume, and that is well given by the level 3 space of this laser contained with almost infinite data points! The data points are more than Avogadro that is beyond big data of today's estimate.

3. Big data is characterised by velocity that is to say that the configuration of the data points in the space is changing instantaneously and that it is dynamic in nature. It is set by the energy parameter in our case. The photons of different intensities are incident on the rhodium with time and hence changing the configuration of the space instantaneously.

4. Big data is characterised by complexity, and there is a complex relations and interaction between the electrons in the level 3 space.

5. Big data is also characterised by the veracity and that is the noise factor that is introduced in the data set along with the information. It has been defined in chapter 3 of the thesis about how the quantumness of the system was lost due to open system and inculcating noise and it's decoherence.

All the above points are sufficient to know the data in level 3 spaces is big in nature or "Bigger Data". At time $t_1$ we incident energy $E_1$ exciting the qubits or changes its configuration in this space and at time $t_2$ we incident energy $E_2$ to the qubits that changes the state and the configuration of this space.

## 5.3 Time complexity for Global optimization via Classical Methods

Whatever is the classical method for finding the optimization in the data sets; it always relies on the basic fact that we traverse all the data points in the space of data and comparing at each step and via iteration calculates the maximum or minimum of the data point in the array or the space. The time complexity varies from $O(n \log n)$, $O(n)$, $O(n^2)$ towards more drastic cases. There are three main problems in classical methods:

1. The space of big data is designated by volume, velocity, complexity and veracity so it is more complex to handle. The number of data points can be beyond Avogadro number or in Yottabyte. It becomes infeasible for the classical methods of optimization to have any say for finding points with minimum energy.

2. There is only iteration of the algorithm and the comparison, what if data points are huge, in an approximation of big data, then the methods cease.

3. There is inefficiency in the memory, CPU and the other resources of the classical system so it can leap beyond the maximal point.

Sahil Imtiyaz (882/FBAS/MSCS/F15)                                                                                              Page 77

## 5.4 Quantum Adiabatic Algorithm

 The overall algorithm of the method is basically divided into four main parts:

1. Reduction Phase
2. Mapping Phase
3. Optimization Phase
4. Evolution phase
5. Simulation Phase

### 5.4.1 Reduction Phase

In this phase we reduce the proposed problem in terms of quantum formulation so as to use the techniques of the later in the former. We define the hyper rectangular space coherent with Hilbert space with data points as vector spaces. There is some configuration so it is taken as potential of the vectors, paving path to deploy Hamiltonian to this system. It was defined and reduced in Hubbard model and the dynamics has already been discussed in chapter 2.

### 5.4.2 Mapping Phase

It constitutes the understanding of coherence and concordance between the parameters of the system and that of the simulated platform. We mapped the lasing parameters like energy of the photons and that of the pulse width with optimisation problem. The information was encoded as qubit in excited electron of the rhodium dye. There is a mapping clearly between Quantum adiabatic evolution and that of the laser. It has been comprehensively defined in chapter 3 of the thesis.

### 5.4.3 Optimization Phase

Once the system of quantum adiabatic evolution is well mapped to the laser parameters then the other main aspect is to be sure that the working area is optimized in nature that is to say the tradeoffs between gap and time as manifested from quantum adiabaticity should be accounted. Though from a simple analysis it becomes clear, there is an inverse relation between the gap and the time of evolution. The equation is as:

$$\tau = \frac{E}{g^2} \ldots\ldots\ldots\ldots\ldots\ldots\ldots\ldots\ldots\ldots (\mathbf{5.1})$$

From the above equation it says that more is the time given for the evolution of the system then there needs to be a least working area. It means that the gap should be least and we know from Bohr model of atomic structure that te energy levels closer to the nucleus tend to have least gap between levels. Now two main questions arise by this fact. One is that how come if we want to increase the time of evolution then we need a smallest possible working region- "The Gap"? The other question is if now we know the optimized gap for the evolution of the system as in our case must be the level 1 and level 2 but it is not the case. What justifies this?

First the system is quantum in nature it is not that we have a particle in some potential barrier and it acts as we can think of. It is a qubit! It acts weird! From the results given in chapter no 3 we changed the gap to see the effect in tunnelling and it was seen from the results that the tunnelling decreased as we increased the gap. But by common sense it must have increased as the gap is maximized and may increase the degree of freedom of the data point. But it is qubit and hence the gap should be minimum to amplify the quantumness of the system. Secondly as an implication of the former statement it is quite natural deduction that the gap for operation is the Level 1 and Level 2 as it is the smallest possible area for working adiabatically. But it is not the case as the dependence to choose any working gap (The gap) is on many aspects which are as:

1. Transitional time: This is the main aspect that governs the working area of the evolution. In our case the level 2 is not the stable level as the electrons that are excited to level 2 doesn't tend to remain there for a longer time rather they come and decay soon via fast transitions and hence we don't get time to evolve the space adiabatically.

2. Stability factor: it is important to understand the geometry of the molecule under analysis. It forms a well defined structure due to its alignment to the least repulsive regions towards stabilizing the structure. Level 2 is not a stable state and hence is not good for our working algorithm. The electrons or the qubits find the lower energy state more stable designated as level 3 for the evolution to be adiabatic.

3. Adiabatic Conditions: While choosing the gap for the working algorithm it is important to note that we have to evolve the system adiabatically so the time must be maximized. Keeping in view the level and other parameters the $\tau$ must be aligned and evolved accordingly.

## 5.4.4 Evolution Phase

The evolution is carried out adiabatically keeping in view the mapping parameters of the laser as discussed in previous chapter.

## 5.4.5 Simulation Phase

The simulation was conducted by changing the parameters of the system and utilizing the FOTRAN program to know the dynamics of the data point and to optimize the algorithm. Now the main aspect is to find the time period of the algorithm for finding the optimization of the data set.

| | |
|---|---|
| 100000 | 4.6 |
| 90000 | 4.57 |
| 80000 | 4.54 |
| 70000 | 4.50 |
| 60000 | 4.46 |
| 50000 | 4.41 |
| 40000 | 4.36 |
| 30000 | 4.28 |
| 20000 | 4.18 |
| 10000 | 4.00 |
| 5000 | 3.41 |
| 3000 | 3.28 |

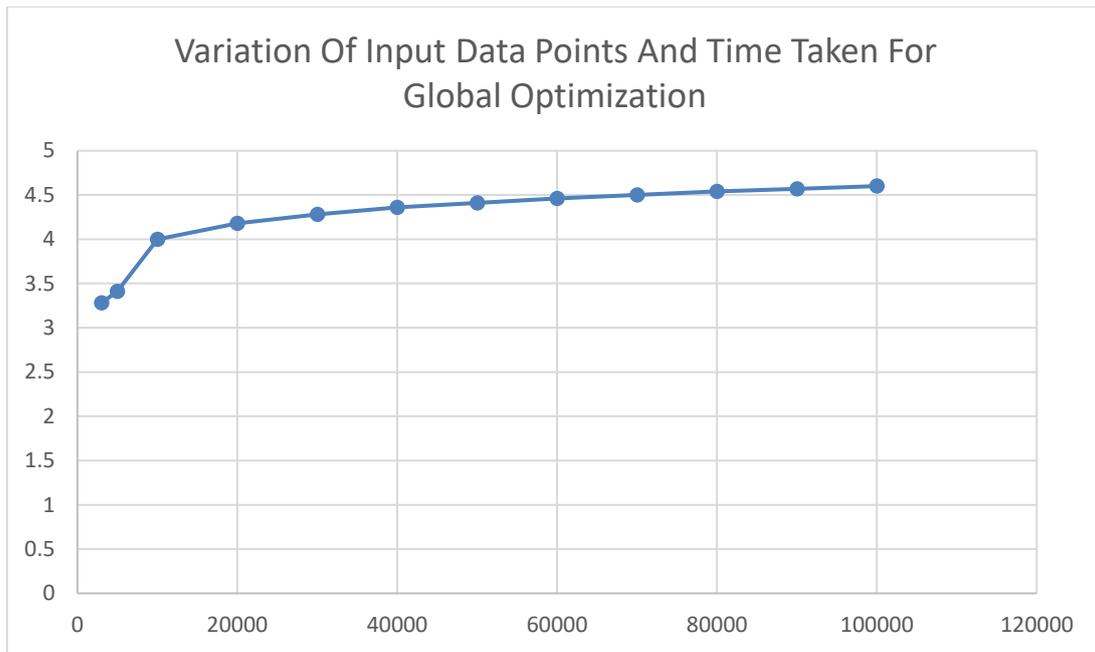

**Graph 4:** Time complexity for Global Optimization with varying data points (Time for optimum Vs Data Points)

From the graphs and tables it is quite manifest that the changing rate of the graph is that of log so the time complexity of our approach and simulation is **O (logn).**

## 5.5 Code

The simulation algorithm has been executed in FORTRAN:

```fortran
PROGRAM SIMULATION
Implicit none
REAL:: C, LPUMP, R1, R2, LO, AA, P, T0, N1, WL,
EL,LL,DT,hm,pulwid, delta
Real:: h, Ep,Eta1, Eta3,Ein, Est,x,vol, Temp, kb,A1, L_LASER
REAL,DIMENSION(99999999):: N,A,Eout, T, G0, XX, Y , Z, u
INTEGER::   K,I
OPEN(UNIT=4,FILE='ONLY1.DAT',STATUS='UNKNOWN')
K      =10000000
C      =3.0E8
LPUMP  =337E-9! wavelength of pump laser in nm!
L_LASER=582E-9
R1     =0.08!reflectivity of output coupler!
R2     =0.99!reflectivity of end mirror!

WL     =(-C /(2.0* LL ))*LOG( R1 * R2 )!loss rate
EL     =( h* C )/ L_LASER !energy of laser photon!
N(1)   = N1
A(1)   = A1
DO I=1, K
N(I+1) = N(I)-2.0* N(I)* P * C * A(I)* DT
A(I+1) = A(I)+( N(I)* P * C * A(I)- WL * A(I))* DT
Eout(I) = A(I)*(1.0- R1 )* EL * AA * C * DT
```

| | |
|---|---|
| AA       =1E-6!cross section area of output beam! | G0(I)= P * N(I) |
| LL       =0.095 !length of the cavity! | T(I)= ( real(i-1) * DT ) + DT |
| P        =3.5E-20!stimulated cross section! | WRITE(4,*) T(I), Eout(I), |
| DT       =0.01E-12 | G0(I),hm |
| Eta1     =0.34!fraction of pump energy that goes as output! | End do |
| Eta3     = LPUMP / L_LASER | |
| H        =6.626E-34!planks constant! | Hm =0.5 * maxval(Eout) |
| Ein      =140E-6!energy of pump laser! | Close(unit=4) |
| X        =0.01!length of cavity! | Open(unit=4, file='only1.dat' |
| VOL      =1E-8!pump volume! | ,status='unknown') |
| Kb       =1.38E-23!Boltzmann constant! | Open(unit=5, file='only.dat' |
| A1       =9.7E-41!initial photon density! | ,status='unknown') |
| Ep       = h * c /LPUMP !energy of a pump photon! | |
| Est      = eta1 * eta3 *Ein!stored energy! | Do I=1,K |
| N1       =( Est/ Ep )*(1.0/vol)*( x / LL )!initial population inversion density! | Read(4,*) T(I), Eout(I), G0(I) |
| | Write (5, *) T(I),Eout(I),G0(I), HM |
| | End do |
| | |
| | Write(*,*) 'Maximum Output=' ,maxval(Eout),N1 |
| | |
| | Close(unit=4) |
| | Close(unit=5) |
| | END PROGRAMSIMULATION |

## 5.6 TUNNELING PROBABILITIES FOR THE RHODIUM DYE ELECTRONS

Rhodium dye can be considered as a potential barrier for the electron. Since the electron has some energy, we find that by considering the average values of the dye laser, the number of electrons in the rhodium dye and the approximate self energy of the electron this energy is far less than the potential of the barrier.

Average power of the dye laser = 60 joules

Average energy of the electrons = $\dfrac{3.75 \times 10^{20} \text{ eV} \times 0.20 \text{ eV}}{N}$

$= \dfrac{7.5 \times 10^{19}}{N}$

N being the total number of electrons.

Consider a case where N = $10^{20}$ electrons

Such that

E = 0.75ev

Transmission probability T = $e^{-2k_2 L}$

Where $k_2 = \sqrt{\dfrac{2m(U-E)}{\hbar^2}}$

Where U = Potential of the barrier

E = Energy of electrons

m = mass of energy

$\hbar^2$ = 1.054 x $10^{-34}$ Js

L = Length of barrier

Here L = 0.50nm = 5 x $10^{-10}$m

To find the potential of the barrier we use Bohrs Formula since the confgram of Rhodium is ……. 5s'

E = 13.6 $Ze^2 \left( \dfrac{1}{n_1^2} - \dfrac{1}{n_2^2} \right)$

E = 10.2ev

Now $k_2 = \dfrac{\sqrt{2 \times 9.1 \times 10^{-31} (10.2 - 0.75)(1.6 \times 10^{-19})}}{1.054 \times 10^{-34}}$

$$k_2 = 1.57 \times 10^{10} \text{ m}^{-1}$$

$$+2k_2L = +2 \times 1.57 \times 10^{10} \times 5 \times 10^{-10}$$

$$= +15.7$$

i) $T = e^{-2k_2L}$
   $= e^{-15.7}$
   $= 1.52 \times 10^{-7}$

**Doubling energy of electron**

**0.75ev x 2 = 1.5ev**

$$k_2 = \frac{\sqrt{2 \times 9.1 \times 10^{-31} (10.2 - 0.15)(1.6 \times 10^{-19})}}{1.054 \times 10^{-34}}$$

$$k_2 = 1.51 \times 10^{10}$$

$$2k_2L = 2 \times 1.51 \times 10^{10} \times 5 \times 10^{-10}$$

$$= 15.1$$

ii) $T = e^{-2k_2L}$
    $= e^{-15.1}$
    $= 2.76 \times 10^{-7}$

**Barrier width doubled to 1.0nm**

**For 0.75 eV**

$$k_2 = 1.57 \times 10^{10} \text{ m}^{-1}$$

$$2k_2L = 2 \times 1.57 \times 10^{10} \times 10^{-9}$$

$$= 31.4$$

iii) $T = e^{-31.4}$

$= 2.30 \times 10^{-14}$

**For 1.50 eV**

$k_2 = 1.51 \times 10^{10}\,\text{m}^{-1}$

$2k_2L = 2 \times 1.51 \times 10^{10} \times 10^{-9}$

$= 30.2$

iv) $T = e^{-30.2}$

$= 7.66 \times 10^{-14}$

The algorithm efficiency lies in the tunnelling probability as the algorithm is not based on the iterative methods or hill climbing. If it compares each data point at each interval of time then it takes almost exponential time for the case of big data. It makes use of greedy approach and of all the ground state Hamiltonians and via tunnelling it optimizes the algorithm. Tunnelling probability depends on varying input energy to pulse width ratio. T is more sensitive to the width of the barrier than to particle energy here. These calculations and simulations clearly depict that the efficiency parameter "tunnelling" of information carrier- qubits encoded as excitation of rhodium atom is unrealizable and is beyond implementation. We don't have tools and simulation methods to make such a precise evolution and the tunnelling parameter has around $10^{-25}$ s probability to tunnel that on our simulation tool truncates to zero. It also instincts us for an open question and that is "How to simulate Big space of data on classical computer. If it makes tunnelling of above precision then it means it takes almost "no time" for global optimization that is quite illogical and hence beyond our technology.

# Chapter 6

# Applications in Neural Networks and Random Number Generation

*"When you don't know the nature of the malady, leave it to nature: don't strive to hasten matters. For either nature will bring about the cure or it will itself reveal clearly what the malady really is"- Avicenna (Ibn Sina)*

Random Number Generation implemented through Quantum- Classical integration. The system includes a plural source of light with coherent states such that each state has an indeterminate number of photons. This varying Photon number produces varying current when input to an avalanche photodiode and the characteristics of this hardware element (Avalanche Photodiode) is changed by varying the temperature, pressure and electric field of the electronic system. The varying characteristics introduce Classical noise of Quantum origin that forms the basic idea for Random Number Generation. The varying electric field on the other hand increases the reverse voltage and hence acts as a gain for the photon incident on its hardware Every state has different photon number and corresponding photodiode characteristic that is being fetch to analogy to digital converter that eventually generates an absolute random number. The gain value of photon is multiplexed with the actual message and acts as a modulation technique.

Communication security in complex computer networks is one of the central and fundamental issue of most information and communication technology and consequently have had increased the interest and scope of cryptographic encryption techiniques.The security of a given encryption technique relies on the crucial assumption that user have access to secret keys. The secret key is the fundamental aspect for the security of some sensitive information over a communication channel. A common way to generate secret key is to use a random number generator. There are three main fundamental properties required for an absolute random number generation. The first main fundamental requirement for a random number generation is equiprobability that is it should have an ability to produce random sequences at uniform distribution having equal probabilities. The mathematical notion for analysing the quality in this aspect can be defined as difference in the apparent probabilistic distribution to the actual uniform distribution. The second fundamental requirement for random number generation is the ability to produce actual randomness, meaning ability to generate unpredictable numbers, since any correlation among the generated numbers is detrimental. If there is any sort of valid rule or equation between the generations of two consecutive random numbers then it would be easy for the third party to easily predict the next random number. The final aspect of the random number generation is the security and this parameter is actually a derived aspect of equiprobability and unpredictability. These three parameters are fundamental and significant of random number generation as if there would be any divergence from the discussed paradigm then eventually the third party could predict the secret key easily. The phases and transitions in the progress of random number generation is from using simple classical methods and then introduction of quantum paradigm proved to be an effective approach for fast generation of numbers with random equation between the consecutive outputs and hence adding speed and security to it. Classical approach being based on the expected phenomena with macroscopic nature so its often predictable on the other hand quantum approach deals with the microscopic phenomena that is efficient from security and speed perspectives. Many methods and approaches have been implemented on the above discussed paradigms and the results are extremely significant. When the speed of the random number generator is increased there is often a trade off with the security of the concerned machine as when speed is increased then it becomes hard for the random generator to output unpredictable equiprobable random numbers with at least no equation governing them. As a result of this fact the classical method inherits the secure platform for the random generation and the quantum platform manages the speed of generation of random numbers. Although both of the approaches have been physically implemented on the circuit and logical level and in some cases both approaches may have lest introduced but the present proposal design is being implemented by using Quantum-Classical integration for random number generation so that we could model a controlled random number machine that would be secure and fast in a logical fashion. The above Idea for generation of random number that can act as secret key in cryptography is being implemented via principle of classical and quantum mechanics A Plural coherent source emitting photon but every individual coherent state being variable in terms of photon number is being fetched to a photo diode that changes it to current as per photoelectric effect. There is a constant fluctuation in intensity of light as there is constant phase difference in each state as a result there is variable indeterminate emission of photon number in each coherent state. Each coherent state with some photon number is fetch to a photodiode and it is converted to current before it is output to ADC the characteristics of photodiode is changed in each state by changing the temperature or pressure of the diode due to some transistor action connected to it. The output is analogy in nature and is digitalised with a ADC. Each digital signal can be decoded in some decimal value and that value is our Random number. The varying Characteristic of Avalanche diode produces three types of classical noise and that will be utilized in actuality for Random number generation. The varying Electric Field changes voltage in diode and it varies the depletion region and hence gain of photodiode. It is being implemented for multiplexing this output with the actual message along with Random number generation. The proposed research maintains the unpredictable nature of the machine by escalating the concepts of Classical Approach and speed is maintained correspondingly by the Quantum Approach. Our goal is to minimize the factor that can apprehend and anticipate the equation between the random number generations. As there is immense unpredictability in the coherent states then adding a variable characteristic transistor will add and embed inherently to the scope of unpredictability and on side grounds the speed of generation remains effective and fast. The main Physical element that is being introduced as a major part of the proposal is utilization of avalanche photodiode for detection of photons from the coherent source. The main focus of paper is to make the generation of Random number secure without restricting the rate of generation along with multiplexing the gain with actual message as a technique for modulation.

There are mainly two main approaches for the generation of random numbers depending upon the type of mode and their method. The two approaches being Pseudo-Random Number Generation (PRNG) and the other is Hardware based Random Number Generation (HRNG). The PRNG is a deterministic algorithm for generating a sequence of uniformly distributed numbers that approximates to the actual random number as the generated numbers are not purely random in nature as it is basically determined by the set of initial parameters and eventually repeats itself after some time and is due to the finiteness of the computer. Initialization of the algorithm is done by an internal state of computer such as current time, mouse click, or even keyboard stroke. The algorithm will produce same sequence after a meanwhile when initialized with the same input. The merit of this approach is speed but security is being compromised. The other safer way for the generation of Random Number Generation is based on Hardware from a physical process including thermal noise, avalanched noise and even time drift. The main demerit of this technique is the rate of generation of random numbers. HRNG can be divided into three main categories depending on the physical phenomena used as a source of randomness. The first category used classical macroscopic approach and they are often predictable as macroscopic system obeys deterministic laws of classical physics. The other category is based on deterministic phenomena of microscopic world, the constituents being electrons or protons [126-128] This approach is also predictable although microscopic world is unpredictable but it is not purely quantum as there are classical instincts contained in it, exploiting deterministic means to distort electronic noise. The third category for the generation of Random number is based on Quantum Phenomena and this physics is actually the integrated view of randomness and moreover the two properties of equiprobability and unpredictability are naturally inherited in it. Recently this ideas was harnessed for Random Number generation using a pulse consisting on average of a single photon travelling through a semi transparent mirror. The mutually exclusive events reaction and transmission are detected and associated with one of the binary outcomes but it lacks speed or rate of generation although the generated values were unpredictable and equiprobable. The other quantum number generator involves utilization of quantum noise from an optical homodyne detection apparatus.

The system utilizes quantum noise generated by splitting laser light. But it had some practical issues and further improvement was devised by employing a laser source of coherent and plural nature which had indeterminate photon number, fetch to photo diode to produce analogy current which is digitalised, and generating random number. Recently the research of generation of random number has reached 80mbps and after resolving it with other constraints it's found to be 68mbps but although the speed or rate of generation of quantum number using quantum spontaneous emission of laser is fast but prone to predictable and hence security may be compromised. The present research proposal makes tradeoffs between security and speed or rate of generation of quantum number using Classical-Quantum Integrated system.

## 6.1 Physical Model Implementation

The main underlying concept that may be physically implemented on hardware comprises of some main components. The apparatus consists of a laser source that is coherent in nature with multiple states having constant phase difference between the consecutive light source and this phase difference for physical interpretations can be changed into intensity so consequently the coherent source has intensity differences. Each state has an indeterminate number of photons and as a direct consequence there is difference in intensity of each state. The output is fetch to a photo detector and in this proposal this is the main area of concern as there is classical integration in the methodology of generation of Random number and that is being deployed in the photo detector. The photo detector that is being implemented in the system is a special type of photodiode named as Avalanche photodiode. The Avalanche photodiode is maintained at varying characteristics by varying the electric field and consequently noise of classical and quantum nature is produced and has a random nature which will be added to the intensity of the photon from coherent source with specific gain. Then afterwards the output is given a analogue to digital converter then consequently generates Random number. The Random Number generation is a direct consequence of the Classical noise that is being produced and controlled by Quantum Phenomena. There are three types of classical noise being pro- -duced in the Avalanche photodiode and they are randomness in the number and in the positions at which dark carrier pairs are generated, randomness in the photon arrival number, and randomness in the carrier multiplication process. The first is not of much significance but the other source of Classical Noise is controlled by Quantum emission of photons by Laser. The last noise is varied by implementation of some specific hardware that might be able to restrict the reverse bias area and change it to higher values so as to broaden the scope of unpredictability. The gain in the photon as a consequence of the varying electric field is multiplexed with the actual message.

### 6.1.1 Mechanism of action

First of all the practical schema will comprise of a laser that will be coherent in nature and as a result there will be phase difference constant in nature and also will have corresponding intensity shifts. The laser source is maintained at constant pressure and temperature as if these parameters would vary then it would inherit a known equation in photon number that is emitted by the laser as the pressure and temperature constraints are classical in nature. When the laser emits its photons, it is found that all the photons will have constant phase difference and hence each coherent state will produce indeterminate number of photons in each state that will be totally having no interference of classical paradigm. The laser is operated at threshold before stimulated emission. So in each state there will be indeterminate number of photons and hence it will act as quantum interference in our proposed model as the emission is clearly a Quantum driven process. After the laser emits indeterminate number of photons in each stage, the Output is fetch to a photo detector and in the proposal an Avalanche photodiode is implemented. The specific photo detector is chosen as it has a random way of producing noise and also the gain factor can be varied due to the electric field. The Avalanche diode changes light signal into electric signal with gain factor component adding to it. The output current of the Avalanche diode fluctuates in the absence of light as well as in its presence. The noise in this current arises from three sources: randomness in the number and in the positions at which dark carrier pairs are generated, randomness in the photon arrival number, and randomness in the carrier multiplication process. All the randomness in noise are classical in nature and hence there is an integrated system of Quantum-Classical approach. The dark carriers are random and they are produces when even no light is incident on it, there is also randomness due to number of photon incident on it and as it has been already proposed that the source produces indeterminate number of photons in every stage so the second type of noise in the Avalanche is also purely random in nature. The last component of noise is due to multiplication gain process which is to add gain factor to the signal. Nowadays we can add a gain of greater than 1500 under the reverse voltage. Since we know that the reverse bias voltages is limited as above its specific value its minority carrier's avalanche off and break the depletion region and it becomes a sort of forward bias. So the limited reverse bias is a practical impediment in our proposal but we deploy a controlled bias voltage by never making the device to exceed the reverse voltage threshold. The configuration that I discuss is simple but practically it may be deployed with more circuit schema. The problem is sought out by adding a transistor with it and it is used for modulation of the actual message so as to make the message to travel with high energy carrier wave.

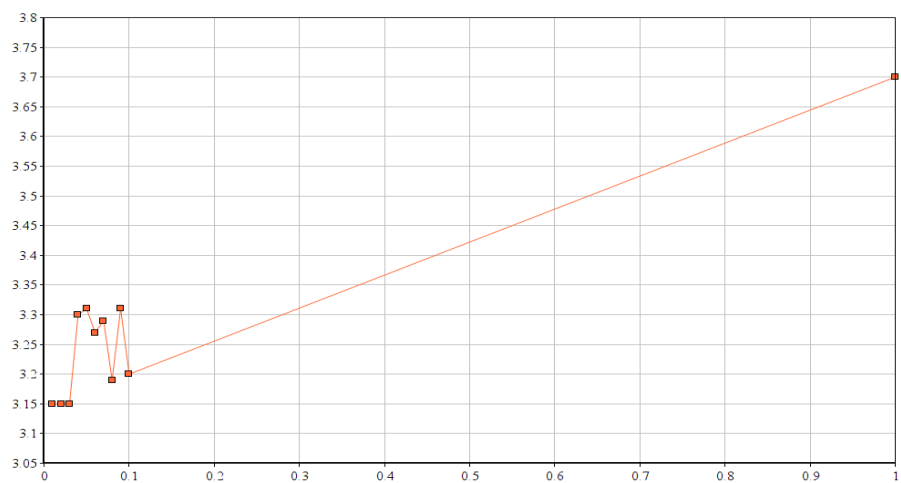

**Graph 5:** Random Nature of Qubit between 0-0.1 fs (Rate of change in classical-quantum transition)

The above results show the loss of quantumness to the system and its random behaviour and hence are best for random number generations.

## 6.2 An insight on Quantum field formulation of neural networks

Ising and spin model has already been utilized in describing the dynamics of neural networks but till now there is no literature to my knowledge on correlations of the neural networks and condensed matter physics. The neuronal activity and its interactions seem in concordance to Hubbard Model utilizing the many body innovation, due to the fact that in former case there are possibly two states and in later there are fermions that can be in only two states, even if it is a Qubit, can be mapped to fermions via Jordan-Wigner transformation. Moreover, In such a system there is same 'free will' in interactions between any two points of the lattice points as that of neuron. Quantum Condensed matter physics provides a platform to redefine the neuronal structure and reduce its essential parameters in the formalism of quantum matter, conceptually driven by quantum field theory. Once the system of neurons is reduced in terms of Quantum field terms then we expect new understandings about functionalities of brain, consciousness and cognition, though it may need more hard work on long distant correlations that may not be provided in this framework..

### 6.2.1 Neural State Network to fermionic State

In neural network theory the state of a system is described by $q(r, t)$, denoting the activity of the Indi- vidual neuron at time $t$. In quantum theory state of system is denoted by $\psi(r, t)$, a superimposition of states. This can be well described by the Fermionic states via Jordan Wigner transformation. The n Qubit state can be envisaged as a n fermions states with $2n$ modes. This is because one could pair those $2n$ modes to get $n$ pair and then use one mode of each pair to fill with fermions. '$n^j$ sites each of which can be empty or occupied by a Fermionic particle, such a site is called as Fermionic modes (LFM's). The extended Hilbert space of this system is Fock space, spanning $2^n$ basis vectors. Everything related to fermions can be expressed in terms of creation and annihilation operators. This transformation addresses the superimposed states of data and decreases the complexity

### 6.2.3 Neural state superposition to Fermionic Braiding

A neuronal configuration may be described as linear combination of neural patterns. And among this set of neuron, there is a set $x$ that defines the neural patterns, a special configuration representing some information. In the similar fashion as in superposition of qubits, Majorana fermions do interact with each other via braids. It is the Majorana Fermionic computations that can store information in either one or zero and in the superimposed states in what is called as braiding. Just as the quantum spin, particles can traverse paths or braids traced by the particles, can also be in superimposed states. The particles interact via braiding, by drawing out sequence of swaps between neighbouring pairs of anyons. In this formalism, information is stored in the 'topological property' that is a collective property of the system.

### 6.2.4 Synergetic Order Parameter to Partition function

A relation has been shown between neural order parameter and quantum probability coefficients. In linear combination each pattern is represented by a corresponding coefficient. The coefficients describe how much a specific pattern is represented in the actual state of the system, or how probable it is that the corresponding pattern will be recalled. So time dependent coefficient encodes quantitatively the meaning of patterns, mapped to partition function $Z = \sum_0^\infty \exp(\frac{-E}{kT})$ that is also a probability function of determining which among the states the pattern can be. Once we know the Hamiltonian (Hubbard) then the partition function can be easily evaluated.

### 6.2.5 Spacio-temporal integration of neuronal signals to Feynman Schrodinger Equation

It is empirical that the dynamics of a neuron is the Spacio temporal integration of all the input neurons to this specific neuron. Since the formalism is in terms of creation and annihilation formalism, so it is best described by Feynman's equation. In case of neural networks of a neuron at position $r_2$ and $t_2$

$$q(r_2, t_2) = \iint J(r_1\ t_1\ r_2\ t_2\ (q)(r_1\ t_1\ dr_1\ dt_1) \quad \text{...................................... (6.1)}$$

This is mapped to Feynman form of Schrodinger equation as

$$\psi_{r_2\ t_2} = \iint G(r_1\ t_1\ r_2\ t_2\ )\psi(r_1\ t_1\ dr_1\ dt_1) \quad \text{......................................... (6.2)}$$

### 6.2.6 Learning Rule and Green's Propagator

The transmission of individual synapse is determined by Hebb learning rule and in similar corresponding fashion green's propagator is the learning rule in this mapping. Since Green propagator contains the action term as well so it understands the change in two infinitesimal intervals and hence learns by this evolution. The action of the neural dynamics can be evaluated from *S*: for field *w(x,t)*, we have

$$S(w) = \int L(z, t, w, \partial_{zt}) \, \partial_x \partial_t \quad \text{............................................} \, (\mathbf{6.3})$$

Where L = T- V is the Lagrangian, difference between the potential and the kinetic term. The probability distribution of w (x, t) is

$$p(w) = e^{-S(w)}/z \quad \text{....................................................} \, (\mathbf{6.4})$$

The dynamics can be formulated in terms of Klien-Gordon equation, pertinently defining the correlation between two neurons at w ($z_1$, $t_1$) and w ($z_2$, $t_2$)

$$\langle w(z_1 \, t_1) w(z_2 \, t_2) \rangle = \frac{1}{z} \int w(z_1 \, t_2) \, w(z_1 \, t_1) e^{-S(w)} Dw \quad \text{............................} \, (\mathbf{6.5})$$

Where Dw denotes integral over all paths from w ($z_1$ $t_1$) to w ($z_2$ $t_2$). The way neural networks can be related to Feynman propagators and diagrams so that correlation between two neurons in space and time.

$$\langle w(z_1 \, t_1) w(z_2 \, t_2) \rangle = G(z_1 \, t_1 \, z_2 \, t_2) \quad \text{.........................................................} \, (\mathbf{6.6})$$

Where G is the Green's Propagator and encodes all action potential of the microtubules of neuron

# Chapter 7

# Topological Field Theory of Data Science and Brain Dynamics

**Discussions from Dr Mario Rasetti, Dr Qin Zhao, Kazuharu Bamba and Dr Mir Faizal**

*Dear friend, your heart is a polished mirror. You must wipe it dean of the veil of dust that has gathered upon it, because it is destined to reflect the light of divine secrets – Imam Gazali*

## 7.1 A Way towards a Theory

Gödel incompleteness theorem gives an insight about the nature of a theory that can well describe the dynamics of neurons, keeping in consideration the complexity and incapacity of reductionism. Though there have remained many controversies about his viewpoint- a tussle between optimists and pessimists but there are few philosophical attributes associated with the theorem that I believe can help us to interpret the nature of theorizing neural dynamics due to its complex nature. David Hilbert recognized that axioms tend to be self consistent if we cannot prove that a statement S and its negation ~S are both true theorems. It can be considered complete if for every statement S we can prove either S or ~S is a true theorem (in terms of language) This statement of Professor Hilbert When taken along with Kurt Gödel we can have an intuition of the theory or a mathematical structure completely defining the neural dynamic. Naturally there is a two different dynamics being active at tubular level of neuron where voltage between membrane define the action and at behavioural and cognitive level the stimuli seems different or I may say almost complementary! We have always sought this diverging formalism as a real impediment but I expect that it may prove a greater significant edge in developing or initiating a definitive mathematical structure for brain dynamics. Let somehow we define a theory **T** that defines the action potential and the dynamics at the tubular level and then formulate somehow some other theory **H** that define the behaviour, cognitive and psychological parameters at aggregate level then there needs to be two condition whose completeness can 'to some extend' initiate a theory and a mathematical formalism of neural dynamics. Let **G** be the group (in group theory formalism) that forms all object of theory **T** and **R** be a group that forms all objects of theory **H**. Then,

- If the intersection of the two groups is an identity group (for not violating group theory), technically a null set, T ∩ H =Φ meaning the two theories or the two groups are adjoint groups. and
- Despite of the ergodicity in the system, there is manifestation of only one theory (Say H) with some special set (Later may be defined as braiding Category ) (Hilbert view on Completeness)

Then it may be believed that the inaccessible law of reality as posed by the theorem may 'to some extend' be solved. Our epistemology shifts from arthematics (posed by Gödel) towards groups or more specifically category. Gödel proved mathematics to be inexhaustible in that finite set of axioms cannot encompass the whole mathematical world. I believe that it is due to our inaccessible conduction to know what an 'intrinsic property' of matter really is! [129-132]. we mostly understand the responses of matter to different scenario and statistically or based on coherent and hierarchical structure of literature we make a step in explaining it. May be it is just another way towards reality! Mathematics may now leap from integration, summation and arthematics towards structures and diagrams like arrows in category theory. Now in order that we define what theory can be devised for brain dynamics I feel the two theories that almost satisfies the above posed condition and discussion is Quantum field theory and Category theory. The manifestation tough at philosophical level seems equivalent but these are two adjoint theories satisfying the first posed condition and secondly that at functional level topology gets manifested satisfying the Hilbert view of completeness.

- The dynamics at the tubular level can be well defined by Quantum field theory (I had discussed it in my previously sent proposal) we use Quantum field mathematics to deal with action, learning and dynamics at the membrane of the cells. The Hebb's rule can be well defined by Green's propagator, neural state network to fermionic state, neural state superposition to fermionic braiding, synergetic order parameter to partition function and Spacio-temporal integration of neuronal signals to Feynman Schrodinger equation. It defines theory **T** in our case.
- The behaviour of the neural at the nerve conduction level is in its very essence topological in nature. It 'illuminates' the specific nerves among the bunch of fibres. To every emotion, understanding, response and cognition there is 'something' happening at the tubular level in terms of QFT but 'whatever happens there' it shows a 'special' topology at conduction level that shall be defined by topological field theory as proposed by MR and EM. It defines theory **H** in our case. So whatever happens in fundamental using QFT the theory but at the conduction our analysis regime it is theory **H** that dominates (Hilbert view of Completeness)

Now to link these two adjoint perspectives, we need to device a comprehensive mathematical structure or at least initiate the mapping the parameters of QFT with that of topology. It may not be possible if we talk in terms of pure arthematics, I feel this mapping and corresponding can be done via Category theory. As an intuitive feeling, what is persistent homology (PH) in terms of QFT? In PH we let the structure evolve with time and at the end get structures for analysis, at fundamental level it may be regarded as creation-annihilation operators in specific time span, yielding a structure after some time! (It is a radical statement to be made at this point but it was for feeling the third point)

If this Idea works then we have to prove the two stated theories must be adjoint in nature and complementary to each other and that too one of the theory proving right at some level of analysis. There is need to map the parameters of two theories and yielding a 'new' mathematical structure. Though it is not an easy task but may be taken for initiation of understanding brain.

There are three phases of this essay that will at least give an idea of the working methodology. The three steps are as:

1. Quantum Field formulation of neural networks
2. Topological interpretation
3. Mapping and Mathematical formalism

From the papers of Mario Rasette and Emaneula Merelli specially Topological field theory of data: Mining data beyond complex networks, Topological mean field theory of data, Survey of TOPDRIM application of topological Data Analysis, Non locality, topology, formal languages: new global tools to handle large data sets and homology scaffolds of brain functional networks there are few insights that need to be taken into consideration so as to define topological step in our step wise categorization.

- The affinity of each simplicial complex form higher dimensional structure, gluing to other structures forming hyper structure. It makes reduction complex and impossible to reduce and can't be reconstructed via reduction.
- In an attempt to simplify the composing elements and their interactions, the global properties are typically very hard to predict.
- There seems to be a domain that is metric free, having strong implications in artificial intelligence and neural networks.
- Complex systems are non identical interacting non-linearly making the reduction and the construction of mathematics harder.

## 7.2 Topological Field Theory of Data

A paper that I have gone through on initiation of new theory on topology for understanding data in new paradigm has been few of Mario Rasetti and Emaneula Merelli titled as "Topological mean field theory of data: a program towards a novel strategy for data mining through data language" in which the main points worth discussing are discussed as and I try to interpret them for brain dynamics at least to give my little perceptions about the aspects:

- The data needs to be metric free and the space forming is neither a Hilbert space nor a vector space but a topological space. Instead of assigning coordinates a parameter 'n' that is subjected to various constraints and range 'distance' and based on this formalism define a hyper-graph that take vertex and edges as signals if the constraints are satisfied else its characterized as noise. The system is reconstructed topologically with its simplicial complexes with its 'n' approximations. These hyper structures glue to lower dimensional simplicial making the system hard to reduce. This metric free is philosophically inheritance of intelligence to the system where each point in space 'feels' the existence of the other points and its neighbourhood. So there is need of a 'sensitive' reduction of the system. Whatever states system acquires, seems to be having free will between its initial form and the final state, any path can be traversed. It can be related to stability factor of the system and principle of least action in terms of Lagrangian. So the 'action' parameter can greatly 'control' the complexity of the system in terms of 'gluing' to lower dimensions at free will. In this regard a link or the mathematical formalism between persistence homology and creation-annihilation operators can make the system simpler. We may designate two points initial and final in terms of temporal parameter and define the mathematics of in operator formalism and persistence. It can to a greater answer the question that if the composing elements and their interaction are highly simplified the global properties are typically hard to predict.

- The other aspect that is discussed is the non identical and heterogeneous nature of the system. The brain being a complex system, at its fundamental level is a complex heterogeneity in terms of action and permeability of membranes. If somehow we are able to reduce the system in some identical particles and understand the dynamism of at least of its dynamics between two identical particles. Additionally, if we can know almost all the possible ways of its communication of the system then we may reduce its complexity. In our proposed idea since the neural dynamism at the fundamental level is defined by QFT and under its considerations there are only fields associated with nature and behaviour of the system at the fundamental level. At elementary level particles are quantised and continuum is just at other perspective of same thing (global) in the same manner though the system is non-identical but if they can be described by field and governed by field equations then system can be reduced.

- The metric free system is also a central discussed part and to obtain that we need to define such formalism that sustains symmetry unlike other aspects of ideas about brain via symmetry breaking. We state our proposed Idea in terms of action and link in terms of topology. We propose that the thoughts and thinking process is circular and equations seem to be linear and correlate it with gauge theory and non abelian formalism. It makes the system metric free and hence we can achieve more as per objectives.

- Though topology is a useful tool to understand the large data sets and the neural dynamics but there seems to be a compromise. While it focuses much on the patterns it naturally makes a sort of 'compromise' in terms of the local dynamics of each data point. Like if we see some 'special pattern' topologically may at fundamental local level be the dynamics of 'many' different paths outputting the special pattern. So how can we know what 'actually' has happened at the local level? With the inculcation of two different theories in the proposed idea it may take more care about this aspect. The analysis of electroencephalographic activity (EEG) has shown that cortical activity doesn't change continuously with time but by multiple spacial patterns in sequence during each perceptual action.

- Abstraction is a powerful tool and by defining the parameter n for constructing the simplicial complex makes it a metric free. The epistemology is in terms of topology where patterns are focussed than the currency of information. So if the system can be locally understood in terms of information unit like bit, qubit or fermion than the property of abstraction is more significant. Moreover 'n' may differentiate between noise and information still without information unit we may not be fully differentiated this 'unimportant'.

This constitute the theory **H** in our proposed case and now as we have already pointed out the nature of the proposed theory for neuronal dynamics should essentially be two complimentary theories and that only one must be true at the global aspects. Whatever and however the fundamental level behaves like the changing action potential at membrane and chemical potential at synapses there is a significant order at global level. How can we map the dynamically changing nature of neuronal dynamics with global topology? To me more precise, the dynamics at the 'boundary' is changing and many different action potential can have the same structure at the global level, so how can we differentiate? But through EEG and other data, it has been known that the cortical region doesn't change continuously but very specifically, so that the structure at topological level is significantly interpretable.

So from the two steps of our discussion, it is quite manifest that we are almost successful in devising two different theories that seems complementary to each other (though may define the same reality). The theory **T** implemented at the neuronal fundamental level is the Quantum field theory and we have already mapped its platform with that of learning and neural networks. The opinion is that the simplicial of these complexes is aggregated at the global level, it is the pattern and the structure that dominates and as per Hilbert's perception of completeness for a theory (Among S and ~S, one must be true for completeness) it gets clear that topology seems true at the global level. So this Gödel-Hilbert definition tries to construct a theory for defining brain dynamics to some extent. So whatever happens at the fundamental level we have no concern now with its dynamics, what forms our understandings is the structure and patterns. But at the same time while constructing the theory of Cognition, perception, understandings and emotions, theory H (QFT) is essentially important. So both theories are thoughtful to devise a comprehensive theory for understanding the dynamics of brain at different levels (fundamental or structural) and at different phases (brain during sleep, anger and other stimuli). Now there is a bigger question to be answered. How can we map the two theories to each other? What must be the formalism that can relate the aspects seen at global level with that of local level? The linking cannot be directly in terms of explanations at this point of time but I expect that it can be correlated in terms of mathematics. [135]. The system cannot be added due to its continuum nature, integration also seems to be not a good tool as the system is highly complex and heterogeneous in nature. This mathematical structure is beyond sets towards group theory. In a similar way as physics evolved from group theory towards gauge invariance and the gauge-gravity correspondence where there are new insights of mathematics explaining reality. In order to device a mathematical structure, the formalism should have few aspects as:

- Due to diverse behaviour of system the mathematics should be in terms of 'belongingness' rather than absolute.

- It must possibly give understanding of all the possible path (almost infinite) that a neuron can traverse for signalling so as to have a comprehensive say about the system
- It must address the complex system as brain is highly complex system, include dynamic nature and long correlations
- It must formulate a 'universal parameter' for knowing the system dynamics more closely.
- It must relate the microscopic levels with the macroscopic level. A way to link neuron dynamics with behaviour.

Such type of mathematics shall not be confined to integration and summations as the system is more dynamical and complex. We can rest our opinion that all the above points and the implications of cognition and perception can be well addressed by Category theory and Algebraic Topology. Category Theory can bridge the gaps between the different levels of brain activity, from microscopic to macroscopic aspects. It proposes radically a new way in system modelling focusing on the relationships and interactions of the elements so easily addresses the complexity issue of the brain system. The definition of category theory can be studied from literature.

## 7.3 Brain as a Complex System

The basic fundamental unit of information processing are neurons and cortical area comprises of approximately 100 billion neurons connected intrinsically to form a highly complex system. Around $10^8$ tubulins in each neuron switch and oscillate in range of $10^7$ per second, giving information capacity on single cell value at microtubule level of $10^{15}$ operations per second per neuron with the total brain capacity at this level translating to $10^{26}$ operations per second. It can give an intuition of how complex the network is! The network formed by this region is highly dynamical in nature, in terms that it instantaneously changes with time. The following points are worth noting about the complexity of brain dynamics:

- The macroscopic nature of human behaviour at its fundamental level is the dynamics at the fundamental neuron level. It can be thought of some 'special' action potential at the membrane (boundary) communicating with any number of other related neurons forming a 'special structure in MRI, giving rise to specific responses in behaviour. It reflects the distributed nature of the computations. Behaviour is 'entangling' of some set of neurons.
- Each neuron has a 'free will' from our perspective to take part in any computation of brain. Any neuron has a free will in terms of interactions. The range and nature of each interaction changes with spacial and temporal aspects. Any neuron at some point of time may take part in the computation but in that part other neuron may not take place. This makes the system more complex and poses a question of how to structure this so called free will?

- Due to uncertainty in the interactions between neurons, instantaneously changes the neuron actions at some frame of time and hence makes this system dynamical in nature.
- Long range correlation in brain is also of central importance. It depicts the non local nature of neuronal dynamics. It also makes the system more complex to understand as for what is the cause of its non locality.
- The non equivalence between the dynamics of macroscopic behaviour and the dynamics in terms of neuronal dynamics. What might be the bridge between microscopic atomic and molecular unit and the macroscopic neural activity as we observe it?

These five points makes it almost clear about the complexity of this system. It is a complex system.

The proposed idea is at its core broadly based on three levels of functionalities. There is a hierarchical binding of biological units to form behaviour, a quantum of emotion or understanding and perception. It starts from a tubules to neurons and then synapses, dendrites and at the end brain till it evaluates stimulus via understandings, cognition and emotions. As a part of topological field theory of data we know that instead of describing a scenario in terms of graphs we define a sub graph that is working as per a proximity parameter α, making it metric and coordinate free. Depending upon this parameter a structure is formed for further analysis using topology. The simplicial complexes forming the structure further joins with the higher dimensions and it in turn can join further with lower dimensions. It makes reduction a partial process for 'freezing' the system in components. The idea rests on the following four main steps:

- The lowest dynamics at the fundamental level is described via quantum field theory that defines the neuron dynamics in terms of action and propagators.
- The macroscopic impulses are topological in nature and we will utilize tools in understanding this aspect
- There is a need of defining a coherent mathematical structure so that we may devise a platform for interpreting the dynamics in terms of category theory, braiding groups.
- There is a mapping of QFT dynamics with that of topology; it can reframe the intrinsic motion in terms of topology and group theory. Action can be reformulated in manifold formalism, annihilation and creation operators in terms of persistent homology.

So far there is no coherent and comprehensive mathematical structure that can define the neural dynamics at fundamental level and its manifestations in the behaviour, cognition and understanding of humans. This statement seems rather more philosophical that can be asked in interrogation, is there any mathematical theory defining neuronal dynamics? This fundamental question gets more complex by Gödel's Incompleteness theorem of our incapacitative attempt for a theory, understanding and defining reality of formal systems. So in order to understand the correctness of some theory T, it is necessary to build a stronger theory X, inducing an infinite regression. Moreover in regards to the initiation of such a mathematical structure of neural dynamics and theory of cognition, the other related complexity lies in nature of brain dynamics. The philosophy of action potential at the membranes and tubules is quite different from the aggregate 'behaviour and response of human psychology. It clearly depicts brain doesn't have an aggregate nature. Technically, the dynamics is not an integrated action at the fundamental neural level. This poses a question of reductionist approach of dealing with complex systems. When the components and working of the aggregate system is not reflection in component level then how to device a theory for such a system! Neuronal dynamics is fundamentally irreducible! Two main questions posing while dealing with reductionism are: What are the elementary components belonging to the complex object? And how well the lower components of a complex object reflect its nature and essential properties? Galileo pointed out that the language of nature is a mathematical in nature but biology is still unnoticed because of more complexity and non linearity. On other hand Schrödinger writes in his book, what is life? That maths won't play any role in biology not because of its simplicity but because it was much to get involved into it that is inaccessible to mathematics. So the mathematical formulation of life is an open philosophical and theoretical question.

## 7.4 Mapping via Category Theory: A Start-up of mathematical structure

In above section we have defined to some extend the dynamics the local and fundamental dynamics at tubular a neural level that we have defined s H theory. Now topology has its role to play as EEG studies show that cortical activity doesn't change continuously so the 'patterns' at global level have a significant interpretations. The mathematical structure defined in terms of Partition function, action and green's propagators have now defined the dynamics of the neurons. Now the most important part of our discussion is the mapping and correspondence between the QFT and the topological interpretation that we believe that the former is being implemented at fundamental microscopic level and later at microscopic level. It can be related if we will be able to link and map the two concepts via a mathematical framework known as Category theory. Though in this essay I may not be able to map the two seemingly different realities but at least I may present the fragments of my thoughts so that you may get my slotted reality about the concepts.

There is a continuous action potential maintained at the membrane of neuron, chemical potential at synaptic nerves and transmission of signals. Almost there are infinite possible ways of this diffusion and dynamics. Following below is the main points of how we may map the problem and that category theory is the possible mathematical framework:

- The fundamental parameter whose dynamics decide the fate of the neuronal functionalities is "action". As defined mathematically via Lagrangian as the difference in the potential and kinetic terms of the system. At the boundary of the membrane the ions flow in and out via semi permeable membrane that changes the overall energy of the system and hence action may be associated. An ion outside the membrane can hop inside due to difference in potential but there is no well defined path that it may traverse during its cycle. There are almost infinite paths from two points that is in similar with the Feynman's propagators. The path formulation of this can be defined as:

$$Z = \int D\Phi e^{-S[\emptyset]} \quad \text{...................................................} (7.1)$$

Φ: M→X is a smooth map between two manifolds M, X
S[Φ]: action functional typically depending on value Φ at each point of M

$$S[\emptyset] = \int L(\emptyset, d\emptyset) * Some\ other\ parameters\ insignificant\ here\ \text{..........} (7.2)$$

Now this path can also lead to some correlation function, intuitively a fact that while drift the elements or ions are themselves correlated. If we have some defined observable. If $< O_1, O_2, ..... O_n >$ be some observable then the correlation function

$$< O_1, O_2, .... O_N >_g = \frac{1}{Z} \int D\emptyset\ O_1 ..... O_N\ e^{-S[\emptyset]} \quad \text{.........................................} (7.3)$$

Where g indicates metric in M and if it is independent of g then we can have a topological field theory. Metric independence implies diffeomorphism invariance, should leave the path integral invariant. The action can be mapped in terms of topological formulation as a 3D compact oriented manifold M, a compact Lie group G, SU (N), a principle G bundle P→M and a connection 1 form A ϵ R

$$S[A] = \gamma \int_M^n Tr\ (\ Formulation\ in\ terms\ of\ Topological\ construction)\ \text{.............}$$
(7.4)

The above structural mathematical form of action is in terms of topological formalism. We analyze the QFT from an angle where it appears as a way of transporting the geometric and dynamical structure of space time into algebraic form of physical states and observables that contain structures. This is an initiation of mapping action in topological platform. So the moment has an inherent geometry and topology associated with its dynamics and we shall attempt to focus on them.

- Between each point at the interface of the membrane there are possibly infinite infinitesimal ways to hop inside as described by Feynman's diagrams but due to stability parameter it must have a finite set of paths that it can traverse. As per principle of least action this probabilistic nature is further narrow down and there is a well defined path. If we mark all the dynamism of action at the boundary of the membrane then it forms a diagram with points and arrows and then comes the play of category theory. The inside being a part of one category and the outside being the other category with a relation in term of brain functors. The diagrams commute signifying that the symmetry of the system remains intact.

- If we are able somehow to group the overall dynamics of the objects inside cells then implementing group theory can further ease the case complexity. If 'n' objects inside forms a group then however it plays its role but the symmetry always remains intact. It further simplifies the brain system. Since the path is already specified and now it forms a group then we can have few universal flips that form the fundamental moves through which we can have almost any configuration of the system, as in case of square we have only 8 moves and it can yield any configuration of the system. So once we are able to implement groups it gets convenient to understand and reduce the complex system.

- The group will be defined under some commutation operation (**.**) that define rules of the action and may be explored further.

- We may also try to link persistent homology in terms of Creation-annihilation operator. The philosophy of persistence is dependent on the overall final structure, where at each interval of time the structure 'glues' together and at final time we get a coherent structure defined via persistence. Once there is a possible link then we may also assign Hamiltonian to the system and then can derive many other aspects of neural system.

- Theories of Schwarz type may be defined through a local action principle independent of metric structure as the underlying manifold, Like TQFT of Schwarz type is 2+1 dimension pure Chern-Simon theory, whether for abelian or non abelian gauge group. I got the idea of linking dynamical chiral symmetry breaking in quantum non abelian Yang Mills Theory from MR and EM paper on topological mean field theory of data. So at the quantum level such topological contribution of action has important physical consequences for gauge field configuration of non trivial topology. On a 4 D manifold and defining a Lie group G, the gauge invariant dynamics of YM theory is defined by Lagrangian density, having an independent part of metric and a dependent part with lowering and raising of manifold. So this definition of Action principle induces gauge field $A^{\alpha}_{\mu}$ scaled by gauge coupling constant $\epsilon$, namely $A^{\alpha}_{\mu} = \epsilon A^{\alpha}_{\mu}$ and with $\epsilon \to \infty$ leads to purely topological field theory of which action is metric independent and purely of topological character. This mapping can make our brain dynamics as metric independent as brain and neuronal dynamics is self felt and deductive system and secondly as per Hilbert viewpoint this condition makes topological character prevalent and hence signifies the completeness of the theory.

- The group formalism of the theory under a defined composition can be used to answer the notion of adaptability. An element is related in terms of belongingness to some category under some composition and then however the dynamics change but it still has to traverse the same arrow. On changes it may have its belongingness to other group and hence adaptability.

- So far three ingredients had been defined to a very little extend: the action, base space and the gauge group that can be utilized for a self-consistent TDFT.

- An action : having typically two parts: a free part ( the combinatorial Laplacian, Generating topological invariants) and the interacting part (representing data Manipulating program: Simulation and Deconstruction)
- A base space and a fibre bundle over it, fibres are replica of the moduli space of Group of operations (X) of the algebra.
- a gauge group, being the simplicial equivalent of the mapping class group, set of transformations into itself preserving topology (as defined above in continuum case with diffeomorphism).

- The correlation function above can be classified by TDFT through topological invariants (genus g of the data manifold) of the space and the Homotopy class of propagators.

- In my previous sent proposal as stated by MR that iff the QTDFT could be mapped on a physical system then idea of mapping data to condensed matter physics could work. In this realm, Hubbard Model was not sufficient due to complexity of brain dynamics. We may also see if we can reformulate the case in our new platform and may add more to brain dynamics

- From the gauge metric free independence above we may propose that natural though is actually circular in nature and when we have the base case it may help us in designing the higher hierarchical levels of our proposed idea. The action and topological attribute to the system with circular dynamics can be really 'felt' at the quantum level and is significant in this area. It is an abelian case of 0+1 dimensions where holonomy parameter $\lambda$ combines with topological term normalization giving unique contributions, possible only via circle topology. It may need more to explore so neglected here.

# Chapter 8

# Conclusions and Open Problems

*"I can't change the direction of the wind, but I can adjust my sails to always reach my destination."*——Jimmy Dean

## 8.1 Conclusions

There are many conclusions and need to be drawn out of the graphs, tables and mathematical framework. The reason of the diverse conclusions is that till now there is no work almost done directly to rephrase the data analytics in terms of physical systems. Many of the methods had been utilized so far from cloud computing to high software and hardware systems so that the challenges of the big data can be addressed. There is a huge gap between physics and computer science and that there is no coherent mathematical structure for computer science or data analytics. The main tool that has been utilized in this thesis for having a new understanding of data from a new prism is that of reductionism. Some of the points and conclusions are discussed as:

1. We reduced the system or space of big data in a physical scenario, termed as a non linear, dynamic and complex system. We initiate a theoretical prototype for Big data analytics based on its nature and characteristics, conceptually driven by Quantum Field theory and utilizing the physics of Condensed Matter Theory. Volume, Velocity and Complexity of Big data is reduced to a complex system in the language of Hamiltonian formalism. The dynamics of each data point and its interrelation in this complex system is studied, eventually defined in a framework that is in concordance with Quantum Condensed Matter - The Hubbard Model. A first step towards future of Information and new paradigm and epistemological shift in dealing with big data. Dimensional reduction of big data from infinite points in space to a single body problem via Hubbard-Anderson framework has been proposed. The algorithm is expected to decrease the time complexity from exponential to polynomially solvable in $O(n \log n)$, depending upon self consistency condition. This system evolves with time and its evolution and dynamics is expressed in terms of Green's function and action with slight mathematical rigor. We formulate a new way of understanding Information being physical and a roadway where Information meets Quantum Condensed matter, expected to open new understanding in Machine Learning, Data Mining and Neural networks.

2. It opens a way for interdisciplinary interactions; in our case we gave some instinct of utilization of quantum condensed matter physics, many body generalization and quantum field theory in understanding the information and implementation in big data analytics. It can be an initiation for new understanding of big data and its related aspects.

3. Once we defined the quantum formulation of big data and that it can be understood via quantum mechanics and quantum field theory then it instilled many aspects to it. The central question was that of simulation. How could we simulate this system and that too when it is quantum in nature? It is a chaotic and frustrated system with non linearity, so Monte Carlo was not sufficient. We carried out laser simulation and encoded information in electron and its excitation as qubit and studied the effect and responses of the changing parameters. There is a correspondence between the laser and the quantum adiabatic evolution and we mapped their correspondence in a well coherent manner and when the system was able to get emulated and simulated then it became easier to deduce the time complexity for big data. Gap condition of quantum adiabatic evolution was analyzed and then we optimized the algorithm by setting all the parameters for optimization.

4. The results give a way towards a new trend that is presently being explored much in the research: a trend towards topology. Topology is encoding information in the topology of space that remains invariant to perturbations, unlike qubits that looses quantumness via interaction with the environment and noise factor. So the quantum computer philosophy is yet in the way to evolve and that topology is going to play a greater role in making qubit more dynamic, realizable and increasing its life time without getting prone to noise and other disturbances.

5. After simulation it was shown that the time complexity of the algorithm used takes about $O(\log n)$ time for global optimization and that tunnelling truncates to zero having no realization in our system.

6. We also stated two applications of our approach: one is in random number generation that was concluded from the randomize behaviour of the qubit in the pulse width. The randomization can be utilized in the RNG and can make the system more equiprobable and unpredictable. The other aspect is that in neural networks, mapping quantum matter to neural networks. The neural state network was mapped to fermionic state, neural state superposition to fermionic braiding, synergetic order parameter to partition function, Spacio-temporal integration of neural signals to Feynman-Schrödinger equation and green's propagator to learning Hebb's rule.

7. It concludes the need of a mathematical framework for data analytics and once it can be formulated can lead us to new solutions to Artificial intelligence, neural networks and data mining.

## 8.2 Open Problems

- The basic notion is that of 'data space'; a space where each point is a datum [N.B. : not a single (qu-)bit]. Let me denote this space as DS.

- Contrary to the state spaces of quantum mechanics (typically Hilbert spaces, possibly dimensional), DS is not a vector space: even though a datum can be represented as a string of symbols belonging to some numerical field K, reminding us of vectors, inner products, norm and components of such 'vectors' have no meaning whatsoever, and – above all – a vector space is a space in which a linear combination of two elements (vectors) is an element of the space itself: a linear combination of two data is not a new datum, but simply the two data again. DS however can be shown to be a topological space.

- Therefore DS must be metric-free and coordinate-free.

- For DS a topological space (in the sense of Grothendieck, namely equipped with a given notion of 'proximity'), information is contained in its correlation patterns, that, for example, a topological field theory is able to pick-up.

- Indeed topology encodes most of the information that data hide, which is global, not local. Therefore a global method is necessary: techniques such as 'message passing' or 'believe propagation' are not sufficient; they provide only local information, related to nearest neighbour relations.

- DS is discrete, compact, and finite-dimensional; it can be naturally embedded in a simplicial complex (possibly by a persistent homology filtration).

- Topological Data Field Theory (TDFT) does not need to be quantum (of course it would acquire efficiency if it was, but quantization may be an almost impossible conceptual task).

- A self-consistent TDFT requires a number of ingredients:
    1. an 'action' – formally analogous to the actions of physics – typically with two parts: a free part (the 'combinatorial Laplacian', that generates the topological invariants) and an interaction part (representing a data manipulation program P);

    2. a base space B – that we identify with DS – and a fibre bundle F over it, whose fibres are noting but replicas of the moduli space of the group of operations, say GP, of the process algebra associated with P;

3. a 'gauge group' G = GP ∧ GMC ; GMC being the (simplicial) equivalent of the 'mapping class group', i.e., the set of all transformations of DS into itself that preserve the topology (in the continuum case, it would be the group of diffeomorphism modulo isotropy of the base space) and ∧ denoting 'semi direct product'. G turns F into a G bundle.

- The correlations in DS are classified by the TDFT through topological invariants (e.g., genus $g$ of the data manifold) of the data space and the Homotopy class of the propagators.

- One could in principle construct a quantum TDFT (Q-TDFT) in which the ground state gives the result of the data manipulation (≡ calculation) performed by P (e.g., mining).

- Iff the Q-TDFT could be mapped on a physical system your idea of quantum matter meeting data analytics could work. The Hubbard-like model you propose is for sure not sufficient, because – as all condensed matter systems – it is based on two-body interactions, whereas in data space the coupling is of 'many-body'-type, as each simplex of the complex implies, and moreover each electron is at most each a single qu-bit. Even describing a single point in data space would require an extremely large number of electrons, mutually interacting in a nonlinear way essentially impossible to model.

- In the context of TDFT one can think of a 'mean-field' approach simply keeping into account the homology groups only up to a given order.

- There is a need to encode information physically and its implementation is far back than theory

- Simulation tools and mapping of big data to some emulated model is still not well defined.

- Tunnelling parameter is not feasible in the simulation as it truncates to zero for sensitive values so it may be open question to understand the efficiency via tunnelling.

# Bibliography

[1] Krishna V. Palem and Lingamneni Avinash, *What to do about the end of Moore's law, probably*, DAC Design Automation Conference 2012, **924-929**

[2] Laszlo B. Kish, *End of Moore's law: thermal (noise) death of integration in micro and nano electronics* Physics Letters A 305 (2002) **144–149**

[3] John D. Sterman, *All models are wrong: reflections on becoming a systems scientist*, System Dynamics Review Vol. 18, No. 4, (Winter 2002): **501–531** DOI: 10.1002/sdr.261

[4] Xindong Wu, Xingquan Zhu, Gong-Qing Wu, Wei Ding, *Data Mining with Big Data, IEEE Transactions on Knowledge and Data Engineering* Volume: 26, Issue: **1**, Jan. 2014, **97 – 107**

[5] Genoveva Vargas-Solar, Javier Alfonso Espinosa Oviedo, Jose-Luis Zechinelli-Martini. *Big continuous data: dealing with velocity by composing event streams, Big Data Concepts, Theories and Applications*, Springer Verlag, 2016, 978-3-319-27763-9.

[6] Ben W Reichardt the Quantum Adiabatic Optimization Algorithm and Local Minima Berkley University

[7] Pulin and B J Berne Quantum Path Minimization: *An efficient method for Global Optimization* Journal of Chemical Physics Vol **118**, (15 Feb 2003)

[8] Kuk- Hyun-Rau and Jung Hwan *Kim Quantum Inspired Evolutionary Algorithm for class of Combinatorial Optimization* IEEE Transaction on Evolutionary Computer Vol **6**: No: 6, (Dec 2002)

[9] Matthias Stefen, Win Von Dam et al *Experimental Implementation for Adiabatic Quantum Optimization Algorithms* Phys. Rev. Lett 90,067903 (14 Feb **2003**)

[10] Laszlo Pal PhD thesis bibitem20 Ben W Reichardt *the Quantum Adiabatic Optimization Algorithm and Local Minima*

[11] Jean Michel et al *Hybrid Methods using Genetic Algorithms for Global Optimization* IEEE transactions on system, man and Cybernetics Part B Vol **26** (2 April 1996)

[12] C GE Bender et al *A Stochastic Method for Global Optimization Mathematical Programming* **22** North Holland Publishing Company

[13] R Horst. *On the Global Minimization of Global Function Introduction and Survey* Springer

[14] Edward Farhi, *Quantum Adiabatic Evolution Vs Simulated Annealing* Phy Rev X **6** 031010

[15] Djurdje. Cnjonic Jacek. Klinowski1 *Taboo Search: An Approach to multiple minima problem* Science Vol.267, Issue **5198** (03 Feb 1995)

[16] Yang. Han Lee et.al Global *Optimization: Quantum annealing with path integral Monte Carlo* J. Phys. Chem. A 2000, 104 (1), **86**{95 (Dec 16, 1999)

[17] Pulin and B. J Berne Quantum Path Minimization: *An efficient method for Global Optimization* Journal of Chemical Physics Vol **118** (Feb 15 2003)

[18] R. v Pappu *Analysis and Application of Potential energy smoothing and Search method for Global Optimization* J. Phys. Chem. B **102**, 9725-9742 (1998)

[19] Ji Qiang and Chad Mitchell, *an Adaptive Unified evolution Algorithm for Global Optimization* Berkley University

[20] Momin Jameel et al *A Literature survey of benchmark functions for global optimization problems* arXive: 1308.4008v1 (19 Aug 2013)

[21] Edward. Farhi, *Quantum Computation by Adiabatic Evolution* https://arxiv.org/abs/quant-ph/0001106 (Jan 28, 2000)

[22] Edward. Farhi Michael. Sipser *Quantum, Adiabatic Evolution Algorithm applied to random instances of NP hard problems* Science 292, **5516** (20 April, 2001)

[23] Edward Farhi, Jeffrey Goldstone, Sam Guttmann, *A Numerical Study of the Performance of a Quantum Adiabatic Evolution Algorithm for Satisfiability*, [arXive: quant-ph/0007071]

[24] Edward Farhi, Jeffrey Goldstone, Sam Guttmann, *Quantum Adiabatic Evolution Algorithms versus Simulated Annealing*, [arXive: quant-ph/0201031]

[25] Karger, D., Lehman, E., Leighton, T. Panigrahy, R., Levine, M., & Lewin D. (1997). *Consistent hashing and random trees.* Proceedings of the twenty-ninth annual ACM symposium on Theory of computing - STOC '97 **654–663** New York, New York, USA: ACM Press. doi:10.1145/258533.258660

[26] Bloom, B. H. (1970). *Space/time trade-offs in hashing coding with allowable errors.* Communications of the ACM, 13(7), **422–426**. doi:10.1145/362686.362692

[27] Hewitt, C., Bishop, P & Steiger, R. (1973). *A universal modular ACTOR formalism for artificial intelligence*, **235–245**.

[28] Dean, J., & Ghemawat, S. (2008). *Map Reduce: Simplified Data Processing on Large Clusters.* (L. P. Daniel, Ed.)*Communications of the ACM*, *51*(1), **1–13**. doi:10.1145/1327452.1327492

[29] Jaskelainen, P. O., De La Lama, C. S., Huerta, P., & Takala, J. H. (2010). *Open CL-based design methodology for application-specific processors*. 2010 International Conference

on Embedded Computer Systems: Architectures, Modelling and Simulation **223–230**. IEEE. doi:10.1109/ICSAMOS.2010.5642061

[**30**] Casella, G., & George, E. I. (1992). *Explaining the Gibbs Sampler*. The American Statistician, 46(3), **167**. doi: 10.2307/2685208

[**31**] Cattell, R. (2011*). Scalable SQL and NoSQL data stores. ACM SIGMOD Record*, *39*(4), **12**. doi:10.1145/1978915.1978919

[**32**] Han, J. (2011). Survey on NoSQL database. 2011 6th International Conference on *Pervasive Computing and Applications* **363–366**, IEEE. doi:10.1109/ICPCA.2011.6106531

[**33**] Brewer, E. A. (2000). *Towards robust distributed systems (abstract). Proceedings of the nineteenth annual ACM symposium on Principles of distributed computing -* PODC **7** New York, New York, USA: ACM Press. doi:10.1145/343477.343502

[**34**] Stonebraker, M. (2010). *SQL databases v. NoSQL databases.* Communications of the ACM, 53(4), **10**. doi:10.1145/1721654.1721659

[**35**] DeCandia, G., Hastorun, D. Jampani, M., Kakulapati, G., Lakshman, A., Pilchin, A., Sivasubramanian, S., et al. (2007). Dynamo. *ACM SIGOPS Operating Systems Review*, *41*(6), **205**. doi:10.1145/1323293.1294281

[**36**] Sumbaly, R., Kreps, J., Gao, L., Feinberg, A., Soman, C., & Shah, S. (2012). *Serving large-scale batch computed data with project* Voldemort. FAST'12 Proceedings of the 10[th] USENIX conference on File and Storage Technologies

[**37**] Hewitt, E. (2011). Cassandra: *The Definitive Guide*. Cambridge: O'Reilly Media

[**38**] Chang, F., Dean, J., Ghemawat, S., Hsieh, W. C., Wallach, D. A., Burrows, M., Chandra, T., etal. (2008). big table*: A Distributed Storage System for Structured Data. Sports Illustrated,* 26(2), **1–26**. doi:10.1145/1365815.1365816

[**39**] Bonnet, L., Laurent, A., Sala, M., Laurent, B., & Sicard, N. (2011). Reduce, You Say: *What NoSQL Can Do for Data Aggregation and BI in Large Repositories*. 2011 22[nd] International Workshop on Database and Expert Systems Applications **483–488** IEEE. doi:10.1109/DEXA.2011.71

[**40**] Banker, K. (2011). *MongoDB in Action. Greenwich, CT, USA*: Manning Publications Co.

[**41**] Angles, R., & Gutierrez, C. (2008). *Survey of graph database models*. *ACM Computing Surveys*, *40*(1), **1–39**. doi:10.1145/1322432.1322433 Robinson, I., Webber, J., & Eifrem, E. (2013). *Graph Databases* (Early Release). O'Reilly Media.

[**42**] Fowler, M., & Sadalage, P. J. (2012). *Introduction to Polyglot Persistence: Using Different Data Storage Technologies for Varying Data Storage Needs* (p. **192**). Boston, MA: Addison-Wesley Professional.

[**43**] Thomas. H. Cohen et al *Introduction to algorithms,* Cambridge University Press

[**44**] Mohammed J. Zaki, Wagner Meira JR, *Data Mining, Fundamental Concepts and Algorithms*, Cambridge University Press **2014**

[**45**] Stephen Kaisleri, Frank Armour J. Alberto Espinosa William Money, *Big Data: Issues and Challenges Moving Forward* **2013** 46th Hawaii International Conference on System Sciences

[**46**] Cristian Calude, Giuseppe Longo, Lois desdieux, *The Deluge of Spurious Correlations in Big Data*, des hommes et de la nature, Nantes, France. Foundations of Science, **1** - **18**, 2016

[**47**] Volkan Cevher, Stephen Becker, and Mark Schmidt, *Convex Optimization for Big Data* arXive: 1411.0972v1 [math.OC] 4 Nov **2014**

[**48**] Nada Elgendy and Ahmed Elragal, Big Data Analytics: A Literature Review Paper In: Perner P. (eds) *Advances in Data Mining. Applications and Theoretical Aspects*. ICDM 2014 Lecture Notes in Computer Science, Vol **8557**. Springer, Cham

[**49**] Albert Bifet, Wei Fan, *Mining Big Data: Current Status, and Forecast to the Future*, SIGKDD Explorations, Vol **14** Issue **2**

[**50**] Abhishek Pandey, Ramesh V. *Quantum computing for big data analysis*. Indian Journal of Science, 2015, **14(43)**, **98-104**

[**51**] C.G.E Boender, A.H.G Rinnooy Kan, G.T Timmer, L Stougie, *A Stochastic Method for Optimization, Mathematical Programming* **22** (1982) **125-140**

[**52**] Ibrahim Abaker, Targio Hashem, IbrarYaqoob, NorBadrulAnuar, Salimah Mokhtar Abdullah Gani, Samee Ullah Khan, *The rise of "big data" on cloud computing: Review and open research issues,* Information Systems 47 (2015) **98-115**

[**53**] C.L. Philip Chen, Chun-Yang Zhang, *Data-intensive applications, challenges, techniques and technologies: A survey on Big Data Information Sciences* 275 (2014) **314–347**

[**54**] Jianging. Fan *Challenges of Big Data Analysis* Natlsci Rev 1 (2) **293-314**, (2004)

[**55**] Caesar Wu, Rajkumar Buyya, Kotagiri Ramamohanarao, *and Big Data Analytics = Machine Learning + Cloud computing*; [arXive: 1601.03115]

[**56**] Bogdan Oancea, Raluca Mariana Dragoescu, *Integrating R and Hadoop for Big Data Analysis*, Romanian Statistical Review, **2**(2014); [arXive: 1407.4908]

[57] A. G. Ramm, C. Van, *Representation of big data by dimension reduction*, [arXive: 1702.00027]

[58] Larry Wassermann *Topological Data Analysis*, [arXive: 1609.08227]

[59] Edelsbrunner, H., Letscher, D., and Zomorodian, A. *Topological persistence and simplification*, Discrete Comput. Geom., 28:511–533

[60] M Rasetti, E Merelli *the Topological Field Theory of Data: a program towards a novel strategy for data mining through data language*, J.Phys:Conf.Ser. **626** 012005

[61] Patrick Rebentrost, Masood Mohseni and Seth Lloyd. *Quantum Support Vector Machine for Big Data Classification* phys.Rev.lett

[62] Jacob Biamonte, Peter Wittek et al *Quantum Machine Learning* 1611.09347v1 (28 Nov 2016)

[63] Scott Aaronson *Quantum Machine Learning: Read the fine art* The European Physics Journal

[64] Subhash Kak *Quantum Mechanics and Artificial Intelligence* DOI10.10071978-1-84628.943-95

[65] Marvin Weinstein Strange Bedfellow: *Quantum Mechanics and Data Mining* phys.Rev.lett

[66] Peter Wittek Quantum Machine Learning, Cambridge University Press

[67] X D Cai et al Entanglement based Machine Learning on a Quantum Computer Phys.Rev.lett 1409.7770v3 (18 Feb 2015)

[68] Seth Lloyd et al Quantum Algorithm for Supervised and unsupervised Machine Learning Phys.Rev.lett **117** (4 Nov 2013)

[69] Peter. Wittek, *Quantum Machine Learning: What Quantum Computing Means to Data Mining*, Academic Press (2014)

[70] Michael. A. Nielson and Isaac. L. Chuang, *Quantum Computation and Quantum Information*, Cambridge University Press (2010)

[71] Patrick. Rebentrost, Masoud. Mohseni, Seth. Lloyd, *Quantum support vector machine for big data classification*, Phys. Rev. Lett. **113**, 130503 (2014)

[72] Peter. W. Shor, *Polynomial-Time Algorithms for Prime Factorization and Discrete Logarithms on a Quantum Computer*, SIAM J. Sci. Statist. Comput. **26**, 1484 (1997)

[73] Edward. Farhi and Jeffrey. Goldstone *A Quantum Approximate Optimization Algorithm*, [arXive: 1411.4028]

[**74**] Edward Farhi and Jeffrey Goldstone *A Quantum Approximate Optimization Algorithm Applied to a Bounded Occurrence Constraint Problem*, [arXive: 1412.6062]

[**75**] Elizabeth Crosson, Edward Farhi, Cedric Yan-Yu Lin, Han-Hsuan Lin and Peter Shor, *Different Strategies for Optimization Using the Quantum Adiabatic Algorithm*,[arXiv:1401.7320]

[**76**] Jacob Biamonte, Peter Wittek, Nicola Pancotti, Patrick Rebentrost, Nathan Wiebe, Seth Lloyd *Quantum Machine Learning*, Nature 549, 195-202 (2017)

[**77**] Marwin. Weinstein *Strange Bedfellows: Quantum Mechanics and Data Mining*, SLAC-PUB-13832

[**78**] M. Schuld, I. Sinayskiy, F. Petruccione, *The quest for a Quantum Neural Network*, [arXive: 1408.7005]

[**79**] Charles H. Bennet, Gilles Brassard *Quantum cryptography: Public key distribution and coin tossing*, Theoretical Computer Science **1**, 560(2014) 7-11

[**80**] S. Roy, L. Kot, and C. Koch, *Quantum Databases*, In Conference on Innovative Data Systems Re- search (CIDR), 2013

[**81**] Kristen L. Pudenz, Daniel A. Lidar, *Quantum adiabatic machine learning*, [arXive: 1109.0325]

[**82**] A.P. Kirilyuk, *Complex Dynamics of Real Quantum, Classical and Hybrid Micro-Machines*

[**83**] Seth Lloyd, Silvano Garnerone, Paolo Zanardi *Quantum algorithms for topological and geometric analysis of big data*, Nature Communications **7**, 10138(2016)

[**84**] Olga Kurasova, Virginijus Marcinkevicius, Viktor Medvedev, Aurimas Rapecka and Pavel Stefanovic, *Strategies for Big Data Clustering*, IEEE 26th International Conference on Tools with Artificial Intelligence 2014

[**85**] Dieter Vollhardt, *Dynamical Mean-Field Theory of Electronic Correlations in Models and Materials*, AIP Conference Proceedings vol. 1297 (American Institute of Physics, Melville, New York, 2010), p. 339

[**86**] Elena Agliari, Adriano Barra, Andrea Galluzzi, Daniele Tantari, Flavia Tavani, *A walk in the statistical mechanical formulation of neural networks*, [arXiv:1407.5300]

[**87**] J Hopfield, *Neural networks and physical systems with emergent collective computational abilities*, PNAS**79**, 8(1982) pp.2554-2558,

[**88**] Liu, S., Ying, L. and Shakkottai, S., *Influence maximization in social networks: An Ising-model-based approach*, In Proc.48th Annual Allerton Conference on Communication, Control, and Computing (2010)

[**89**] Rolf Landauer, *The physical nature of information*, Phys.Lett.A **217** (1996), 188-193

[**90**] Bei Zeng, Xie Chen, Duan-Lu Zhou, Xiao-Gang Wen, *Quantum Information Meets Quantum Matter– From Quantum Entanglement to Topological Phase in Many-Body Systems*, [arXiv:1508.02595]

[**91**] Michael A. Nielsen, *the Fermionic canonical commutation relations and the Jordan-Wigner trans- form*, Notes on Jordan-Wigner Transformation (Jul 29, 2005)

[**92**] Alan Morningstar, Roger G. Melko, *Deep Learning the Ising Model near Criticality*, [arXive: 1708.04622]

[**93**] Vá clav Janiš, *Introduction to Mean-Field Theory of Spin Glass Models*, [arXive: 1506.07128]

[**94**] H. J. Kappen, F. B. Rodr´ıguez, *Boltzmann Machine learning using mean field theory and linear response correction*, . In M. S. Kearns, S. A. Solla, and D. A. Cohn, editors, Advances in Neural Infor mation Processing Systems 11, pages 280–286. MIT Press, Cambridge, MA, 1999

[**95**] Haiping Huang, Taro Toyoizumi, *Advanced Mean Field Theory of Restricted Boltzmann Machine*, Phys. Rev. E 91, 050101 (2015)

[**96**] Haiping Huang, *Mean-field theory of input dimensionality reduction in unsupervised deep neural networks*, [arXive: 1710.01467]

[**97**] Victor Bapst, Laura Foini, Florent Krzakala, Guilhem Semerjian, Francesco Zamponi, *The Quantum Adiabatic Algorithm applied to random optimization problems: the quantum spin glass perspective*, Physics Reports 523, 127 (2013)

[**98**] Davide Venturelli, Salvatore Mandra`, Sergey Knysh, Bryan O'Gorman, Rupak Biswas, Vadim Smelyanskiy, *Quantum Optimization of Fully-Connected Spin Glasses*, Phys. Rev. X 5, 031040 (2015)

[**99**] Gasper Tkacik, Elad Schneidman, Michael J. Berry II, William Bialek, *Spin glass models for a net- work of real neurons*, [arXiv:0912.5409]

[**100**] Louis-Franc¸ois Arsenault, O. Anatole von Lilienfeld, Andrew J. Millis, *and Machine learning for many- body physics: efficient solution of dynamical mean-field theory*, [arXive: 1506.08858]

[**101**] Juan Carrasquilla, Roger G. Melko, *Machine learning phases of matter*, [arXive: 1605.01735]

[**102**] Jeff Byers, *The physics of data*, Nature Physics **13**, 718-719 (2017).

[**103**] Toshiyuki Tanaka *Mean-field theory of Boltzmann machine learning*, Phys. Rev. E **2**, 58(1998)

[**104**] Mitja Perus, *Neural Networks as a basis for Quantum Associative Networks*, Neural Network World, vol 10, (2000) pp.1001-1013.

[**105**] P C Bressloff, *a Green's function approach to analysing the effects of random synaptic background activity in a model neural network*, J. Phys. A: Math. Gen. **27**, 4097(1994)

[**106**] Bertrand Clarke, Ernest Fokoue, Hao Helen Zhang, *Principles and Theory for Data Mining and Ma- chine Learning*, Springer-Verlag New York (2009), [DOI: 10.1007/978-0-387-98135-2]

[**107**] Gentle, Simon, Adam (2013) *Holography, blackholes and condensed matter physics*, Durham theses, Durham University. Available at Durham E-theses Online: http://etheses.dur.ac.uk/7286/

[**108**] M. Gams, Marcin Paprzycki, Xindong Wu, *Mind Versus Computer - were Dreyfus and Winograd right?*, Volume 43 of Frontiers in artificial intelligence and applications, ISSN 0922-6389, IOS Press, 1997.

[**109**] Philip Van Loocke, *The Physical Nature of Consciousness* (Advances in Consciousness, Series a, v.29), John Benjamins Publishing Company (2001)

[**110**] P. W. Anderson *Localized Magnetic States in Metals*, Phys. Rev. **124**, 41(1961)

[**111**] Jaime David Gómez Ramírez, *New Foundation for Representation in Cognitive and Brain Science: Category Theory and the Hippocampus*, Departamento de Automática, Ingeniería Electrónica e Informática Industrial, **2010** Thesis

[**112**] Peter Smith, *an Introduction to Gödel's Theorems*, Faculty of philosophy, University of Cambridge, Dec 12 2005

[**113**] Aaronson, S., 2015. *Quantum machine learning: read the fine print*. Nature Physics, 11(4), pp. **291-293**

[**114**] *Quantum entanglement between an optical photon and a solid-state spin qubit*, E. Togan, Y. Chu, A. S. Trifonov, L. Jiang, J. Maze, L. Childress, M. V. G. Dutt, A. S. Sørensen, P. R. Hemmer, A. S. Zibrov1 & M. D. Lukin

[**115**] *Millisecond Coherence Time in a Tunable Molecular Electronic Spin Qubit*, Joseph M. Zadrozny et al ACS Cent Sciv.1(9); 2015 Dec 23 PMC4827467

[**116**] *A short introduction to topological quantum computation* Ville T. Lahtinen1 and Jiannis K. Pachos Sci post phy , 3 021,2017

[**117**] Barry Coyle, D., Guerra, D. V. & Kay, R. B., 1995. *An interactive numerical model of diode-pumped, Q-switched/cavity-dumped lasers.*. Journal of Physics D: Applied Physics, 28(3), p. 452.

[**118**] Kay, R. B. & Waldman, G. S., 1965. *Complete Solutions to the Rate Equations Describing Q-Spoiled and PTM Laser Operation.*. Journal of Applied Physics, 36(4), pp. 1319-1323.

[**119**] Djurdje, Jacek, C. & Klinowski, 1995. *Taboo search: an approach to the multiple minima problem. Science,* 267(5198), pp. 664-666.

[**120**] Kak, S., 2007. *Quantum Mechanics and Artificial Intelligence.* London.

[**121**] Liu, S., Ying, L.and Shakkottai, S., Influence maximization in social networks: *An Ising-model-based approach,* In Proc.48th Annual Allerton Conference on Communication, Control, and Computing (2010)

[**122**] Alan Morningstar, Roger G. Melko, *Deep Learning the Ising Model Near Criticality,*[arXiv:1708.04622]

[**123**] Peter Smith, *Introduction to Gödel's Theorems* University of Cambridge Press

[**124**] Matteo Rucco, Adane Letta Mamuye, Marco Piangerelli, Michela Quadrini, Luca Tesei and Emanuela Merelli, *Survey of TOPDRIM applications of Topological Data Analysis* 2016

[**125**] Gurjeet Singh, Facundo Memoli, and Gunnar E. Carlsson *Topological Methods for the Analysis of High Dimensional Data Sets and 3D Object Recognition*. In Symposium on Point Based Graphics, Prague, Czech Republic, 2007, 91-100, 2007.

[**126**] Mario Rasetti and Emanuela Merelli, *Topological Field Theory of Data: Mining Data beyond Complex Networks,* 978-1-107-12410-3 — Advances in Disordered Systems, Random Processes and Some Applications, 27 March 2017 doi.org/10.1017/9781316403877.002

[**127**] Emanuela Merelli and Mario Rasetti*, Non locality, topology, formal languages: new global tools to handle large data sets*, International Conference on Computational Science, ICCS 2013, Procedia Computer Science 18 ( 2013 ) 90 – 99

[**128**] Emanuela Merelli, Marco Pettini and Mario Rasetti**,** *Topology driven modelling: the IS metaphor*, Nat Comput (2015) 14:421–430, 24 June 2014, DOI 10.1007/s11047-014-9436-7

[**129**] Luca Chirolli, Guido Burkand, *Solid State Qubit*, [arXive: **0809.4716**]

[**130**] A. Kitaev and J Preskill, *Topological entanglement entropy*, Phys.Rev.lett**, 96,110404** (**2006**)

[**131**] I Bloch et al, *Quantum Simulation with ultra cold Quantum Gases*, Nat.Phys **8**, **267 (2012)**

[**132**] Niklas M Gerges, Larz Firtz and Drik Schuricht, *Topological order in the Kitaev/Majorana Chain in the presence of disorder and interaction*, [arXive: **1511.02817**]

[**133**] P. Y. Lum, G. Singh, A. Lehman, T. Ishkhano, M. Vejdemo-Johansson, M. Alogappan, J. Carlsson and G. Carlsson, *Extracting insights from the shape of complex data using topology*, Scientific Reports, DOI: 10.1038/srep01236

[**134**] Casey Bowman, *Big Data Simulation: Traffic Simulation based on sensor data*, Proceedings of the **2014** Winter Simulation Conference

[**135**] Art Hobson, *There are no particles, there are only fields*, American Journal of Physics

[**136**] Juha M Kreula et al, *Few Qubit Quantum- Classical simulation of strongly correlated lattice fermions*, EPJ Quantum Technology, 3-11, 2016

[**137**] ]Robert B Winder, *Can UML Model quality be quantified?*, January 14 2015